\documentclass[notitlepage,a4paper,twocolumn,superscriptaddress,nofootinbib,longbibliography]{revtex4-1}
\RequirePackage[colorlinks,citecolor=blue,urlcolor=blue,linkcolor=blue]{hyperref}
\usepackage{amsmath}
\usepackage{amsfonts}
\usepackage{amssymb}
\usepackage[latin1,utf8]{inputenc}
\usepackage{xcolor,cancel}
\usepackage[normalem]{ulem}
\newcommand{\stkout}[1]{\ifmmode\text{\sout{\ensuremath{#1}}}\else\sout{#1}\fi}
\RequirePackage{flushend}
\usepackage{mathptmx}
\usepackage{graphicx}
\hypersetup{pdftitle=Cosmological dynamics and bifurcation analysis of the   general non-minimal coupled scalar field models,pdfauthor={Wompherdeiki Khyllep, Jibitesh Dutta}}

\usepackage[all]{hypcap}

\usepackage{latexsym}
\usepackage{graphicx}
\usepackage{color}
\usepackage{subfigure}

\newtheorem{stmt}{Statement}

\newtheorem{defn}{Definition}

\begin{document}
	
	\title{Cosmological dynamics and bifurcation analysis of the   general non-minimal coupled scalar field models}

	\author{Wompherdeiki Khyllep}
	\email{sjwomkhyllep@gmail.com}
	\affiliation{Department of Mathematics, North-Eastern Hill University, Shillong, Meghalaya 793022, India} 
	\affiliation{Department of Mathematics, St. Anthony's College, Shillong, Meghalaya 793001, India}

	\author{Jibitesh Dutta}
	\email{jibitesh@nehu.ac.in}
	\affiliation{Mathematics Division, Department of Basic Sciences and Social Sciences, North-Eastern Hill University,  Shillong, Meghalaya 793022, India}
	\affiliation{Inter-University Centre for Astronomy and Astrophysics, Pune 411007, India }

\begin{abstract}
Non-minimal coupled scalar field models are well-known for providing interesting cosmological features.  These include a late-time dark energy behavior, a phantom dark energy evolution without singularity, an early-time inflationary Universe, scaling solutions, convergence to the standard $\Lambda$CDM, etc. While the usual stability analysis helps us determine the evolution of a model geometrically, bifurcation theory allows us to precisely locate the parameters' values describing the global dynamics without a fine-tuning of initial conditions. Using the center manifold theory and bifurcation analysis, we show that the general model undergoes a transcritical bifurcation, predicting us to tune our models to have certain desired dynamics.  We obtained a class of models and a range of parameters capable of describing a cosmic evolution from an early radiation era towards a late time dark energy era over a wide range of initial conditions. There is also a possible scenario of crossing the phantom divide line. We also find a class of models where the late time attractor mechanism is indistinguishable from a structurally stable general relativity-based model; thus, we can elude the big rip singularity generically. Therefore,  bifurcation theory allows us to select models that are viable with cosmological observations.

\end{abstract}

\maketitle
	
	\section{Introduction}
	
		Non-minimal coupled scalar field models are often used to explain various cosmological observations.   These models naturally arise from the quantum corrections to the scalar field theory and motivated by high energy physics such as superstrings and grand unified theories \cite{Capozziello:1993tr}.  Further, these models provide a natural solution to the problem associated with the energy scale difference between inflation and the Universe's dark energy (DE) era \cite{Cardone:2005aa}.

		Most of the cosmological model's governing equations are nonlinear and pose a severe impediment to extract exact analytical solutions. However, one can infer the global asymptotic behavior described by the cosmological equations using the advanced tools of dynamical systems. The main advantage is that we can represent the Universe's history geometrically. One can also predict the sensitivity of the solution to  initial conditions. Dynamical system methods have been used extensively in cosmology; see \cite{Bahamonde:2019urw, Basilakos:2019dof, Alho:2019jho,Dutta:2017wfd, Zonunmawia:2018xvf, Carloni:2018yoz, Dutta:2017fjw, Dutta:2018xcz,Khyllep:2019odd,Kerachian:2019tar,Leon:2018skk,Khyllep:2021pcu,Paliathanasis:2021egx} for relevant work and \cite{Bahamonde:2017ize} for a  comprehensive review.

		The dynamical system of most cosmological models usually contains parameters. One can determine the system's global dynamics for fixed values of parameters using the formal stability analysis. On the other hand, to understand how the global dynamics changes with a change of parameters, the bifurcation theory plays a crucial role  (see Refs. \cite{seydel2009practical,kuznetsov2013elements,perko2013differential} for detailed information). The dynamical system's nonlinear nature usually leads to vital structures of the solutions, such as bifurcations and chaos.  A more in-depth analysis of such forms is interesting from an observational perspective (e.g., see \cite{Klen:2020kdb}).

		One of the bifurcation theory's novelties is that we can use it to classify the Universe's evolution into two categories: generic and non-generic evolution \cite{Humieja:2019ywy}. While the former occurs for various solutions over a wide range of initial conditions, the latter corresponds to a particular solution for a given initial condition.  The parametric relation associated with a non-generic scenario forms a bifurcation boundary between regions of different generic cases in the parameter space. Non-generic evolution is also exciting but requires fine-tuning of initial conditions. In some cases, generic evolution emerges from non-generic one in the form of bifurcation.  Therefore, bifurcation theory can help extract a class of models describing the observed dynamical evolution irrespective of initial conditions for a wide range of parameters. To have a clear picture of how the bifurcation phenomenon depends on model parameters, one has to use bifurcation diagrams. These diagrams stratify the parameter space into different regions, each with distinct dynamical behavior. The steps involved in the bifurcation analysis are two-fold. Initially, we extract the range of parameters for both the generic and the non-generic evolution. Then, we analyze different qualitative behaviors that arise from each scenario. 
		
		Another novelty of bifurcation theory is that it allows us to identify a structural stable model's emergence from a structurally unstable one. For instance, Szydlowski and Tambor showed that the notion of bifurcation and structural instability could be instrumental in detecting the emergence of the structurally stable $\Lambda$CDM model from the structurally unstable CDM model \cite{Szydlowski:2008rz}. Kokarev further extended a similar analysis to various Friedmann-Robertson-Walker (FRW)-models \cite{Kokarev:2008ba}. Usually, structurally stable models are physically viable and hence fit with most observations. In a 2-dimensional system, Peixoto's theorem completely characterizes the structurally stable vector fields, which guarantee their generic behavior.  Identifying structurally stable models is useful when the prediction of model parameters from the empirical analysis is unsettled.  Thus,  bifurcation theory helps in finding observationally viable models and further endows the usual stability analysis.
		
		In most of the dynamical analysis for the non-minimal coupled scalar field models, the dynamical variables are constructed for a specific case of coupling or potential functions in flat or curved spacetime \cite{Uzan:1999ch, Gunzig:2000ce, Faraoni:2006sr, Carloni:2007eu,Szydlowski:2008in, Maeda:2009js, Jarv:2010zc, Hrycyna:2015eta,Kerachian:2019tar}. However, the analysis for a general non-minimal coupled scalar field model will certainly help us to identify classes of viable models. The extension to a broad class of scalar field potentials and couplings might help us to connect the phenomenological models with some high-energy physical theories. Therefore, it will be scientific and economical to carry out the dynamical analysis for a broad class of coupling functions and potentials. 
		
		We find in the literature that bifurcation phenomena arise naturally in cosmological models. For instance,   Ref. \cite{Kohli:2018lsg} shows that in FRW-models with perfect fluids and the cosmological constant, the expanding and contracting deSitter Universe arise as bifurcation. It is worth mentioning that interesting bifurcation scenarios were reported in the  Randall-Sundrum braneworld model \cite{Goheer:2002bq}, interacting Veneziano ghost DE \cite{Feng:2012wx}, Brans-Dicke model  \cite{Hrycyna:2013yia}, non-minimal coupled scalar field model \cite{Szydlowski:2013sma,Hrycyna:2015eta} etc. Recently,  bifurcation scenarios and chaos were discussed in the context of Ho\v{r}ava-Lifshitz gravity \cite{Hell:2020yyo}, non-minimal coupled scalar field with Ratra-Peebles potential \cite{Humieja:2019ywy},  interacting $f(T)$ gravity \cite{Mishra:2019vnv} and bulk viscous cosmology \cite{Azim:2020yce}. These recent work show that the study of bifurcation is important in cosmology, giving rise to interesting scenarios.
		
		In the non-minimal coupled scalar field context,  interesting bifurcation scenarios were reported for a specific coupling and potential function. For instance, Hrycyna et al.  \cite{Hrycyna:2015eta} obtained a particular bifurcation value of a coupling constant for the case of a constant potential.  Then,   Szydlowski et al. in   \cite{Szydlowski:2013sma} analyzed the phase space's structural stability of a specific coupling model for a broad class of potentials. They found that an exponential potential constitutes a structurally stable model.  Using the bifurcation methods, Humieja et al. in \cite{Humieja:2019ywy}  extract the conditions of model parameters under which a specific non-minimal coupling with Ratra-Peebles potential generically evolve from an early de Sitter to a  late time de Sitter state. The analysis in \cite{Humieja:2019ywy} was performed in the absence of a matter component. We extend the analysis for general coupling and potential functions along with the matter component in the present work.  To meet our objective, we consider a different choice of dynamical variables to encompass a broad class of models. By employing bifurcation methods, we obtain a class of models and pinpoint the range of parameters capable of describing a  cosmic evolution over a wide range of initial conditions from an early radiation era towards a late time DE era. We also found that the system undergoes a transcritical type of bifurcation, which predicts how to tune our models to have certain desired dynamics.
		
		The bifurcation theory's concrete tools have been used extensively in various fields. However, they have not been applied systematically in many cosmological systems, particularly for the non-minimal coupled scalar field. Thus, it is imperative to use bifurcation theory to identify a class of scalar field models describing some of the main generic cosmic evolution. Therefore, the present work serves as an introductory analysis for scalar field models required to test against interesting observational signatures.

		The paper's order is as follows: In Sec. \ref{sec:brief}, we briefly discuss the framework of a non-minimal coupled scalar field model. We follow this by a dynamical system analysis of a non-minimal coupled scalar field model for a broad class of coupling function and potential in Sec. \ref{sec:DS_gen}.  In Sec. \ref{sec:example}, we demonstrate the dynamics by an example using the quadratic coupling functions and the power-law form of potentials. Within this section, we perform the stability analysis of critical points in Sec. \ref{sub:fin} and the discussion on bifurcation scenarios in Sec. \ref{sub:bif}. Lastly, we summarized the work in Sec. \ref{sec:conc}.
	
	\section{Non-minimal coupled scalar field model}	\label{sec:brief}
	We consider a model of a non-minimal coupled scalar field and a barotropic fluid in the present work. Here the non-minimal coupled scalar field is playing the role of DE, while a barotropic fluid is the matter component of the Universe. The total action  is given by \cite{1968IJTP....1...25B,Nordtvedt:1970uv,1970PhRvD...1.3209W} 
	\begin{align}\label{action}
	S=\frac{1}{\kappa^2}\int d^4 x \sqrt{-g} \Big[ \frac{F(\phi)}{2} R-\frac{1}{2}\nabla^a \phi \nabla_a \phi-\ell^{-2}\,V(\phi) \nonumber\\ +\kappa^2 \mathcal{L}_m \Big]\,,
	\end{align}	
	where the integration is taken over a 4-dimensional Lorentzian curved spacetime manifold. In the above action, $\kappa^2$ is the gravitational constant, $R$ is the Ricci scalar, $g$ is the determinant of the spacetime metric $g_{ab}$ ($a, b=0, 1, 2, 3$),  $F$ is the coupling function, $V$ is the potential of a scalar field $\phi$ and $\mathcal{L}_m$ is the matter Lagrangian. We  have used the units where $c=1$ and $\ell$ is the positive  parameter having the dimension of length.   For a fixed scalar field, the above action reduces to the case of general relativity (GR) with a potential playing the role of a cosmological constant. While the case $F(\phi)=1$ of \eqref{action} corresponds to the minimal coupled scalar field, $F(\phi)=\frac{\phi^2}{\omega}$ reduces to the Brans-Dicke gravity limit, with $\omega$ as the Brans-Dicke parameter \cite{Capozziello:2006dj,Carloni:2007eu,Capozziello:1993vs,Brans:1961sx}.  On varying the action \eqref{action} with respect to the metric $g_{ab}$, one can obtain the modified Einstein's field equation as
	\begin{gather}
	F(\phi) G_{ab}+\frac{1}{2} \nabla_a \phi \nabla_b \phi-\frac{1}{4} g_{ab}\nabla^c\phi\nabla_c\phi+\frac{1}{2} g_{ab}\, \ell^{-2}\,V(\phi)\nonumber\\
	-\nabla_a\nabla_b F(\phi)+g_{ab} \square F(\phi)=-\kappa^2 T_{ab} \, , \label{EFE}
	\end{gather}
	where $G_{ab}$ is the Einstein tensor, $\square\equiv \nabla_a \nabla^a$ with $\nabla_a$ as the covariant derivative with respect to the metric and  $T_{ab}$ is the matter energy-momentum tensor given by
	\begin{align}
	T_{ab}=p_m\,g_{ab}+(\rho_m+p_m) u_a u_b\, .
	\end{align}

In the above equation, $\rho_m$ and $p_m$ are respectively the energy density and pressure of the barotropic fluid, and $u_a$ is a four-velocity vector of the fluid. One  interesting feature of the action \eqref{action}  is that  the effective Newton's gravitational parameter depends on the coupling function $F$, i.e., on the scalar field as
\begin{align} \label{Geff}
G_{\rm eff}=\frac{\kappa^2}{F(\phi)}\,.
\end{align}
The negative values of $G_{\rm eff}$ and hence of $F$ indicates the ghost instability in the theory \cite{EspositoFarese:2000ij}. On varying the action with respect to the scalar field $\phi$, we get
\begin{eqnarray}\label{phi_eqn}
\square \phi + \frac{1}{2}RF_{,\phi}- \ell^{-2}\,V_{,\phi}=0\,,
\end{eqnarray}
where the notation $(\cdot)_{,\phi}$  denotes a derivative with respect to $\phi$. Note that the second term of \eqref{phi_eqn} arises from the non-minimal coupling of scalar field to gravity. At a very large scale, consistent with the observed data, we assume the homogeneous and isotropic Universe whose evolution is determined by the scale factor $a(t)$ associated with the FRW metric
\begin{align}\label{metric}
ds^2=-dt^2+a^2(t)(dx^2+dy^2+dz^2)\,,
\end{align}
where  $t$ is the coordinate time and $x, y, z$ are the Cartesian coordinates. Under this metric, the field equations \eqref{EFE} and the Klein Gordon equation \eqref{phi_eqn} can be reduced to the ordinary differential equations
\begin{eqnarray}
3H^2 F+3H \dot{\phi}F_{,\phi}-\frac{1}{2} \dot{\phi}^2-\ell^{-2}\,V&=&\kappa^2 \rho_m\,, \label{FRD_EQ} \\
2 F \dot{H}+\dot{\phi}^2F_{,\phi\phi}+(\ddot{\phi}-H\dot{\phi})F_{,\phi}+\dot{\phi}^2&=&-\kappa^2 \rho_m (1+w)\,,\label{RC_EQ} \\
\ddot{\phi}+3H \dot{\phi}+\ell^{-2}\,V_{,\phi}-3 F_{,\phi}(2 H^2+\dot{H})&=&0\,. \label{KG_EQ}
\end{eqnarray}
In the above equations, $w$ is an equation of state (EoS) defined by a relation $p_m=w \rho_m$ and the upper dot denotes derivative with respect to $t$.   The scalar field describes DE and for simplicity, we shall consider for matter a single  barotropic fluid with a constant $w$, constrained to be between 0 and 1.  While a non-relativistic dust fluid corresponds to $w = 0$,  relativistic radiation fluid corresponds to	$w =\frac{1}{3}$. We note here that one needs to consider a two-fluid model containing  radiation and dust fluids for a more phenomenologically interesting case.   Further, on assuming  the conservation of the matter energy-momentum tensor i.e., $\nabla_a T^{ab}=0$, under a metric \eqref{metric}, one can obtain the  conservation equation
\begin{align}\label{cons}
\dot{\rho}_m+3 H (1+w)  \rho_m=0\,.
\end{align}
To determine the energy density contribution of each component, we introduce the relative energy densities of the  scalar field and that of the barotropic fluid, respectively as

\begin{eqnarray}
\Omega_\phi &=&\frac{\rho_\phi}{3H^2 F}=\frac{\dot{\phi}^2}{6F H^2}-\frac{F_{,\phi} \dot{\phi}}{H F}+\frac{\ell^{-2}V}{3H^2 F} \,, \label{Omp}\\	
\Omega_m &=& \frac{\kappa^2\rho_m}{3 F H^2}\,. \label{Omm}
\end{eqnarray}
These  energy densities  are connected by the Friedmann constraint \eqref{FRD_EQ} as
\begin{align}\label{Om_reln}
\Omega_m+\Omega_\phi=1\,. 
\end{align}
While we can identify the first two terms of \eqref{Omp}  as the kinetic component of the relative energy density of the scalar field, the last term corresponds to the relative potential energy density component of the scalar field. From \eqref{Om_reln}, we can define the matter domination as a scenario where $\Omega_m \approx 1 $ and $\Omega_\phi \approx 0$. Similarly,  we can also define from \eqref{Om_reln}, a kinetic dominated solution or potential dominated solution when the first two terms or the last term in \eqref{Omp} dominate over the others, respectively. 

As expected, the above quantities \eqref{Omp} and \eqref{Omm} reduce to the familiar relative energy densities of a scalar field and matter for the minimal coupled case (i.e., $F(\phi)=1$) respectively. Further, the quantity \eqref{Omp} reduces to the relative energy density of the cosmological constant for a non-dynamical scalar field (i.e., the GR case). In the context of minimal coupling, the above relative energy densities are usually bounded within the interval $[0,1]$. However, this is not necessarily true in non-minimal coupling due to coupling function $F$. Under the physical assumption $\rho_m \geq 0$ and an attractive gravitational force, $F(\phi)> 0$, the problem of negative $\Omega_m$ does not arise.  Nonetheless, due to the second term in the right-hand side of the equation \eqref{Omp} for  $\Omega_\phi$, there is also a possibility for $\Omega_\phi$ to be negative. Thus taking into account the above conditions, the relation \eqref{Om_reln} implies that in the matter domination, the dust fluid can be relatively overdense (i.e., $\Omega_m>1$) compared to the corresponding $\Lambda$CDM case.

 The cosmological equations \eqref{FRD_EQ} and \eqref{RC_EQ} can be rewritten in the form
	\begin{eqnarray}
	3H^2 &=& \kappa^2 \rho_{\rm eff}\,,\\
	3H^2+ 2\dot{H}&=&-\kappa^2 p_{\rm eff}\,,
	\end{eqnarray}
	where $\rho_{\rm eff}$ is the effective energy density and $p_{\rm eff}$ is the effective  pressure  of all the components which are respectively given by
	\begin{eqnarray}
	\rho_{\rm eff}&=&\frac{\rho_m+\rho_\phi}{F(\phi)}\,\nonumber\\
	&=& \frac{\rho_m}{F(\phi)}+ \frac{1}{\kappa^2 F(\phi)} \left[\frac{\dot{\phi}^2}{2} -3H F_{,\phi}(\phi) \dot{\phi} +\ell^{-2} V(\phi) \right]\,,\\
	p_{\rm eff}&=&\frac{p_\phi+p_m}{F(\phi)}\, \nonumber\\
	&=&\frac{1}{\kappa^2F(\phi)} \left[\frac{\dot{\phi}^2}{2}+2 H F_{,\phi}(\phi) \dot{\phi} + F_{,\phi\phi}(\phi)\dot{\phi}^2 + F_{,\phi}(\phi)\ddot{\phi} \right.\nonumber\\ && \left. - \ell^{-2}\,V(\phi) +\kappa^2 p_m\right]\,.
	\end{eqnarray}
Using equation \eqref{FRD_EQ}, the effective EoS of all the components $w_{\rm eff}$ defined as $\frac{p_{\rm eff}}{\rho_{\rm eff}}$ is given by
\begin{eqnarray}
w_{\rm eff} &= &\frac{1}{3  F(\phi) H^2} \left[\frac{\dot{\phi}^2}{2}+2 H F_{,\phi}(\phi) \dot{\phi} + F_{,\phi\phi}(\phi)\dot{\phi}^2  + F_{,\phi}(\phi)\ddot{\phi} \right.\nonumber\\ && \left.  - \ell^{-2}\,V(\phi) \right]+w\, \Omega_m \,. \label{weff}
\end{eqnarray}
While for the accelerated behavior of the Universe, one requires the condition $w_{\rm eff}<-\frac{1}{3}$, super-accelerated Universe or phantom dominated Universe demands $w_{\rm eff}<-1$. It is worth noticing from \eqref{weff}  that in the GR limit, within the matter domination epoch, we have $w_{\rm eff}=w$ and under the scalar field potential dominated epoch (i.e., cosmological constant epoch) $w_{\rm eff}=-1$.

The above equations \eqref{FRD_EQ}-\eqref{cons} are complicated to solve analytically, yet, by recasting them into a dynamical system, one can still obtain important information on the characteristics of solutions. Therefore, in the next section, we shall analyze the dynamics of a general class of non-minimal coupling scalar fields using dynamical system techniques.

	\section{Dynamical system analysis}\label{sec:DS_gen}
	In order to qualitatively analyze the background cosmological dynamics of the present model, we shall convert the cosmological equations \eqref{FRD_EQ}-\eqref{cons} into a dynamical system using the following set of normalized variables  \cite{Bahamonde:2017ize}:
	\begin{gather}
	x= \frac{\dot{\phi}}{H\sqrt{F}}\,,~~~~~~~y=\frac{\ell^{-2}\,V}{3H^2F}\,, \nonumber\\
	\lambda_F=-\frac{F_{,\phi}}{\sqrt{F}}\,,~~~~~~~ \lambda_V=-\frac{V_{,\phi}}{V}\sqrt{F}\,. \label{dyn_var}
	\end{gather}	
	We note here that the chosen variables are well-defined  for $F>0$, i.e., attractive gravity, which is also free from any ghost instability, even though the case $F<0$ may lead to physically interesting scenarios \cite{Capozziello:1996xg}. From the cosmological equations \eqref{FRD_EQ}-\eqref{cons}, we see that there are basically  four variables $H, \phi, \dot{\phi}, \rho_m$. As we have considered the usual $H$-normalized variables, the variable $H$ is being absorbed by other variables, so we are left with three variables \cite{Bahamonde:2017ize}. Since the $H$-normalized variables are connected by the Friedmann constraint \eqref{FRD_EQ}, the number of independent variables reduces to two.  The extra variables $\lambda_F$, $\lambda_V$ are introduced to monitor the  overall effect of coupling function and potential on the dynamics. It is important to note here that the above choice of variables fails for static Universe $H=0$. However, these variables are of physical interest as the energy density of each component can be easily tracked in terms of these variables. In this work, we shall focus on the case of an expanding Universe i.e., $H>0$ as favored by various present observational data. Therefore, we can choose the above normalized variables without any extra concern.	Employing the variables \eqref{dyn_var}, the cosmological equations \eqref{FRD_EQ}-\eqref{cons} can be re-written as the following dynamical system:
	\begin{eqnarray}
	x' &=&\frac{1}{6\,{\lambda_F}^{2}+4}\Big[{x}^{3} \left( 2\,  \Gamma_F \,{\lambda^2_{F}}-w+1
	\right) +{x}^{2}\lambda_F\, \left( 3\,\lambda_F\, \left( 2\, \Gamma_F \,{\lambda_F}\right.\right. \nonumber\\ && \left. \left.  +\lambda_F \right) -9\,w+7 \right) -x
	\left(6(3\,{\lambda^2_F}w-w+1)  +6\,y \left( \lambda_F\,\lambda_V  \right.\right. \nonumber\\ && \left. \left.   +w+1 \right)  \right)+y \left( 12\,\lambda_V -18\, \left( w+1 \right) \lambda_F
	\right) \nonumber\\ && +6\, \left( 3\,w-1 \right) \lambda_F \Big]
	\,, \label{eq:x_ST}\\
	y' &=& \frac{y}{3\,{\lambda^2_F}+2}  \Big[ 12\,{\lambda_F}^{2}+6\,w+6+x \left( \lambda_F\, \left( 3
	\,{\lambda^2_F}-6\,w+4 \right) \right. \nonumber\\ && \left.  - \left( 3\,{\lambda^2_F}+2 \right) \lambda_V
	\right) +{x}^{2} \left( 2\,{\lambda^2_F} \Gamma_F -w+1 \right) \nonumber\\&& -6\,y \left( \lambda_F\,\lambda_V +w+1 \right)  \Big] , \label{eq:y_ST}\\
	\lambda_F' &=& \frac{1}{2} x \lambda _F^2 \left(1-2 \Gamma _F\right) \,,\label{eq:LF_ST}\\
	\lambda_V' &=& -\frac{1}{2} x \lambda _V \left[\lambda _F+2 \left(\Gamma _V-1\right) \lambda _V\right] \,,\label{eq:LV_ST}
	\end{eqnarray}
	where $\Gamma_F=\frac{FF_{,\phi\phi}}{F^2_{,\phi}}$ and $\Gamma_V=\frac{VV_{,\phi\phi}}{V^2_{,\phi}}$.
	 The prime notation denotes the differentiation with respect to the number of $e$-folds $N=\ln a(t)$. We note that the above system \eqref{eq:x_ST}-\eqref{eq:LV_ST} reduces to  the minimal coupling case for $\lambda_F=0$ \cite{Fang:2008fw}.
	
	For the above dynamical system to represents an autonomous system of equations, we consider a class of coupling function $F$ and potential $V$ where $\Gamma_F$, $\Gamma_V$ can be written as functions of $\lambda_F$, $\lambda_V$ respectively \cite{Zhou:2007xp}. If $\lambda_F=\lambda_F(\phi)$ is invertible, then we can express $\phi$ as function of $\lambda_F$. As $\Gamma_F$ is a function of $\phi$, therefore, we can also express $\Gamma_F$  as a function of $\lambda_F$. Similarly, one can express $\Gamma_V$ as a function of $\lambda_V$. In general, the quantities $\Gamma_F$, $\Gamma_V$ may not be a functions of variables $\lambda_F$, $\lambda_V$. In such a case, one has to consider the higher derivatives of the scalar field function \cite{Xiao:2011nh} or consider new dynamical variables \cite{Nunes:2000yc}. We note that the above system has an invariant submanifold $y=0$, as $y'$ vanishes when $y=0$. This submanifold corresponds to a scenario where the scalar field potential vanishes. Physically, it means that if there is no potential source, the scalar field will not evolve.   Further,  depending on $F$ and $V$  (hence in the form of $\Gamma_F$ and $\Gamma_V$), the system also contains $\lambda_F=0, \lambda_V=0$ as invariant submanifolds.  Therefore, a global attractor (if exists) should lie at an intersection of all these invariant submanifolds \cite{Bahamonde:2017ize}. On the other hand, the absence of a global attractor makes the application of bifurcation theory more appealing as the evolution depends on the values of parameters and initial conditions.

	Using the dynamical variables \eqref{dyn_var}, one can express various cosmological parameters viz., the relative energy density parameter of the scalar field ($\Omega_\phi$) and of matter ($\Omega_m$), the EoS of the scalar field ($w_\phi $) and the effective EoS  ($w_{\rm eff}$) respectively as
	\begin{eqnarray}
	\Omega_\phi &=& \frac{x^2}{6}+y+x \lambda_F\,, \label{para1}\\
	\Omega_m &=& 1- x \lambda_F-\frac{x^2}{6}-y\,,  \label{para2}\\
	w_\phi &= &\frac{p_\phi}{\rho_\phi}= \frac {1}{ \left( 3 \,{\lambda_F}^{2}+2 \right)  \left( 6\,x\lambda_F+{x}^{2}+6\,y \right) } \Big(\left( 3\,{\lambda_F}^{2}w  \right.\nonumber\\ && \left. +4\,{\lambda_F}^{2}\Gamma _{{F}}+2 \right) {x}^{2}+6\,{\lambda_F}^{2} \left( 3\,x\lambda_F\,w+3\,wy \right.\nonumber\\&&\left. -3\,w+1\right) -12\,y \left( \lambda_F\,\lambda_V +1 \right) +4\,x\lambda_F \Big)\,, \\
	w_{\rm eff}&=& \frac{p_{\rm eff}}{\rho_{\rm eff}}=  \frac{1}{3 (3\,{\lambda_F}^{2}+2)}\,\left( (2 \Gamma_F {\lambda^2_F}-w+1){x}^{2}+\lambda_F (3\,{\lambda_F} \right.\nonumber\\&&\left.  +2\,x-6\,w x) -6\,y\,(\lambda_V \lambda_F +w+1) +6\,w\right)\,.  \label{weff_para}
	\end{eqnarray}
	Notably the coupling term $H \dot{\phi} F_{,\phi}$ of \eqref{FRD_EQ} can steer the value of $w_\phi$ to $\pm \infty$ during the matter domination epoch. However, this does not cause any physical singularity problem as the effective EoS $w_{\rm eff}$ remains smooth and finite. The divergence behavior of $w_{\phi}$ reduces as the value of $\lambda_{F}$ approaches zero i.e., as the model approaches the minimal coupling case.

	By imposing the physical constraint $\rho_m \geq 0$ on the relation \eqref{Om_reln}, the dynamical variables \eqref{dyn_var} obey the constraint
	\begin{align}
	x \lambda_F+\frac{x^2}{6}+y \leq 1\,.
	\end{align}
	Hence, the phase space of the system is  given by
	\begin{align}\label{phase_space}
	\Psi=\left\lbrace (x, y, \lambda_F, \lambda_V) \in \mathbb{R}^4 ~\vert ~ x \lambda_F+\frac{x^2}{6}+y \leq 1 \right\rbrace.
	\end{align}
	From the cosmological equations \eqref{FRD_EQ}-\eqref{cons}, one can solve the scale factor $a(t)$ evaluated at the critical point by re-writing the equations in terms of the dynamical variables as
	\begin{equation}\label{ray}
	\beta\dot{H}+ H^2=0\,,
	\end{equation}
	where
	\begin{gather*}	
	\beta=\Big[\frac{3}{2}+\frac{1}{2 (3\,{\lambda_F}^{2}+2)}\,\left(({\lambda^2_F}-w+1){x}^{2}+\lambda_F (3\,{\lambda_F}+2\,x \right.\nonumber\\\left. -6\,w x)  -6\,y\,(\lambda_V \lambda_F +w+1) +6\,w\right)\Big]^{-1}.
	\end{gather*}	 
	Integrating equation \eqref{ray}, we get
	\begin{equation} \label{eq:sf_exact_bif}
	a=a_i (t-t_i)^\beta, 
	\end{equation}
	where $a_i$ and $t_i$ are constants of integration. We recall that $0<\beta<1$ corresponds to a decelerated expanding Universe, while  $\beta>1$ corresponds to an accelerated expanding Universe.

	\begin{table*}[!ht]
		\centering
		\small
		\caption{Critical points of the system \eqref{eq:x_ST}-\eqref{eq:LV_ST}.}
		\label{tab:c_pts_general}
			\begin{tabular}{cccccccc}
				\hline\hline
				Point~ &~~ $x$~~ &~~     $y$ ~~& ~~  $\lambda_F$ ~~&~~$\lambda_V$~~ & $w_{\phi}$      &
				$\Omega_m$ &				$w_{\rm eff}$   \\
				\hline
				$A_{1}$ &     $\frac{\lambda_{F_\ast} (3w-1)}{\lambda_{F_\ast}^2-w+1}$     &      $ 0$   & $\lambda_{F_\ast}$ & $\lambda_{V_\ast}$ & $\frac {6\,{\lambda^2_{F_\ast}}w-2\,{\lambda^2_{F_\ast}}-3\,{w}^{2}+7\,w-2}{6\,{\lambda^2_{F_\ast}}-3\,w+5} $ &  $\frac{1}{6}\,{\frac { \left( 3\,{\lambda^2_{F_\ast}}+2 \right)  \left( 3\,{(w-1)}^{2}-2\,{\lambda^2_{F_\ast}} (3w-2) \right) }{ \left( {\lambda^2_{F_\ast}}-w+1 \right) ^{2}}}
				$    
				&    $\frac{1}{3}\,{\frac {{\lambda^2_{F_\ast}}-3\,{w}(w-1)}{{\lambda^2_{F_\ast}}-w+1}}$
				\\[1.5ex]
				$A_{2\pm}$ &     $x_{2\pm}$     &      $0$& $\lambda_{F_\ast}$&$\lambda_{V_\ast}$  &   $1 -\frac{2 \lambda_{F_\ast}}{3}x_{2\pm}$   &  $0$ 
				&    $1 -\frac{2 \lambda_{F_\ast}}{3}x_{2\pm}$
				\\[1.5ex]
				$A_{3}$ &     $\frac{3(w+1)}{\lambda_{V_\ast}}$     &     $y_3$ & $\lambda_{F_\ast}$&$\lambda_{V_\ast}$  &$\Xi$ &  $\frac{1}{2}\,\frac {- \left( 3\,w+7 \right) \lambda_{V_\ast}\,\lambda_{F_\ast}+2\,{\lambda^2_{V_\ast}}-3\,
					\left( {\lambda^2_{F_\ast}}+2 \right)  \left( w+1 \right) }{\lambda^2_{V_\ast}}$    
				&    $-\frac{w (\lambda_{F_\ast}-\lambda_{V_\ast})+\lambda_{F_\ast}}{\lambda_{V_\ast}}$
				\\[1.5ex]
				$A_{4}$ &     $-{\frac {2(2\lambda_{F_\ast}-\lambda_{V_\ast})}{{\lambda_{F_\ast}}^{2}+\lambda_{F_\ast}\,\lambda_{V_\ast} +2}}$     & $y_4$ & $\lambda_{F_\ast}$&$\lambda_{V_\ast}$  & $\frac{1}{3}\,{\frac {{\lambda^2_{F_\ast}}-9\,\lambda_{F_\ast}\,\lambda_{V_\ast}+2\,{\lambda^2_{V_\ast}}-6}{{\lambda^2_{F_\ast}}+\lambda_{F_\ast}\,\lambda_{V_\ast}+2}} $   &  $0$      &    $\frac{1}{3}\,{\frac {{\lambda^2_{F_\ast}}-9\,\lambda_{F_\ast}\,\lambda_{V_\ast}+2\,{\lambda^2_{V_\ast}}-6}{{\lambda^2_{F_\ast}}+\lambda_{F_\ast}\,\lambda_{V_\ast}+2}} $
				\\[1.5ex]
				\hline\hline
		\end{tabular} 
				\end{table*}
				
				\begin{table*}[!ht]
		\centering
		\small
		\caption{Eigenvalues of the critical points of the system \eqref{eq:x_ST}-\eqref{eq:LV_ST} presented in Table \ref{tab:c_pts_general}. 
			Here: $x_k$ denotes the corresponding  $x$-component of a critical point. }
		\label{tab:eigen_general}
			\begin{tabular}{cccccc}
				\hline\hline
				Point & $E_1$ &     $E_2$ &   $E_3$ &$E_4$\\ 
				\hline
				$A_{1}$ &     ${\frac {3\,{\lambda^2_{F_\ast}}(w+1)-\lambda_{F_\ast}\,\lambda_{V_\ast}\,(3w-1)-3\,({w}^{2}-1)}{{\lambda^2_{F_\ast}}-w+1}}
				$    & $\frac{1}{2}\,{\frac {6\,\lambda^2_{F_\ast}w-4\,\lambda^2_{F_\ast}-3\, \left( w-1
						\right) ^{2}}{\lambda^2_{F_\ast}-w+1}}
				$&    $-\lambda_{F_\ast}^2 x_1 \, \Gamma\,^\prime_F(\lambda_{F_\ast}) $ & $- x_1 G(\lambda_{F_\ast},\lambda_{V_\ast}) $   
				\\[1.5ex]
				$A_{2\pm}$ &     $-\lambda_{F_\ast} x_{2\pm} +3(1-w)$    & $-\frac{\left(\left(\lambda_{F_\ast}+\lambda_{V_\ast}\right) x_{2\pm} -2\right)}{2}$&  $-\lambda_{F_\ast}^2 x_{2\pm}\, \Gamma\,^\prime_F(\lambda_{F_\ast}) $ & $- x_{2\pm} G(\lambda_{F_\ast},\lambda_{V_\ast})$    
				\\[1.5ex]
				$A_{3}$ &     $\frac {1}{4\lambda_{V_\ast}} \left( 3\, \left( \lambda_{F_\ast}+\lambda_{V_\ast} \right) w+
				\lambda_{F_\ast}-\lambda_{V_\ast}+\sqrt {{\frac {3\Delta}{3\,\lambda^2_{F_\ast}+2}}}\right)  $    &$\frac {1}{4\lambda_{V_\ast}} \left( 3\, \left( \lambda_{F_\ast}+\lambda_{V_\ast} \right) w+
				\lambda_{F_\ast}-\lambda_{V_\ast}- \sqrt {{\frac {3\Delta}{3\,\lambda^2_{F_\ast}+2}}}\right)  $  &  $-\lambda_{F_\ast}^2 x_3\, \Gamma\,^\prime_F(\lambda_{F_\ast}) $ & $- x_3 G(\lambda_{F_\ast},\lambda_{V_\ast})$    
				\\[1.5ex]
				$A_{4}$ &     $-\frac{5\,{\lambda_{F_\ast}}^{2}+4\,\lambda_{F_\ast}\,\lambda_{V_\ast}-{\lambda_{V_\ast}}^{2}+6}{\lambda_{F_\ast}+\lambda_{F_\ast}\lambda_{V_\ast}+2} $   & $-{\frac {3\, \left( \lambda^2_{F_\ast}+2 \right)  \left( w+1 \right) +
						\left( 3\,w+7 \right) \lambda_{V_\ast}\,\lambda_{F_\ast}-2\,\lambda^2_{V_\ast}}{\lambda^2_{F_\ast}+\lambda_{V_\ast}\,
						\lambda_{F_\ast}+2}}
				$ &   $-\lambda_{F_\ast}^2 x_4\, \Gamma\,^\prime_F(\lambda_{F_\ast}) $ & $-x_4  G(\lambda_{F_\ast},\lambda_{V_\ast}) $       
				\\[1.5ex]
				\hline\hline
		\end{tabular}
				\end{table*}

		In order to extract the dynamics of the above system, we will carry out the standard procedures of the dynamical system analysis \cite{Bahamonde:2017ize}. The critical points ($A_1, A_{2\pm}, A_3, A_4$) of the system \eqref{eq:x_ST}-\eqref{eq:LV_ST} for the general case of $F$ and $V$ are presented in Table \ref{tab:c_pts_general} along with the corresponding values of the cosmological parameters. The corresponding eigenvalues of the perturbed matrix of each critical point are presented in Table \ref{tab:eigen_general}. The existence and stability of each critical point can be determined without specifying the potential and coupling function by treating $\lambda_{F_\ast}$ and $\lambda_{V_\ast}$ as parameters. Therefore, there are as many critical points of the system \eqref{eq:x_ST}-\eqref{eq:LV_ST} as the number of parameters $\lambda_{F_\ast}$ and $\lambda_{V_\ast}$.  Note here that $\lambda_{F_\ast}$ and $\lambda_{V_\ast}$ denote the solutions of the equations $2 \Gamma_F(\lambda_{F})-1=0$ and $\lambda_{F}+2 \left(\Gamma_V(\lambda_V)-1\right) \lambda_V=0$ respectively.  The quantities $\Gamma\,^\prime_F$ and $\Gamma\,^\prime_V$ denote the derivatives of $\Gamma_F$ and $\Gamma_V$ with respect to $\lambda_F$ and $\lambda_V$ respectively. In Tables \ref{tab:c_pts_general} and \ref{tab:eigen_general}, we have 
			\begin{eqnarray}
			&&x_{2\pm}=-3 \lambda_{F_\ast}\pm\sqrt{9\lambda_{F_\ast}^2+6}\,,\nonumber\\
			&& y_3=\frac{1}{2}\,{\frac {3\,{\lambda^2_{F_\ast}}w-3\,\lambda_{F_\ast}\,\lambda_{V_\ast}\,w+3\,{\lambda^2_{F_\ast}}+\lambda_{F_\ast}\,\lambda_{V_\ast}-3\,{w}^{2}+3}{{\lambda_{V_\ast}}^{2}}}\,,\nonumber\\
			&&y_4=  \frac { \left( 5\,{\lambda^2_{F_\ast}}+4\,\lambda_{F_\ast}\,\lambda_{V_\ast}-{\lambda^2_{V_\ast}}+6 \right)  \left(3 {\lambda^2_{F_\ast}}+2 \right) }{3\,({\lambda^2_{F_\ast}}+\lambda_{F_\ast}\,\lambda_{V_\ast}+2)^2}\,,\nonumber\\
			&&\Xi = \frac {3\,w \left( w+1 \right) {\lambda^2_{F_\ast}}+\lambda_{V_\ast}\, \left( 3\,{w}^{2}+5\,w-2 \right) \lambda_{F_\ast}+6\,w \left( w+1 \right) }{3\, \left( w+1 \right) {\lambda^2_{F_\ast}}+\lambda_{V_\ast}\, \left( 3\,w+7 \right) \lambda_{F_\ast}+6\,w+6}\,,\nonumber\\
			&&G(\lambda_{F_\ast},\lambda_{V_\ast}) =\left( {\lambda_{V_\ast}^2} \Gamma\,^\prime_V(\lambda_{V_\ast})   +2 (\lambda_{F_\ast}+2(\Gamma_V-1) \lambda_{V_\ast}) \right) ,\nonumber\\~~~
			&&\Delta =(81\,{\lambda^4_{F_\ast}}{w}^{2}+18\,{\lambda^3_{F_\ast}}\lambda_{V_\ast}\,{w}^{2}-63\,{\lambda^2_{F_\ast}}{\lambda^2_{V_\ast}}{w}^{2}+162\,{\lambda^4_{F_\ast}}w \nonumber\\
			&&~~~+192\,{\lambda^3_{F_\ast}}\lambda_{V_\ast}\,w-210\,{
				\lambda^2_{F_\ast}}{\lambda^2_{V_\ast}}w-72\,{\lambda^2_{F_\ast}}{w}^{3}+48\,\lambda_{F_\ast}\,{\lambda^3_{V_\ast}}
			w \nonumber \\
			&&~~~-72\,\lambda_{F_\ast}\,\lambda_{V_\ast}\,{w}^{3}+81\,{\lambda^4_{F_\ast}}+174\,{\lambda^3_{F_\ast}}\lambda_{V_\ast}+17
			\,{\lambda^2_{F_\ast}}{\lambda^2_{V_\ast}}\nonumber \\
			&&~~~+78\,{\lambda^2_{F_\ast}}{w}^{2}-16\,\lambda_{F_\ast}\,{\lambda^3_{V_\ast}}-300\,\lambda_{F_\ast}\,\lambda_{V_\ast}\,{w}^{2}+54\,{\lambda^2_{V_\ast}}{w}^{2}\nonumber\\  &&~~~
			+372\,{\lambda^2_{F_\ast}}w
			-24\,\lambda_{F_\ast}\,\lambda_{V_\ast}\,w-12\,{\lambda^2_{V_\ast}}w-144\,{w}^{3}+222\,{\lambda^2_{F_\ast}} \nonumber\\  &&~~~
			+204\,\lambda_{F_\ast}\,\lambda_{V_\ast}-42\,{\lambda^2_{V_\ast}}-144\,{w}^{2}+144\,w+144).	\nonumber			 
			\end{eqnarray}

		The behavior of the system \eqref{eq:x_ST}-\eqref{eq:LV_ST} might change dramatically due to a small change of the parameters emerges from potential and coupling. Consequently, it will lead to a change in the phase space's topological structure and gives rise to bifurcation. To have a general information on the effect of various parameters on the dynamics of a system \eqref{eq:x_ST}-\eqref{eq:LV_ST}, we present the bifurcation diagrams for each critical point in Fig. \ref{fig:region_w_0}. These diagrams comprise of finite number of regions in the parameter space $(\lambda_{F_\ast},\lambda_{V_\ast})$. Each region in a bifurcation diagram corresponds to parameter values with distinct dynamical behavior \cite{kuznetsov2013elements}.  Bifurcation curves separating different regions in a diagram corresponding to the specific relation between parameters in which the perturbed matrix evaluated at a critical point has at least one zero real part eigenvalue.  When the perturbed matrix evaluated at a critical point has at least one zero real part eigenvalue, a critical point is said to be {\it non-hyperbolic}. Otherwise, it is {\it hyperbolic}. For a hyperbolic point, we can use linear stability analysis to determine the stability of a point. However, for non-hyperbolic point,  one has to analyze beyond the linear stability analysis via sophisticated tools of dynamical systems such as the center manifold theory (see Ref. \cite{perko2013differential} for details). Thus along the bifurcation curves, one has to investigate the nature of points by the center manifold theory. However, we shall postpone such analysis to a concrete model in Sec. \ref{sec:example} and refrain from the general case analysis as the equations involved are complicated and not very illuminating. In what follows, we summarize each critical point's nature and identify the possible bifurcation scenarios:

\begin{figure*}
	\centering
	\subfigure[]{%
		\includegraphics[width=7cm,height=7cm]{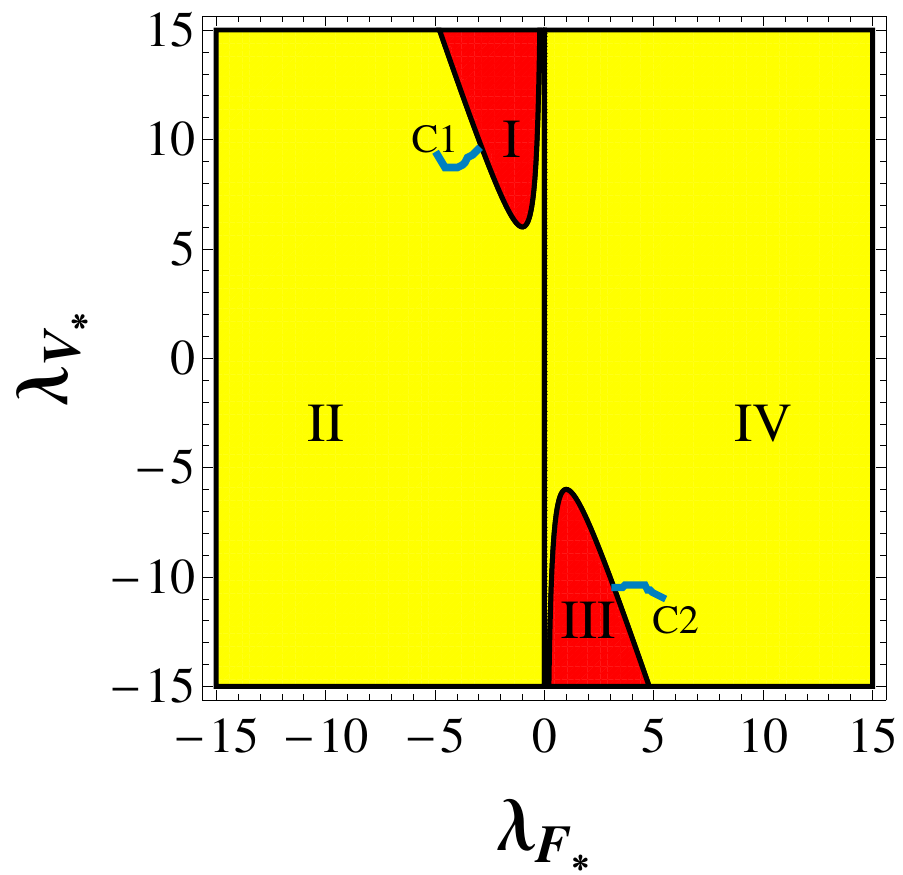}\label{fig:region_A6_w_0}}
	\qquad
	\subfigure[]{%
		\includegraphics[width=7cm,height=7cm]{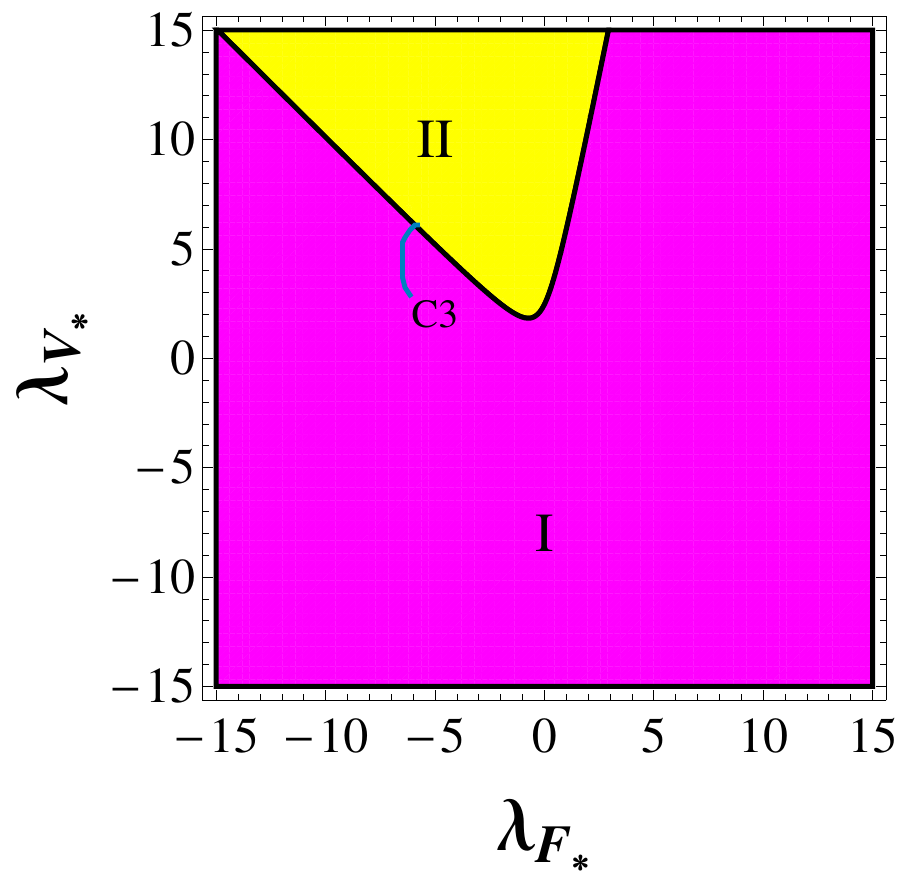}\label{fig:region_A7p_w_0}}
	\qquad
	\subfigure[]{%
		\includegraphics[width=7cm,height=7cm]{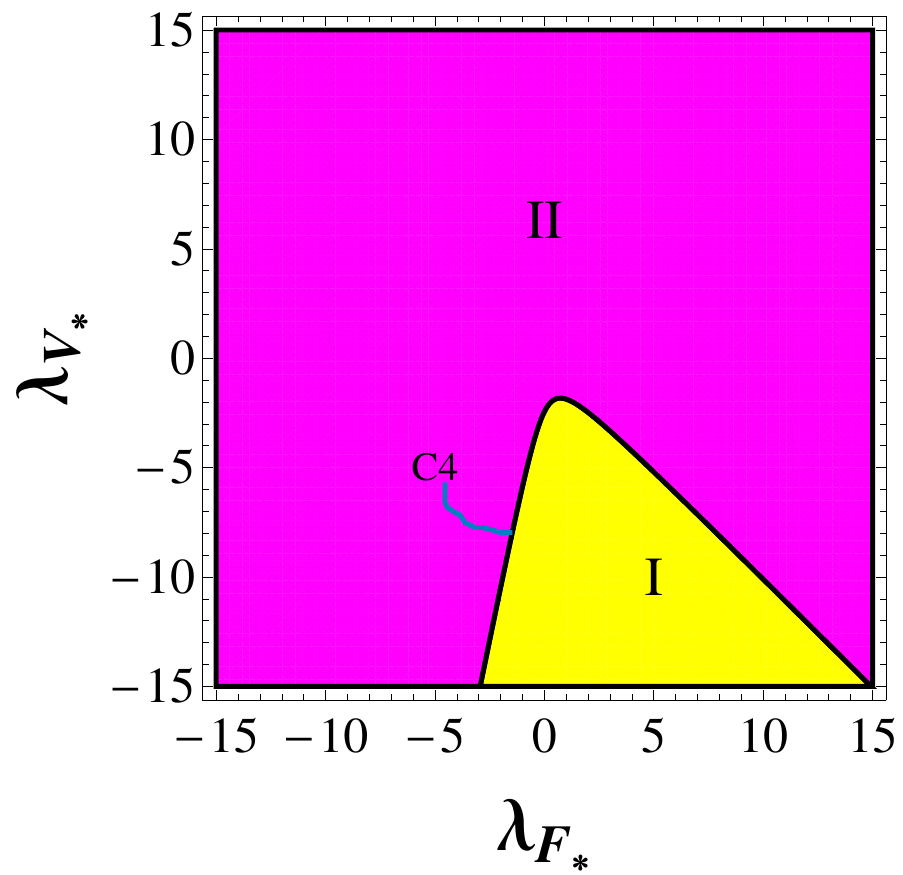}\label{fig:region_A7n_w_0}}
	\qquad
	\subfigure[]{%
		\includegraphics[width=7cm,height=7cm]{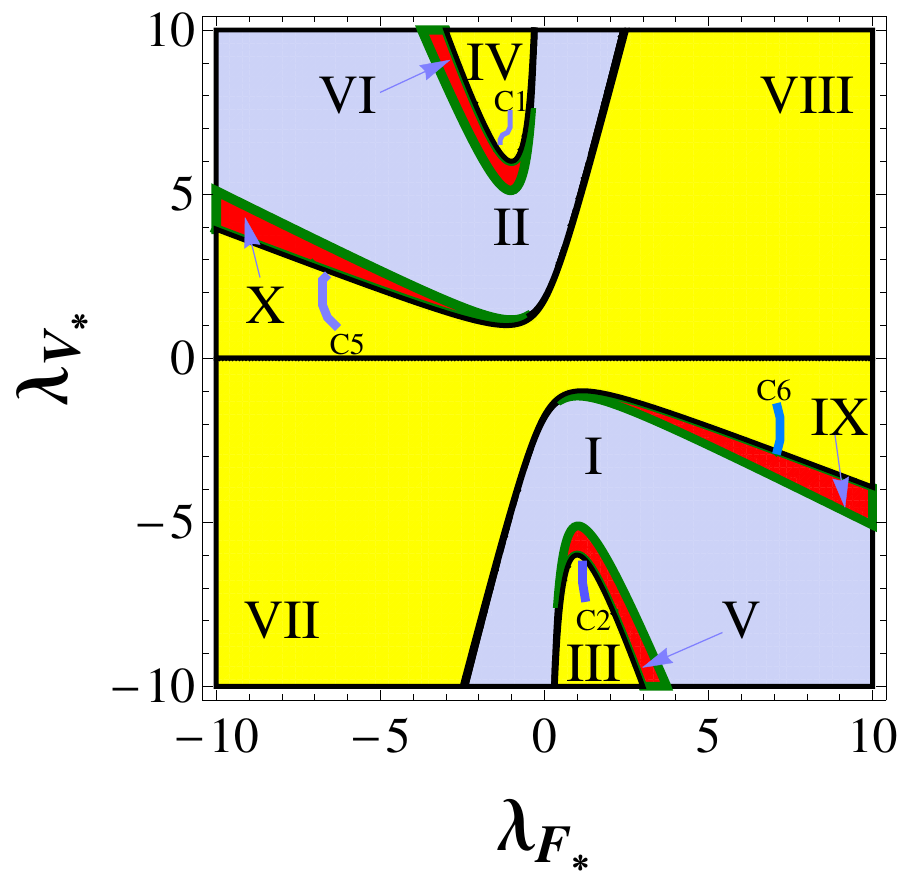}\label{fig:region_A8_w_0}}
	\qquad
	\subfigure[]{%
		\includegraphics[width=7cm,height=7cm]{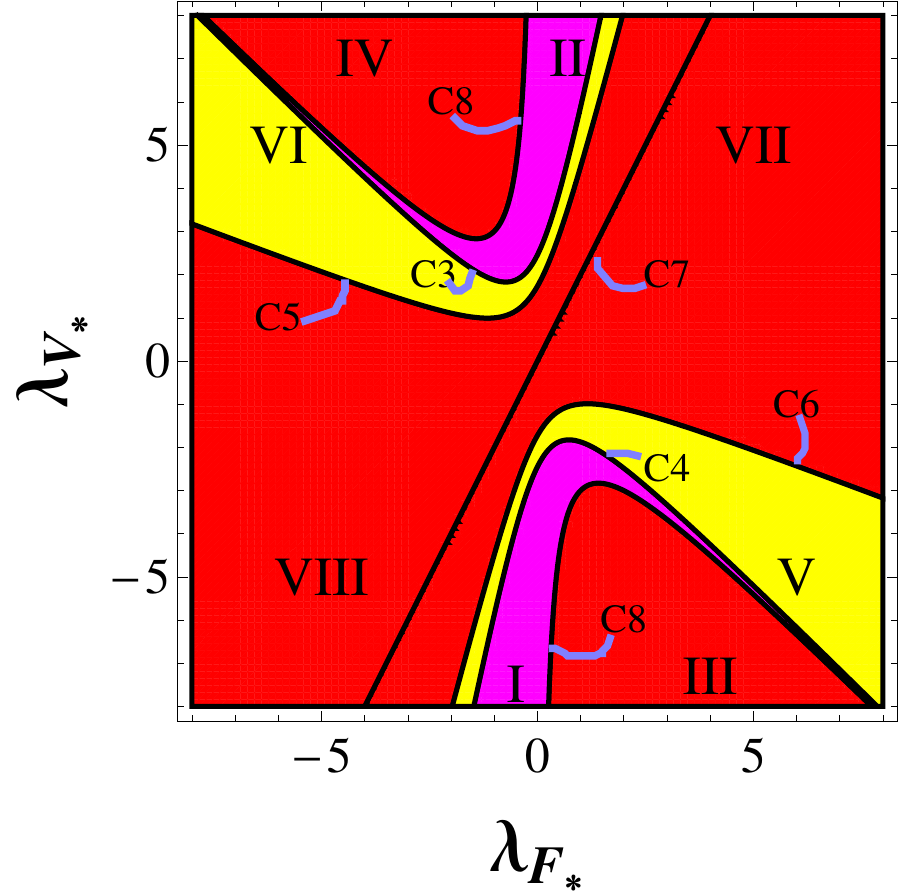}\label{fig:region_A9_w_0}}
	\caption{Bifurcation diagrams in $(\lambda_{F_*}, \lambda_{V_*})$ parameter space exhibiting  the local stability of points $A_1$ in (a), $A_{2+}$ in (b), $A_{2-}$ in (c), $A_3$ in (d) and $A_4$ in (e) for the case of $w=0$.  In all panels, the yellow shaded regions represent the regions where the corresponding critical points are saddle. Black colored line  $\lambda_{V_\ast}=0$ in (d) and curve C8 in  (e) corresponds to the non-existence of a point $A_3$ and $A_4$ respectively. A green colored curve in (d) represents the curve where $\Delta=0$. Black colored curves C1 to C7    separating each region represent  non-hyperbolic curves. The curve C7 is $2\lambda_{F_\ast}=\lambda_{V_\ast}$. For the expressions corresponding to curves C1 to C6, we refer to the  text  in Statement 1.} 
	\label{fig:region_w_0}
\end{figure*}

\begin{itemize}
	\item Point $A_1$ corresponds to a  solution with a vanishing  potential component. For this  point, the exponent for the solution \eqref{eq:sf_exact_bif} is given by
	\begin{equation}
	\beta_{A_1}=\frac {2({\lambda}^{2}_{F_\ast}-w+1)}{4\,{\lambda}^{2}_{F_\ast}-3\,{w}^{2}+3}\,.
	\end{equation}
	It can be checked that $0<\beta_{A_1}<1$ for $0 \leq w \leq 1$  and any choice of $\lambda_{F_\ast}$. Hence, this point corresponds to a decelerated expanding solution  for any choice of model parameters even though the scalar field can possibly behave as the quintessence field ($-1<w_{\phi}<-\frac{1}{3}$). This point can be either stable or saddle depending on the values of $\lambda_{F_\ast}$, $\lambda_{V_\ast}$ and $w$. For instance, as shown in Fig. \ref{fig:region_A6_w_0} for $w=0$ case, this point behaves as  stable node in region I only if $\Gamma\,^\prime_F(\lambda_{F_*})>0$ and $G(\lambda_{F_\ast},\lambda_{V_\ast})>0$. Region III represents the region of stable node only if $\Gamma\,^\prime_F(\lambda_{F_*})<0$ and $G(\lambda_{F_\ast},\lambda_{V_\ast})<0$; otherwise all regions in the $(\lambda_{F_\ast},\lambda_{V_\ast})$ parameter space represent the regions of saddle for any choice of coupling function $F$ and potential $V$. However, for the case of radiation ($w=\frac{1}{3}$), it is always saddle. Further, for $\lambda_{F_\ast}=0$, this point corresponds to a usual decelerated matter dominated solution of the minimal coupled case  ($\Omega_m=1, w_{\rm eff}=w$).
	
	\item Points $A_{2\pm}$ correspond to  kinetic dominated solutions which exist for any choice of model parameters. In this case, we have
	\begin{gather}
	\beta_{A_{2+}}={\frac {\lambda_{F_\ast}\,\sqrt {9\,{\lambda}^{2}_{F_\ast}+6}+3\,{\lambda}^{2}_{F_\ast}+3}{12\,{
				\lambda}^{2}_{F_\ast}+9}}\,,\\
	\beta_{A_{2-}}={\frac {-\lambda_{F_\ast}\,\sqrt {9\,{\lambda}^{2}_{F_\ast}+6}+3\,{\lambda}^{2}_{F_\ast}+3}{12\,{\lambda}^{2}_{F_\ast}+9}}\,,
	\end{gather}
	which are both positive and less than one and hence, correspond to  decelerated expansion. The bifurcation diagrams for these points are given in  Figs. \ref{fig:region_A7p_w_0}, \ref{fig:region_A7n_w_0}. 
	These points can be either unstable node or saddle depending on the form of potential and coupling function. For example, region I in Fig. \ref{fig:region_A7p_w_0} represents region of unstable node of  point $A_{2_+}$ if  $\Gamma\,^\prime_F(\lambda_{F_*})<0$ and $G(\lambda_{F_\ast},\lambda_{V_\ast})<0$; region II in Fig. \ref{fig:region_A7n_w_0} represents  the region of unstable node of $A_{2_-}$ if  $\Gamma\,^\prime_F(\lambda_{F_*})>0$ and $G(\lambda_{F_\ast},\lambda_{V_\ast})>0$. Otherwise, they are saddle in nature for any choice of parameters.  When $(\lambda_{F_\ast},\lambda_{V_\ast})=(0, \sqrt{6})$ and $(0, -\sqrt{6})$, points $A_{2+}$ and $A_{2-}$ correspond to  stiff matter dominated solutions ($\Omega_\phi=1, w_{\rm eff}=1$) of the minimal coupled scalar field.

	\item Point $A_{3}$ corresponds to a scaling solution and exists for all values of the model parameters except when $\lambda_{V_\ast}=0$. This point can either be a stable node or stable focus or behaving as a saddle.  The bifurcation diagrams for this point for $w=0$ case is given in Fig. \ref{fig:region_A8_w_0}. In this plot, when  $\Gamma\,^\prime_F(\lambda_{F_*})>0$ and $G(\lambda_{F_\ast},\lambda_{V_\ast})>0$, region II represents region  of stable focus,  regions VI and X represent regions  of stable node and the remaining regions represent the regions where this point behaves as saddle. However, when $\Gamma\,^\prime_F(\lambda_{F_*})<0$ and $G(\lambda_{F_\ast},\lambda_{V_\ast})<0$, region I  represents region  of stable focus,  regions V and IX represent regions  of stable node and the remaining regions   represent the regions where this point behaves as saddle.   It is important to note that as the parameters values change across the bifurcation curves  C5 and C6 of Fig. \ref{fig:region_A8_w_0}, the property of the Universe changes  from a decelerated scaling solution ($w_{\rm eff}>-\frac{1}{3}, \Omega_m>0$) to a decelerated scalar field dominated solution ($w_{\rm eff}>-\frac{1}{3}, \Omega_m=0$). For this point, the exponent $\beta$ in \eqref{eq:sf_exact_bif} is given by
	\begin{equation}
	\beta_{A_3}=\frac{2}{3}\,{\frac { \lambda_{V_\ast}}{ \left( w+1 \right)  \left(  \lambda_{V_\ast} - \lambda_{F_\ast}\right) }}\,.
	\end{equation}
	Therefore, it represents a decelerated expansion when $0<\frac{\lambda_{V_\ast}}{\lambda_{V_\ast}-\lambda_{F_\ast}}< \frac{3}{2}(w+1)$ and an accelerated expanding Universe when $\frac{\lambda_{V_\ast}}{\lambda_{V_\ast}-\lambda_{F_\ast}}> \frac{3}{2}(w+1)$. When $\lambda_{F_\ast}=\lambda_{V_\ast}$, it corresponds to an effective  cosmological constant behavior ($w_{\rm eff}=-1$),  but  it  is unphysical as $\Omega_m<0$. Further, from the bifurcation diagram of this point, we have checked that within the unstable   (or saddle) accelerated regions of the parameter space, this point is unphysical.

	\item Point $A_4$ corresponds to a scalar field dominated solution ($\Omega_\phi=1$). This point disappears for values of $\lambda_{F_\ast}, \lambda_{V_\ast}$ satisfying ${\lambda_{F_\ast}}^{2}+\lambda_{F_\ast}\,\lambda_{V_\ast} +2=0$. It can either be a stable node or unstable node or saddle depending on the choice of model parameters (see  Fig. \ref{fig:region_A9_w_0}). For  $\Gamma\,^\prime_F(\lambda_{F_*})>0$ and $G(\lambda_{F_\ast},\lambda_{V_\ast})>0$, while region I  represents a region of unstable node, regions III and VIII  represent regions of stable node. The remaining regions represent the regions where this point behaves as a saddle. For  $\Gamma\,^\prime_F(\lambda_{F_*})<0$ and $G(\lambda_{F_\ast},\lambda_{V_\ast})<0$,  region II  represents a region of unstable node,  regions IV and VII  represent regions of stable node and the remaining regions represent the regions where this point behaves as a saddle.   The  corresponding exponent for the scale factor solution \eqref{eq:sf_exact_bif} is given by
	\begin{equation}
	\beta_{A_4}=\frac {{\lambda}^{2}_{F_\ast}+\lambda_{F_\ast}\,\lambda_{V_\ast}+2}{2\,{\lambda}^{2}_{F_\ast}-3\,\lambda_{F_\ast}\,\lambda_{V_\ast}+\lambda^2_{V_\ast}}.
	\end{equation} 
	This point exhibits an accelerated expanding Universe or decelerated expanding Universe for some parameter values. We have checked that this point is stable from the bifurcation diagram when it corresponds to an accelerated expansion. Hence, this point can describe the late time Universe. Further, depending on coupling function and potential, this point can correspond to an accelerating Universe when it is a saddle. Therefore, we can also use this point to model the graceful exit phenomenon.   We have verified numerically that this point exhibits different behavior of the Universe when this point undergoes a bifurcation. For example, this point describes a   decelerated Universe and an accelerated expanding Universe as this point changes from a saddle (some parts of regions V and VI of Fig. \ref{fig:region_A9_w_0}) to a stable node (some parts of regions VII and VIII of Fig. \ref{fig:region_A9_w_0}).  It is worth mentioning that this point corresponds to an effective cosmological constant behavior ($w_{\rm eff}=-1$) when $\lambda_{F_\ast}=\lambda_{V_\ast}$ or $\lambda_{F_\ast}=\frac{\lambda_{V_\ast}}{2}$. The solution corresponds to $\lambda_{F_\ast}=\frac{\lambda_{V_\ast}}{2}$ is of interest as it is identical to the de Sitter solution in GR  (since $\phi$ is constant in time).  From the bifurcation diagram (i.e., Fig. \ref{fig:region_A9_w_0}), one could confirm that there is a possibility that this point is stable when $\lambda_{F_\ast}=\frac{\lambda_{V_\ast}}{2}$.  This point's stable nature explains the possible late time convergence behavior of the scalar-tensor theory towards GR.  In other words, such a model converges towards a structurally stable GR-based model.  For coupling function $F(\phi)=1+\xi \phi^2$ and $V =V_0 \phi^n$, this condition is met for $n=4$. Also when $\lambda_{F_\ast}< \lambda_{V_\ast}< 2\lambda_{F_\ast}$, this point behaves as a stable phantom-like attractor. A similar result for a particular potential case has been reported in \cite{Perivolaropoulos:2005yv}. Therefore, our present work provides a framework for choosing proper coupling and potential functions to get interesting dynamics for a broad class of non-minimal coupled scalar field models.
	
\end{itemize}
As there is no stable critical point lying on the intersection of invariant sub-manifolds $y=0, \lambda_F=0, \lambda_V=0$, therefore, the above system does not contain any global attractor. Further, from the above analysis, one can see that the present model exhibits interesting solutions that can describe various cosmological eras of the Universe. For instance, the decelerated scalar field dominated solution can describe the early radiation era, the matter scaling solution that can describe the intermediate dark matter (DM), and the late time scalar field dominated solution describing the DE dominated era. Numerically, we observe that the local stability behavior of critical points changes for the coupling and potential parameters (Fig. \ref{fig:region_w_0}), and hence the system as a whole undergoes bifurcation.	Using the bifurcation diagrams of various critical points, we can classify the parameters for which the evolution corresponds to the generic evolution and non-generic evolution. Geometrically, when an orbit evolves from an unstable point and then settles in a stable point, it is called generic evolution. However, if the initial or final point is a saddle, it is called non-generic evolution. From the above analysis, we found that only class of models where the parameters $\lambda_{V_\ast}, \lambda_{F_\ast}$  belong to the common region of region I of Fig. \ref{fig:region_A7p_w_0} and region VII of Fig. \ref{fig:region_A9_w_0}; and also in the common region of region II of Fig. \ref{fig:region_A7n_w_0} and region VIII of Fig. \ref{fig:region_A9_w_0} can possibly lead to  physically interesting generic evolutionary scenarios. These correspond to the evolution from a decelerated kinetic dominated unstable node point $A_{2+}$/$A_{2-}$ to an accelerated potential dominated stable point $A_4$ via a matter scaling solutions $A_1$ or $A_3$. However, solutions evolving near an intermediate point $A_3$ correspond to unphysical solutions ($\Omega_m<0$). Therefore, the sequence of generic cosmic viable evolution is given by  $A_{2\pm} \to A_1 \to A_4$. Thus, bifurcation diagrams allow us to classify the coupling and potential functions, describing interesting cosmological dynamics.  

Further from the bifurcation diagrams  of each point, we see that there is an occurrence of transcritical bifurcation  between critical points.  For example, critical points $A_1$ and $A_3$ undergo transcritical bifurcation  (Figs. \ref{fig:region_A6_w_0}, \ref{fig:region_A8_w_0}), $A_{2+}$ and $A_4$ (Figs. \ref{fig:region_A7p_w_0}, \ref{fig:region_A9_w_0}), $A_{2-}$ and $A_4$ (Figs. \ref{fig:region_A7n_w_0}, \ref{fig:region_A9_w_0}), $A_{3}$ and $A_4$ (Figs. \ref{fig:region_A8_w_0}, \ref{fig:region_A9_w_0}). This type of bifurcation occurs when two critical points interchange their stability properties at the bifurcation curve \cite{kuznetsov2013elements}.   We can summarize these bifurcation scenarios as follows:

\begin{stmt}(Existence of transcritical bifurcation)\\
	A system \eqref{eq:x_ST}-\eqref{eq:LV_ST} undergoes a transcritical bifurcation when
	\begin{enumerate}
		\item  Critical points $A_1$ and $A_3$ interchange a saddle and stable node behavior  along the bifurcation curve $\lambda_{V_*}=\frac{3(1+\lambda^2_{F_*}-w)(1+w)}{(3w-1)\lambda_{F_*}}$  for fixed $w$.\\
		In particular,  for $w=0$, the above curve is represented geometrically by two branch curves C1 and C2 of Figs. \ref{fig:region_A6_w_0}, \ref{fig:region_A8_w_0}.
		
		\item Critical points $A_{2+}$ and $A_4$ interchange a saddle and unstable node behavior  along a bifurcation curve  $\lambda_{V_\ast}=2\lambda_{F_\ast}+ \sqrt{9 \lambda_{F_\ast}^2+6}$, represented geometrically by curve C3 of Figs. \ref{fig:region_A7p_w_0} and \ref{fig:region_A9_w_0}.\\
		Also, critical points $A_{2-}$ and $A_4$ interchange a saddle and unstable node behavior  along a bifurcation curve  $\lambda_{V_\ast}=2\lambda_{F_\ast}- \sqrt{9 \lambda_{F_\ast}^2+6}$, represented geometrically by curve C4 of Figs. \ref{fig:region_A7n_w_0} and \ref{fig:region_A9_w_0}. 
		\item  Critical points  $A_{3}$ and $A_4$ interchange a saddle and stable node behavior  along the bifurcation curves
		\begin{gather*}				
		\lambda_{V_\ast }=\frac{1}{4}(3\,\lambda_{F_\ast}\,w+7) \\
		+ \frac{1}{4}\,\sqrt {9\,{\lambda_{F_\ast}}^{2}{w}^{2}+66\,{\lambda^2_{F_\ast}}w+73\,{\lambda^2_{F_\ast}}+48\,w+48}
		\end{gather*}				
		and 
		\begin{gather*}				
		\lambda_{V_\ast }=\frac{1}{4}(3\,\lambda_{F_\ast}\,w+7) \\
		- \frac{1}{4}\,\sqrt {9\,{\lambda_{F_\ast}}^{2}{w}^{2}+66\,{\lambda^2_{F_\ast}}w+73\,{\lambda^2_{F_\ast}}+48\,w+48}
		\end{gather*}	for fixed value of $w$.
		In particular, for $w=0$, the above curves are represented geometrically by  curves C5 and C6 of  Figs. \ref{fig:region_A8_w_0}  and \ref{fig:region_A9_w_0}.
	\end{enumerate}
\end{stmt}
Note that one can prove the existence of transcritical bifurcation analytically by using the Sotomayor's theorem  \cite{kuznetsov2013elements,seydel2009practical,perko2013differential}. However, since the bifurcation parameters $\lambda_{V_\ast}, \lambda_{F_\ast}$ do not appear explicitly on the system \eqref{eq:x_ST}-\eqref{eq:LV_ST}, therefore, we postpone the analytical proof to a concrete example of coupling function $F$ and scalar field potential $V$. Even though we can determine the general properties without specifying the concrete model, to understand the cosmological applications of a general model and better investigate its dynamics, we must assume a concrete example. Therefore, in the next section, we consider a specific model and analyze its cosmological dynamics in detail.

\begin{figure*}
	\centering
	\subfigure[]{%
		\includegraphics[width=7cm,height=7cm]{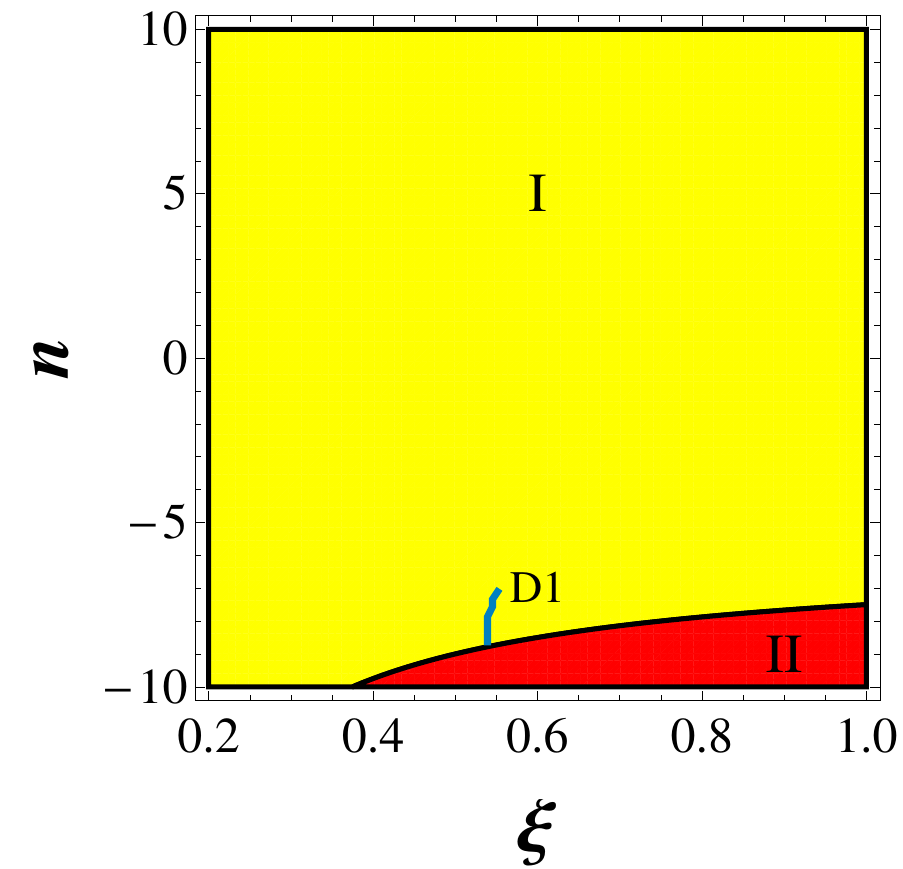}\label{fig:region_A6_w0_pow}}
	\qquad
	\subfigure[]{%
		\includegraphics[width=7cm,height=7cm]{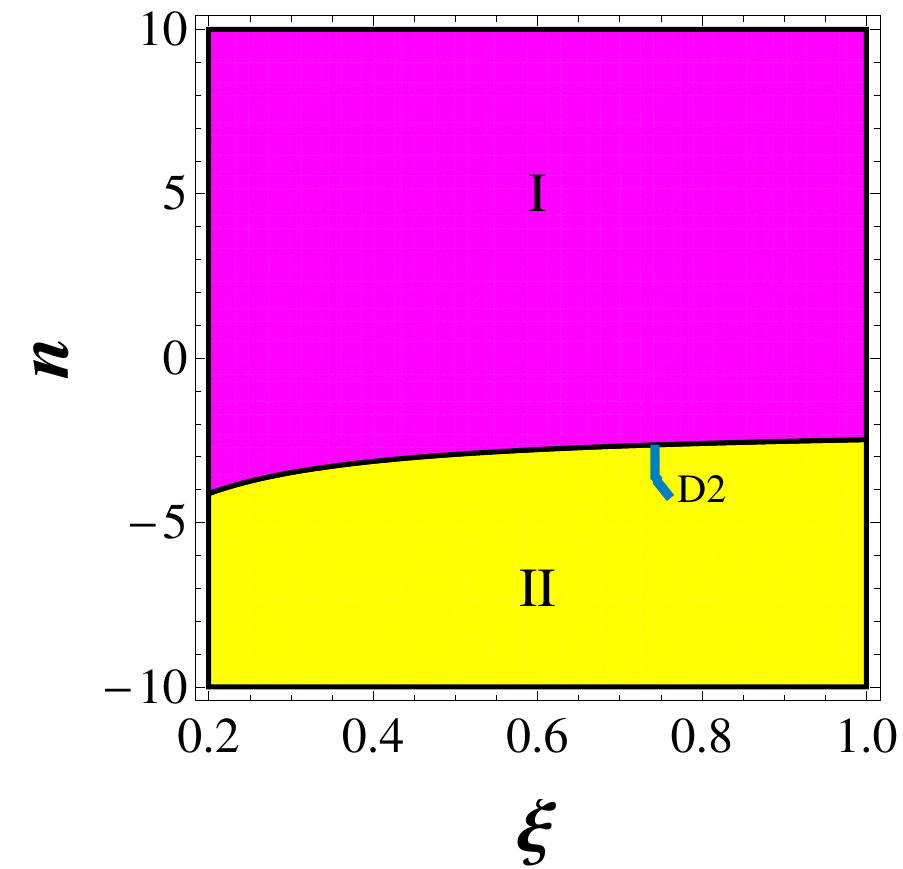}\label{fig:region_A7p_w_0_pow}}
	\qquad
	\subfigure[]{%
		\includegraphics[width=7cm,height=7cm]{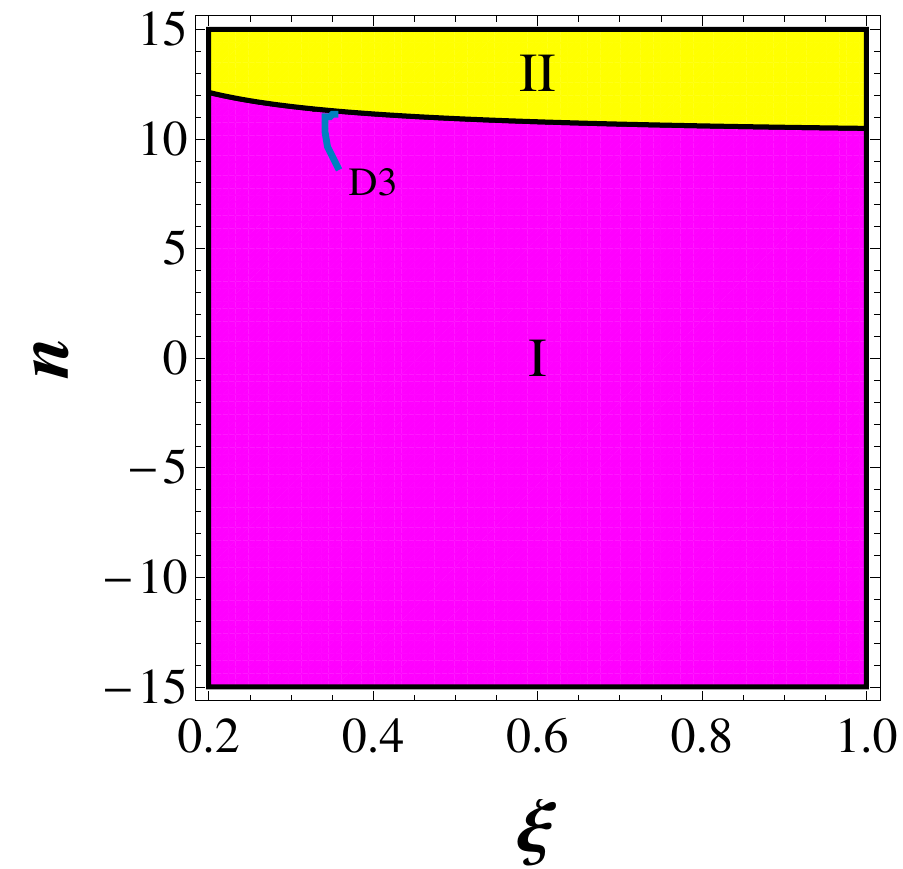}\label{fig:region_A7n_w_0_pow}}
	\qquad
	\subfigure[]{%
		\includegraphics[width=7cm,height=7cm]{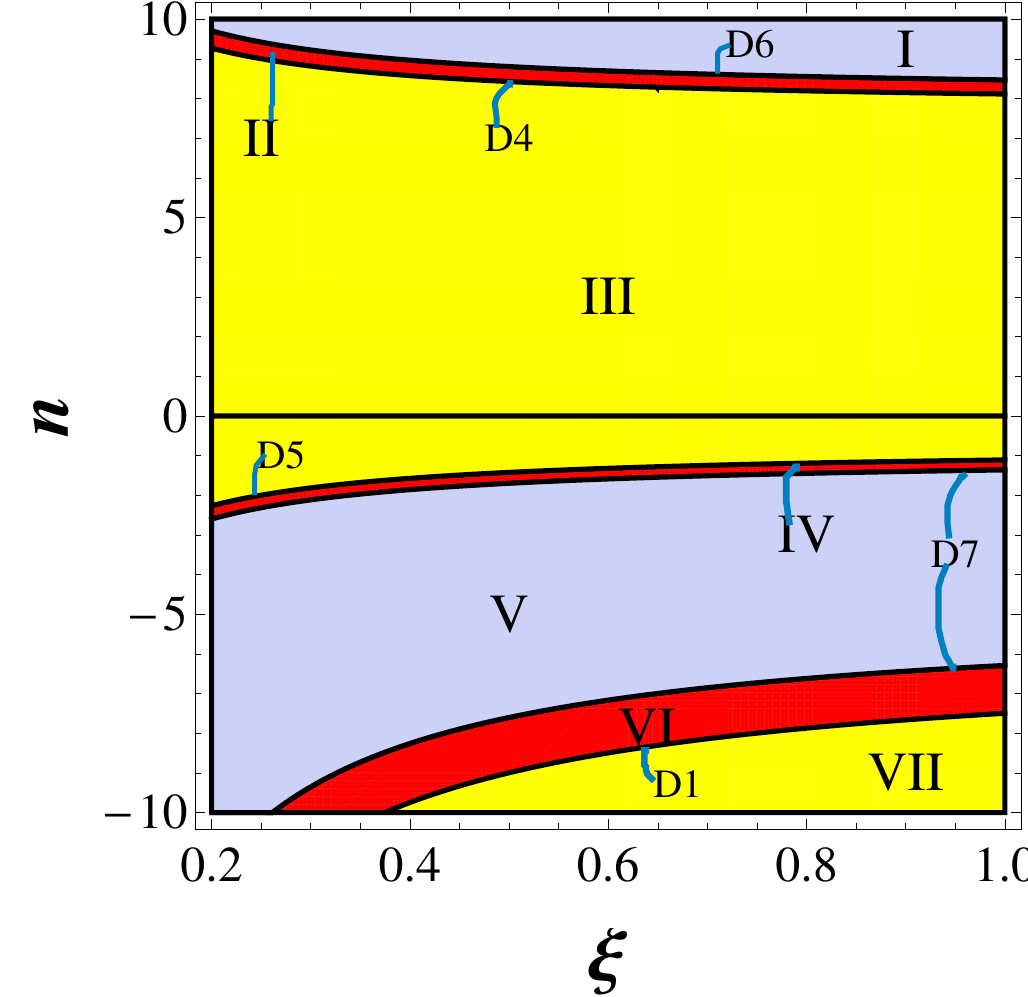}\label{fig:region_A8_w_0_pow}}
	\qquad
	\subfigure[]{%
		\includegraphics[width=7cm,height=7cm]{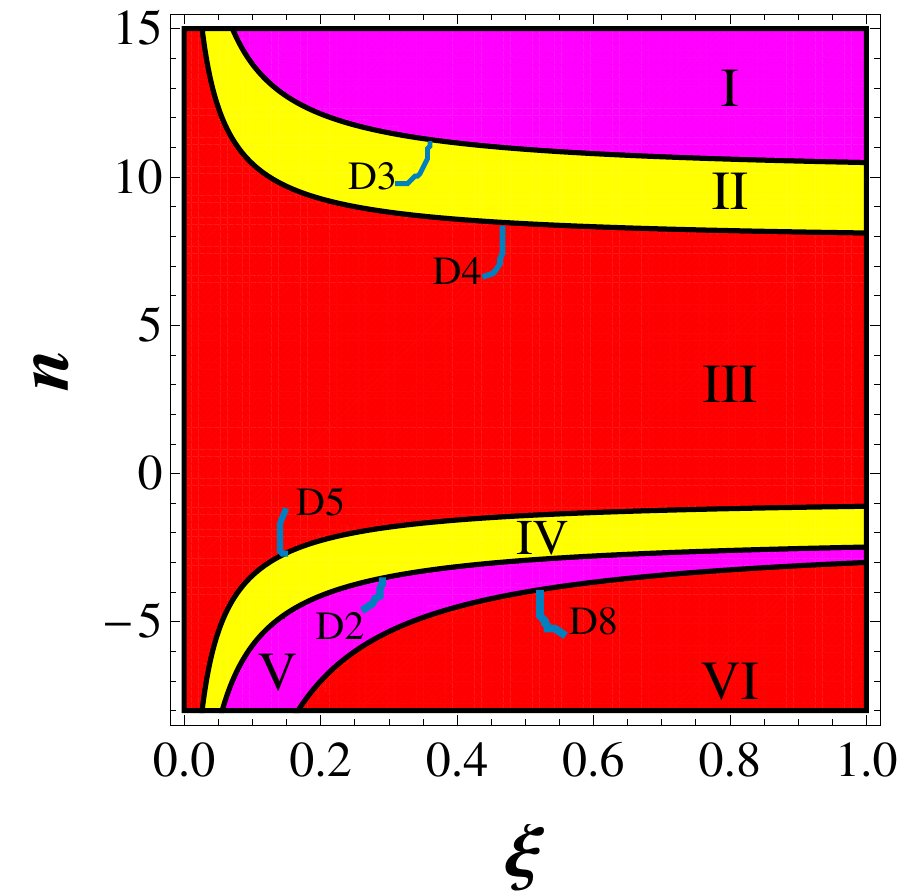}\label{fig:region_A9_w_0_pow}}
	\caption{Bifurcation diagrams in $(\xi, n)$ parameter space exhibiting different local stability regions of points $B_1$ in (a), $B_{2+}$ in (b), $B_{2-}$ in (c), $B_3$ in (d) and $B_4$ in (e) for the case of $w=0$.  In each panel, the yellow color regions correspond to saddle regions, red color regions correspond to  regions of stable node, blue color regions correspond to  regions of stable focus, magenta color regions correspond to  regions of unstable node.  Black colored curves  D1 to D5 (see statements 2, 3, 4 for their expressions)  separating different  regions represent the non-hyperbolic curves.   The curves D6 and D7 represent the curves where point $B_3$ changes between a stable node and stable focus nature. Line $n=0$ in (d) and curve D8 represents non-existence of points $B_3$ and $B_4$ respectively.}
	\label{fig:region_w_0_pow}
\end{figure*}

\begin{table*}[!ht]
	\caption{Critical points of the system \eqref{eq:x_ST_pow}-\eqref{eq:y_ST_pow}.}
	\label{tab:c_pts_pow}
	\begin{tabular}{cccccccc}
		\hline\hline
		Point~ &~~ $x$~~ &~~     $y$ ~~    & Existence & Stability  \\
		\hline
		$B_{1}$ &     $\frac{2 \sqrt{\xi} (1-3w)}{4\xi-w+1}$     &      $ 0$  &$\xi\neq \frac{w-1}{4}$& Saddle/Stable node
		\\[1.5ex]
		$B_{2\pm}$ &     $6\,\sqrt {\xi}\pm\sqrt {36\,\xi+6}$     &      $0$ & Always &  Saddle/Unstable node
		\\[1.5ex]
		$B_{3}$ &     ${\frac {-3(w+1)}{n\sqrt {\xi}}}$     &      $\frac{1}{2}\,{\frac {-6\,\xi\,nw+2\,\xi\,n-3\,{w}^{2}+12\,w\xi+12\,\xi+3}{{n}^
				{2}\xi}}$    & $n \neq 0$  & Stable focus/Stable  node/Saddle
		\\[1.5ex]
		$B_{4}$ &     $-{\frac {\sqrt {\xi} \left( n-4 \right) }{\xi\,n+2\,\xi+1}}$     &      $-{\frac { \left( 6\,\xi+1 \right)  \left( {n}^{2}\xi-8\,\xi\,n-20 \,\xi-6 \right) }{ 6 \left( \xi\,n+2\,\xi+1 \right) ^{2}}}$ &  $\xi \neq -\frac{1}{n+2}$   & Stable node/Unstable node/Saddle
		\\[1.5ex]
		\hline\hline
	\end{tabular}
\end{table*}

	\section{Example: Quadratic coupling along with power-law potential}\label{sec:example}
	\subsection{Stability Analysis}\label{sub:fin}
Here, we shall examine the case where the non-minimal coupling is of the form $F(\phi)= \xi \phi^2$ and the scalar field potential is of form $V(\phi)=V_0 \phi^n$  (power-law). 
For this example, we have  $\lambda_F=-2 \sqrt{\xi}$, $\lambda_V=-n\sqrt{\xi}$ and hence they are constants. This particular example is inspired physically by the string-dilaton,  Brans Dicke actions and several effective quantum field theories \cite{Fujii:2003pa}. Mathematically, this form of coupling and potential agrees with the Noether symmetry approach of the Lagrangian given by \eqref{action} and hence, leads to physically interesting exact solutions \cite{Capozziello:1996bi,Paliathanasis:2014rja}. This model is also compatible with the solar system constraint test \cite {Finelli:2007wb}. Various cosmological data constrain the value of $\xi$ to be  $ \lesssim 10^{-2}$ at 95\% confidence limit. However,  a degeneracy between the value of $\xi$  and the present Hubble constant $H_0$ allows a larger value of $\xi$ \cite{Umilta:2015cta,Ballardini:2016cvy}.

Stability analysis for this particular example has been performed earlier in \cite{Carloni:2007eu} where the main focus is on the hyperbolic points. However, a discussion on the condition for non-hyperbolicity of points has not been performed. The non-hyperbolic nature of the critical point is important to analyze the bifurcation scenarios.  Since for this concrete example, $\lambda_F$ and $\lambda_V$ are fixed, the  system \eqref{eq:x_ST}-\eqref{eq:LV_ST} reduces to following two-dimensional system:
\begin{eqnarray}
x'&=&  \frac {1}{4(6\,\xi+1)} \Big[{x}^{3} \left( 4\,\xi-w+1 \right) -2\,{x}^{2}\sqrt {\xi}
\left( 24\,\xi-9\,w+7 \right) \nonumber \\&&  -6 x \left( 12\,w\xi- \,w+1+y \left( 2
\,n\xi+w+1 \right)  \right) +12 \sqrt {\xi} y   \nonumber \\&& \left( - n+3\, \left( w+
1 \right) \right) -12\, \left( 3\,w-1 \right) \sqrt {\xi} \Big]\,, \label{eq:x_ST_pow}\\
y'&=&\frac {y}{2(6\,\xi+1)} \Big[  6(8\,\xi+ w+1) +2\,\sqrt {\xi}x \left( -2(6\,\xi-3\,w+2) \right. \nonumber \\ && \left. + \left( 6\,\xi+1 \right) n \right) -{x}^{2} \left( -4\,\xi+w-1 \right) \nonumber \\&& -6\,y \left( 2\,n\xi+w+1 \right)  \Big]\,.  \label{eq:y_ST_pow}
\end{eqnarray}

The physical phase space of the reduced system is  
\begin{align}\label{phase_space_eg}
\Psi=\left\lbrace (x, y) \in \mathbb{R}^2 ~\Big\vert ~\frac{(x-6\sqrt{\xi})^2}{6}+y \leq (1+6 \xi) \right\rbrace\,.
\end{align}
The critical points for the above system are given in Table \ref{tab:c_pts_pow}. We note that critical points $B_1, B_{2\pm}, B_3, B_4$ correspond to  points  $A_1, A_{2\pm}, A_3, A_4$ respectively for this concrete example of coupling function and potential.  The existence and stability behavior of these critical points are similar to the general case (see Table \ref{tab:c_pts_pow}).   The bifurcation diagrams exhibiting the stability regions of critical points are given in Fig. \ref{fig:region_w_0_pow}. It is worth noting that each critical point shows a  non-hyperbolic behavior for different values of parameters (represented by black colored curves in Fig. \ref{fig:region_w_0_pow}). As analyzed in the general case (see Sec. \ref{sec:DS_gen}),  critical points $B_1$ and $B_3$ coincide along the curve D1 in which both of them are non-hyperbolic. Therefore, in this case, it is sufficient to analyze the non-hyperbolic nature only for a point $B_1$. The analysis of center manifold theory for this case is performed in the appendix \ref{cmt_b1} and it is found that point $B_1$ behaves as a saddle. For a detailed mathematical background on the center manifold theory, we refer to \cite{perko2013differential}. Further, points $B_{2+}$ and $B_{2-}$ coincide with point $B_4$ and are non-hyperbolic along the curves D2 and D3 respectively.  Also, critical points $B_3$ and $B_4$ are non-hyperbolic and coincide along the curves D4 and D5. Therefore, in each case, we shall analyze the non-hyperbolic property for a point $B_4$ only. The analysis performed in the appendix \ref{cmt_b4} reveals that point $B_4$ behaves as a saddle along each bifurcation curve.

As the phase space \eqref{phase_space_eg} is in general not compact, for the sake of completeness, we analyze the nature of the  system \eqref{eq:x_ST_pow}-\eqref{eq:y_ST_pow} at infinity by employing the Poincar\'e's projection method involving the following transformation \cite{perko2013differential}:
\begin{align}
x_r=\frac{x}{\sqrt{1+x^2+y^2}},~~~~y_r=\frac{y}{\sqrt{1+x^2+y^2}}.
\end{align}
The compactified phase space of the resulting system is therefore given by
\begin{align}
\Psi_r =\left\lbrace   (x_r, y_r) \in \mathbb{R}^2 \Big| -1 \leq x_r, y_r \leq 1, x_r^2+y_r^2\leq 1, \right.\nonumber\\ \left.
\frac{\left( \frac{x_r}{R}-6 \sqrt{\xi}\right)^2}{6}+\frac{y_r}{R} \leq 1+6 \xi \right\rbrace\,,
\end{align}
with $R=\sqrt{1-x_r^2-y_r^2}$ . The critical points near infinity i.e., $\sqrt{x^2+y^2} \rightarrow \infty$ correspond to points on the circle 
\begin{align}
\left\lbrace (x_r, y_r) \in \mathbb{R}^2  : x_r^2+y_r^2= 1 \right\rbrace\,.
\end{align}
We should take due care in applying the Poincar\'e compactification method, resulting in a wrong phase space topology, and hence false properties of the solutions \cite{Alho:2016gzi}. However, in this case, the above choice of the dynamical variables does not destroy the phase space's global structure since  $H>0$ \cite{Coley:2003mj}. The method's advantage is to capture the possible critical points hidden at infinity that cannot be detected by the finite analysis.
On employing this method,  we found that there are four critical points near infinity viz., $(x_r, y_r)=(\pm1,0), (0,\pm 1)$ lying on the equator of the Poincar\'e's sphere. The analysis shows that points $P_{1}^\infty\,(1,0)$,   and  $P_{3}^\infty\,(-1,0)$ are saddle in nature. Points $P_{2}^\infty\, (0,1),  P_{4}^\infty\, (0,-1)$  are non-hyperbolic and hence the stability needs to be checked with an extra effort by the
center manifold theory. However, by using a quick numerical check, we found that point $P_2^\infty$ is saddle and $P_4^\infty$ is stable for various values of model parameters $n, \xi$ and $w=0$. We note that of all critical points at infinity, only a saddle point $P_2^\infty$ corresponds to an accelerated solution, but it is unphysical. Thus, critical points at infinity cannot describe either late time or early time behavior of the Universe and hence are not phenomenologically interesting. Therefore, here we do not present a detailed calculation of the analysis at infinity.  In the next section, we discuss the possible bifurcation scenarios occurring at the non-hyperbolic condition of each critical point.

\subsection{Bifurcation Scenarios}\label{sub:bif}
In this section, we shall discuss the occurrence of  local bifurcation of the system \eqref{eq:x_ST_pow}-\eqref{eq:y_ST_pow} with respect to parameters $\xi$ and $n$. Then we extract the condition on  $\xi$ and $n$ under which the present model describes the generic evolution of the Universe. In the case of a two-dimensional system, we can completely characterize the structural stability by Peixoto's theorem. However, we cannot extend the theorem to a system of dimensions greater than two \cite{perko2013differential}.	 As a result of Peixoto's theorem, a structurally stable system guarantees an open dense subset of initial conditions leading to a generic evolution (see appendix \ref{soto}).

In general, the necessary condition for the occurrence of bifurcation of the system is the non-hyperbolicity of the critical point.  However, in the two-dimensional system, according to Peixoto's theorem,  the existence of non-hyperbolic critical points also implies the structural instability of the system  \cite{perko2013differential}. As we have seen in the previous section,  the system \eqref{eq:x_ST_pow}-\eqref{eq:y_ST_pow} contains  non-hyperbolic points  $P_2^\infty$ and  $P_4 ^\infty$ on the Poincar\'e  sphere, therefore, the vector field of the system is structurally unstable. Moreover, finite critical points can be non-hyperbolic for some values of $\xi$, $n$ and $w$. In what follows, similar to the general case,  we again prepare the bifurcation diagrams (Fig. \ref{fig:region_w_0_pow}) for each critical point in the $(\xi,n)$ parameter space and then apply the Sotomayor's theorem (see the appendix \ref{soto} for the statement). The theorem's main aim is to analytically investigate and specify the types of bifurcation.

The local bifurcation diagram for point $B_1$ is given in Fig. \ref{fig:region_A6_w0_pow}. A topological change occurs as this point changes from stable node to saddle along the curve D1 via a non-hyperbolic saddle node.

Another bifurcation occurs for points $B_{2+}$ and $B_{2-}$ along the curves D2 and D3 respectively, where both the points undergo an upheaval from unstable node to saddle via a non-hyperbolic saddle node  (see Figs. \ref{fig:region_A7p_w_0_pow}, \ref{fig:region_A7n_w_0_pow}).

Point $B_{3}$  changes its stability from a stable node to a saddle along the bifurcation curves D1, D4, D5 of Fig. \ref{fig:region_A8_w_0_pow}.  For a particular case, $n=2$, this point exhibits an effective cosmological constant behavior and for $n=6$, the point's dynamics change from an unaccelerated to an accelerated behavior.

Out of the above mentioned critical points, point $B_{4}$ is an interesting point which can explain the late time behavior of the Universe. This  point undergoes a stability change from an unstable node to a saddle along the curves D2 and D3 of Fig.  \ref{fig:region_A9_w_0_pow}. Again, a change from a stable node to a saddle occurs when the point passes through  the bifurcation curves  D4 and D5 of Fig.  \ref{fig:region_A9_w_0_pow}.  As this point can represent interesting late time Universe, using bifurcation diagram (Fig. \ref{fig:region_A9_w_0_pow}), we summarize the stability property of this point for different range of $\xi,n, w$ as follows:

\begin{itemize}
	\item Point $B_4$ is not stable when 
	\begin{enumerate}
		\item[i)] $n > \frac{3\,w+7}{2} +\frac {\sqrt {9\,{w}^{2}{\xi}^{2}+66\,w{\xi}^{2}+12\,w\xi+73\,{\xi}^{2}+12\,\xi}}{2\xi}\,$\\ (regions I \& II of Fig. \ref{fig:region_A9_w_0_pow}),\\
		or,
		\item[ii)] $\frac{4 \xi-\sqrt{6 \xi +36 \xi^2}}{\xi}<n<$ \\
		$\frac{3\,w+7}{2} -\frac {\sqrt {9\,{w}^{2}{\xi}^{2}+66\,w{\xi}^{2}+12\,w\xi+73\,{\xi}^{2}+12\,\xi}}{2\xi}\,$~~~~~~\\ (regions IV \& V of Fig. \ref{fig:region_A9_w_0_pow}).
	\end{enumerate}
	\item  	It is stable when 
	\begin{enumerate}
		\item[iii)] $4\leq n < \frac{3\,w+7}{2} +\frac {\sqrt {9\,{w}^{2}{\xi}^{2}+66\,w{\xi}^{2}+12\,w\xi+73\,{\xi}^{2}+12\,\xi}}{2\xi}\,,$~~~\\
		\text{or,}\\~~~~~~~~~~~$2\leq n \leq 4\,,$\\
		\text{or,}\\ ~~~ $\frac{3\,w+7}{2} -\frac {\sqrt {9\,{w}^{2}{\xi}^{2}+66\,w{\xi}^{2}+12\,w\xi+73\,{\xi}^{2}+12\,\xi}}{2\xi}\,<n \leq 2\,$~~~\\
		(region III of Fig. \ref{fig:region_A9_w_0_pow}),\\
		or,
		\item[iv)] $n<\frac{4 \xi-\sqrt{6 \xi +36 \xi^2}}{\xi}\,$ (region VI of Fig. \ref{fig:region_A9_w_0_pow}).
	\end{enumerate}
\end{itemize}

It is worth mentioning that independent of the value of $\xi$, this point describes a stable phantom attractor behavior for $2<n<4$. Also, when $n=4$, point $B_4$ belongs to a stable region of parameter space. Therefore, as discussed in Sec. \ref{sec:DS_gen} (i.e., model with the condition  $\lambda_{F_\ast}=\frac{\lambda_{V_\ast}}{2}$) this point corresponds  to a deSitter solution in GR for $n=4$. Similar result also holds for model considered in \cite{Humieja:2019ywy} with biquadratic potential.

From the above discussion on the bifurcation diagrams, we found the possibility of transcritical bifurcation exhibited by the system \eqref{eq:x_ST_pow}-\eqref{eq:y_ST_pow}. In what follows, we summarize the occurrence of bifurcation and mathematically analyze the transcritical bifurcation with $\xi$ as the bifurcation parameter using Sotomayor's theorem.

\begin{stmt}\label{trans_1}
	The system \eqref{eq:x_ST_pow}-\eqref{eq:y_ST_pow} undergoes a transcritical bifurcation  of critical points $B_1$ and $B_3$ along a curve $$\xi=\frac{3}{2}\,{\frac {1-{w}^{2}}{n(3w-1)-6\,(w+1)}}\,,$$ for fixed $w$. \\
\end{stmt}
{\it Proof:}~	In this case, the bifurcation value is $$\xi=\xi_0=\frac{3}{2}\,{\frac {1-{w}^{2}}{n(3w-1)-6\,(w+1)}}\,,$$  in which the points $B_1$ and $B_3$ coincide and the real part of the eigenvalue corresponding to common critical point vanishes.

The eigenvector corresponding to a simple eigenvalue $\lambda=0$ (i.e., multiplicity is 1) of the Jacobian matrix $D{\bf f}(B_1,\xi_0)$ evaluated at a point $B_1$ when $\xi=\xi_0$  is $${\bf v}=\left[\frac{\sqrt{6(1-w^2)} \sqrt{n(3w-1)-6(1+w)}}{(3w-1)(w-1)}~~~~~1 \right]^T\,,$$ where $T$ stands for transpose and {\bf f} is the vector field of the system \eqref{eq:x_ST_pow}-\eqref{eq:y_ST_pow}.

Also, the eigenvector corresponding to a simple eigenvalue $\lambda=0$ of the transpose of the Jacobian matrix   is ${\bf w}=\left[0~~1 \right]^T\,.$

After, few simple algebraic calculations  one could easily verify that when $\xi=\xi_0$, we have
\begin{eqnarray}
&&{\bf w}^T {\bf f}_\xi(B_1, \xi_0) = 0\,, \\[1ex]
&& {\bf w}^T [{\bf Df}_\xi(B_1, \xi_0) {\bf v}] =\frac{1}{\left( 3\,{w}-1 \right) (w-1) {n}^{2}} \Big( \left( (3w-1)n(n+2) \right. \nonumber\\&& \left. -24\,(w+1) \right)  \left( 3\,nw-n-6\,w-6 \right)\Big)  \,,\label{soto1}  \\[1ex]
\text{and}\nonumber\\
&&{\bf w}^T [{\bf D^2 f}(B_1, \xi_0) ({\bf v, v})] = -\,{\frac { 6 \left( w+1 \right) n}{3\,w-1}}\,,\label{soto2}
\end{eqnarray}
where vector ${\bf f}_\xi$ denotes the   partial derivative of ${\bf f}$ with respect to $\xi$.  Note here that the right hand side of \eqref{soto1}, \eqref{soto2} is non-zero for any admissible value of $n$ and $w$,  otherwise $\xi$  is undefined or negative. Hence, by the Sotomayor's theorem, the system \eqref{eq:x_ST_pow}-\eqref{eq:y_ST_pow} undergoes a transcritical bifurcation  when $\xi=\frac{3}{2}\,{\frac {1-{w}^{2}}{n(3w-1)-6\,(w+1)}}$, which is represented by a curve D1 of Figs. \ref{fig:region_A6_w0_pow}, \ref{fig:region_A8_w_0_pow} for $w=0$. In a similar manner, one could also verify the following statements.
\begin{stmt}
	The system \eqref{eq:x_ST_pow}-\eqref{eq:y_ST_pow} undergoes a transcritical bifurcation  of critical points $B_{2\pm}$ and $B_4$ along a curve $$\xi=\frac{6}{n^2-8n-20}\,.$$ 
\end{stmt}
For $w=0$, the above equation is represented   by two branch curves D2, D3 of Figs. \ref{fig:region_A7p_w_0_pow}, \ref{fig:region_A7n_w_0_pow}, \ref{fig:region_A9_w_0_pow}.

\begin{stmt}
	The system \eqref{eq:x_ST_pow}-\eqref{eq:y_ST_pow} undergoes a transcritical bifurcation  of critical points  $B_3$ and $B_4$ along a curve $$\xi=\frac {3(w+1)}{{n}^{2}-3\,nw-7\,n-6\,w-6}\,,$$  for fixed $w$. \\
\end{stmt}
For $w=0$, the above equation is represented  by two branch  curves  D4 and D5 of Figs. \ref{fig:region_A8_w_0_pow}, \ref{fig:region_A9_w_0_pow}.

\begin{table*}
	\caption{\label{tab:gen_evo}Conditions on parameters $\xi, n, w$  for which dynamics of Universe governed by the system \eqref{eq:x_ST_pow}-\eqref{eq:y_ST_pow} undergoes a generic evolution. Here, $Q_{\pm}=\frac {3\,w\xi+7\,\xi\pm \sqrt {9\,{w}^{2}{\xi}^{2}+66\,w{\xi}^{2}+12\,w\xi+73\,{\xi}^{2}+12\,\xi}}{2\xi}\,.$}
	\begin{ruledtabular}
		\begin{tabular}{ccccc}
			Scenarios~~~ &~~~$\xi, n, w$~~~~ &~~~~ Starting point ~~~~ &~~~~     End point ~~  \\ \hline 
			I& $n>Q_+$ &   Unstable node  $B_{2+}$   &    Stable node/focus  $B_3$
			\\
			&  &   (decelerated scalar field expansion)     &  (decelerated matter scaling expansion) \\[1.5ex]
			II &$Q_-<n<Q_+$ &    Unstable node    $B_{2+}/B_{2-}$     &   Stable node   $B_4$    
			\\
			& & (decelerated scalar field expansion)  &   (decelerated/accelerated scalar field expansion)      \\[1.5ex]
			III & $Q_-<n<\frac {3(1-{w}^{2}+4\,\xi(w+1))}{2 \left( 3\,w-1 \right) \xi}$ &    Unstable node $B_{2-}$     &  Stable node/focus    $B_3$    
			\\
			& & (decelerated scalar field expansion)  &   (decelerated matter scaling expansion)   \\[1.5ex]
			IV & $n<\frac{4\xi-\sqrt{6\xi+36 \xi^2}}{\xi}$ &    Unstable node   $B_{2-}$     &    Stable node   $B_4$    
			\\
			& & (decelerated scalar field expansion)  &   (decelerated/accelerated scalar field expansion)   \\[1.5ex]
			V& $n>\frac{4\xi+\sqrt{6\xi+36 \xi^2}}{\xi}$ &    Unstable node   $B_{4}$     &   Stable focus    $B_3$   \\
			&  &   (decelerated scalar field expansion)     &  (decelerated matter scaling expansion) \\[1.5ex]
			VI & $n<\frac {3(1-{w}^{2}+4\,\xi(w+1))}{2 \left( 3\,w-1 \right) \xi}$ &    Unstable node   $B_{2-}$     &   Stable node   $B_1$   
			\\
			&  &   (decelerated scalar field expansion)     &  (decelerated matter scaling expansion) \\[1.5ex]
		\end{tabular}
	\end{ruledtabular}
\end{table*}

\begin{figure}[b!]
	\centering
	\includegraphics[width=6cm,height=6cm]{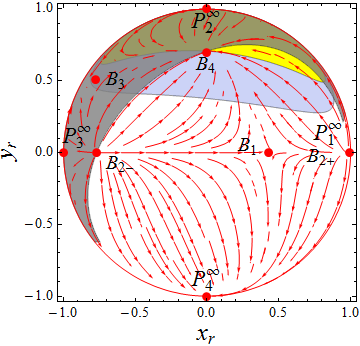}
	\caption{A global phase portrait of the system \eqref{eq:x_ST_pow}-\eqref{eq:y_ST_pow} for the generic scenario (II) with $w=0, n=4, \xi=0.1$. The grey shaded region represents the non-physical region (i.e., $\Omega_m<0$). The yellow shaded region corresponds to the phantom like solution ($w_{\rm eff}<-1$) and the blue shaded region corresponds to a quintessence like solution ($-1<w_{\rm eff}<-\frac{1}{3}$).}
	\label{fig:poin_pow_n_pos}
\end{figure}

\begin{figure}
	\centering
	\subfigure[]{%
		\includegraphics[width=7cm,height=7cm]{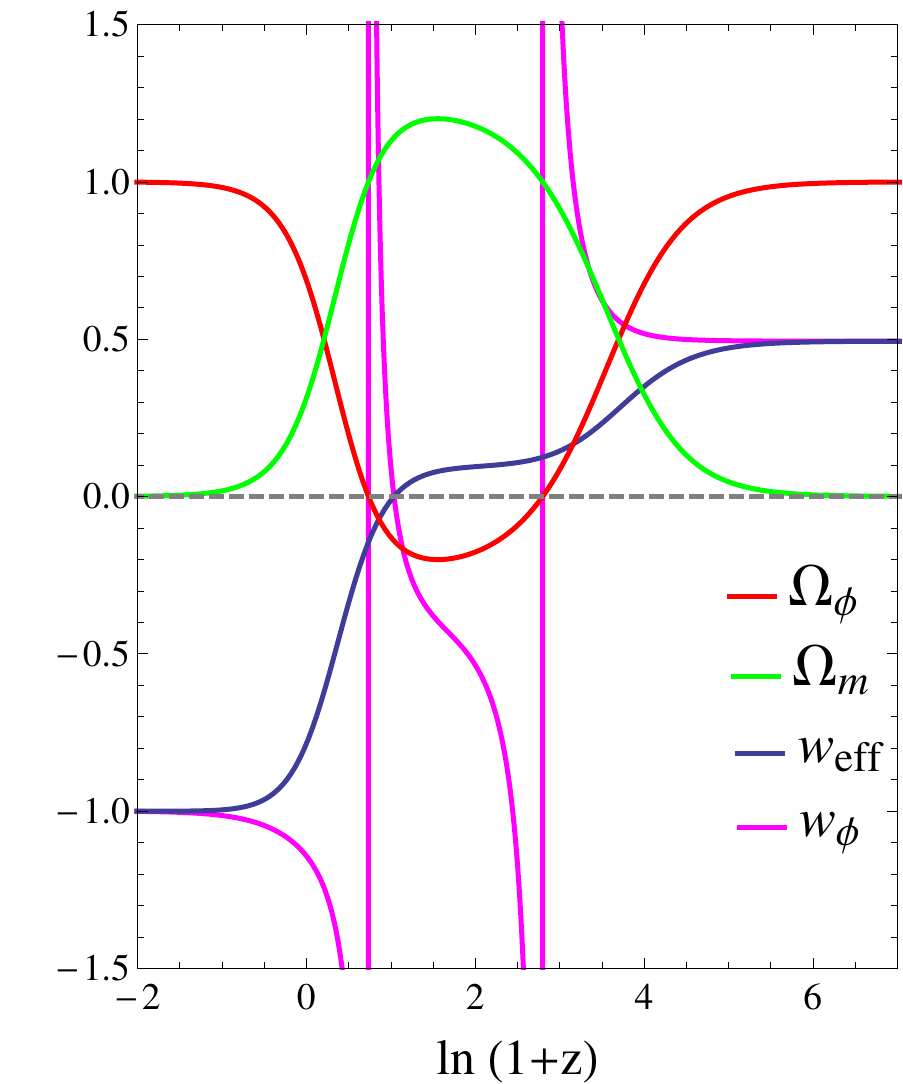}\label{fig:rmde}}
	\qquad 
	\subfigure[]{%
		\includegraphics[width=7cm,height=7cm]{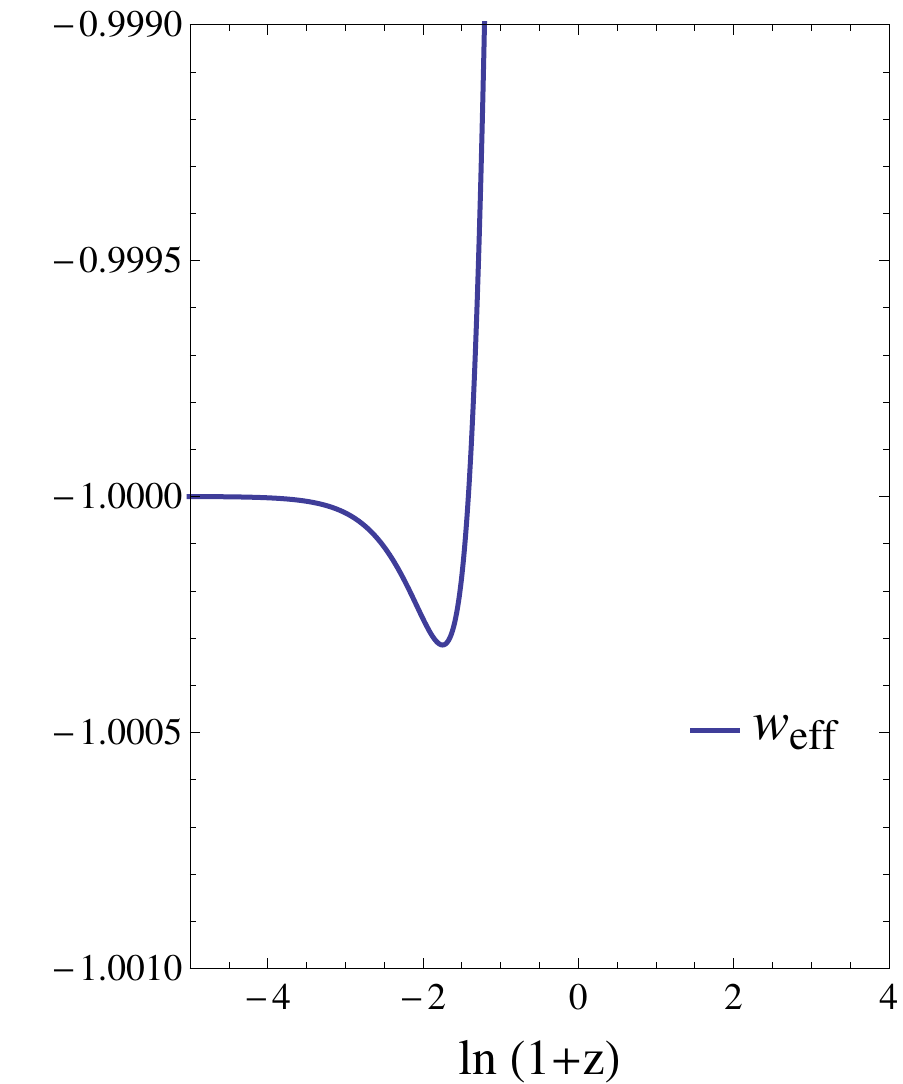}\label{fig:weff_zoom}}
	\qquad
	\subfigure[]{%
		\includegraphics[width=7cm,height=7cm]{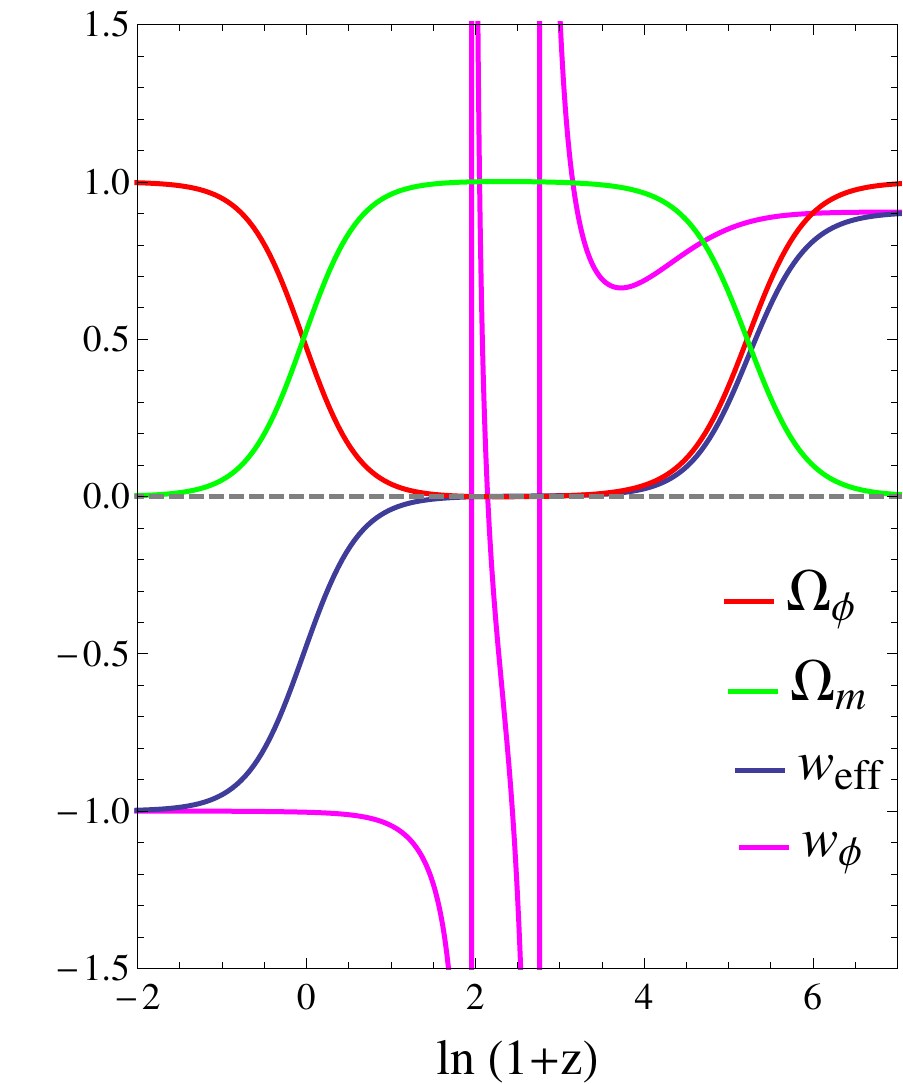}\label{fig:smde}}
	\caption{Time evolution of the scalar field energy density $\Omega_\phi$, matter energy density $\Omega_m$,  effective EoS $w_{\rm eff}$ and EoS of scalar field $w_\phi$ for  a generic scenario (II). In (a) we have taken  $w=0, n=4, \xi=0.1$, (b) A closer look on the evolution of $w_{\rm eff}$ given in (a) at late time, (c)   $w=0, n=4, \xi=0.001$.}
	\label{fig:parameters}
\end{figure}

The above bifurcation scenarios help us to properly separate the Universe's evolution into generic and non-generic evolution. In Table \ref{tab:gen_evo}, we present the conditions satisfied by the model parameters to describe the Universe's possible generic cosmological evolution. This type of evolution is interesting as it can determine the initial phase and the final phase of the Universe's cosmological evolution for a wide range of initial conditions.  Out of all the scenarios presented in Table \ref{tab:gen_evo}, only scenarios II and IV are of cosmological interest as they could explain the late-time behavior of the Universe and fits with various CMB and BAO observational data  \cite{Umilta:2015cta}.  In Fig. \ref{fig:poin_pow_n_pos}, we present the global phase space diagram for a particular choice of $n$ and $\xi$ which corresponds to a generic evolution of scenario II where the model exhibits a  late-time acceleration (similar dynamics is exhibited by scenario IV).  It is important to remark here that for the parameters within the range of generic scenario II, there is no bifurcation, so it is sufficient to present only one phase portrait for this scenario. Furthermore, in Sec. \ref{sec:conc}, we will discuss the evolution of cosmological quantities described by this model for parameters corresponding to scenario II.

	\section{Discussion and Conclusion}\label{sec:conc}

In this work, we studied the global qualitative cosmological dynamics of a non-minimal coupled scalar field for a general class of coupling function and potential. We focused on the bifurcation analysis to investigate the effect of varying the model parameters on the global dynamics.  The main objective of applying bifurcation theory is to determine the existence of generic evolutionary scenarios. It will help us identify the Universe's initial and final phases for a wide range of initial conditions. The bifurcation theory also allows us to understand how physics described by the phase space change with parameters.

The general model described by the system \eqref{eq:x_ST}-\eqref{eq:LV_ST} contains different interesting cosmological solutions for different model parameters. For instance, the present model exhibits a scalar field dominated solution $A_{2\pm}$, resembling a stiff matter Universe or radiation Universe for some choice of coupling and potential parameters.  The scalar field's sole contribution to an early radiation epoch is an interesting scenario missing in a minimal coupled canonical scalar field within the GR context. For some choice of coupling function and potential, the general model also shows the presence of  DM-DE scaling solutions $A_1$ and $A_3$ describing an intermediate matter epoch. Lastly, the present model exhibits a late time acceleration of the Universe via a critical point $A_4$. Thus, from the analysis presented in Sec. \ref{sec:DS_gen}, one can select a class of non-minimal coupled scalar field models describing a viable sequence of cosmic evolution. In particular, for models with $\lambda_{F_\ast}=\frac{\lambda_{V_\ast}}{2}$, point $A_4$ corresponds to a late time de Sitter solution of the GR case. Hence, such a class of scalar-tensor theories converges towards a structurally stable GR-based model, i.e., the $\Lambda$CDM model. This is a common feature of the scalar-tensor gravity which has been verified numerically and analytically \cite{GarciaBellido:1990jz,Damour:1993id,Mimoso:1998dn,Serna:2002fj,Jarv:2010xm,Jarv:2011sm}. Therefore, the present analysis identifies a broad class of scalar-tensor models generically possessing this property.  Further, it is possible that the point $A_4$ corresponds to a late time super-accelerated phase ($w_{\rm eff}<-1$) when $\lambda_{F_\ast}<\lambda_{V_\ast}<2 \lambda_{F_\ast}$. Hence, in the presence of a non-minimal coupling, the model can explain a super-accelerated phase ($w_{\rm eff}<-1$)  without the need for the introduction of a phantom field. Depending on the potential and coupling function,  point $A_4$ is a saddle in nature and hence can also describe the possible inflationary exit phenomenon.

In Sec. \ref{sec:example}, we illustrate the  global dynamics in detail by considering a simple example corresponding to a quadratic  coupling function and a power-law form  potential. For this particular example, we  find the range of parameters $n$ and $\xi$ describing a generic evolution scenario (see Table \ref{tab:gen_evo}). In Fig. \ref{fig:parameters}, we have  plotted the evolution of cosmological parameters against the redshift $\rm z$  for a choice of parameters  corresponding to one of the generic scenarios, i.e., scenario (II) which starts from an unstable decelerated solution ($B_{2\pm}$) towards a deSitter like solution $B_4$ (a similar evolution is obtained for scenario IV). Recall that redshift ${\rm z}=-1+\frac{1}{a}$ where  the present value of the scale factor taken to be unity, with ${\rm z}=0$ corresponds to the present Universe and ${\rm z}=-1$ corresponds to its infinite future. It is worth mentioning here that only the parameters' values corresponding to the generic evolution of scenarios II or IV satisfy various observational constraints coming from {\it Planck} and BAO datasets  \cite{Ballardini:2016cvy}. Therefore,  mathematically, stability analysis and bifurcation theory help us to locate various physically rich models. For instance, we can find the parameter's range which can generically describe the thermal history of the Universe starting from an early radiation domination ($w_{\rm eff} \simeq \frac{1}{3}$) or stiff matter solution  ($w_{\rm eff} \simeq 1$) represented by points $B_{2\pm}$  towards a DE dominated solution  $B_4$  ($w_{\rm eff} \simeq -1$) via a matter like solution   $B_1$ ($w_{\rm eff} \simeq 0$) i.e., $B_{2\pm} \to B_1 \to B_4$ (see Figs.  \ref{fig:poin_pow_n_pos},  \ref{fig:parameters}).

The possibility that $\Omega_m>1$ and $w_{\phi}$ diverges at the onset of a matter-dominated era (as explained in paragraphs after Eqs. \eqref{Om_reln} and \eqref{weff_para}) can also be confirmed  from Fig. \ref{fig:rmde}.  The divergence of $w_\phi$, however, does not cause any issue to the evolution behavior of $w_{\rm eff}$, as this corresponds to the vanishing of scalar field energy density.  The overdensity of matter component is not a surprise in cosmological models where interaction between different components occurs \cite{Quartin:2008px}. Such behavior is more visible by comparing a change in the behavior of $\Omega_m$ and $w_\phi$ to $\xi$ (i.e., $\lambda_F$) from Figs. \ref{fig:rmde} and \ref{fig:smde} (the range of divergence of $w_\phi$ reduces and $\Omega_m \rightarrow 1$ as $\xi \rightarrow 0$).  Interestingly, the behavior is consistent with the result of \cite{Finelli:2007wb}, that the background dynamics of the present model approach GR with the cosmological constant in the limit $\xi \to 0$.  For a generic scenario (II),  there is also a possibility of crossing the phantom divide line and eventually the solution settles down towards a cosmological constant behavior (see Fig. \ref{fig:poin_pow_n_pos}). We can confirm it by taking a closer look at the late time evolution of $w_{\rm eff}$ (see Fig. \ref{fig:weff_zoom}).  In Fig. \ref{fig:parameters}, we choose initial conditions such that the Universe agrees with the current observational data i.e., $\Omega_m \approx 0.3, w_{\rm eff} \approx -0.8$ \cite{Aghanim:2018eyx}.

Our analysis shows that the non-minimal coupled scalar field model describes a rich cosmological history of the Universe. For instance, the model exhibits the transition: radiation $\rightarrow$ matter $\rightarrow$ DE and the possible crossing of phantom divide line but avoiding the big rip singularity. It is worth noting that only by using the stability analysis, one can determine such a transition. However, bifurcation tools help us locate the parameter values describing such dynamics without a fine-tuning of initial conditions. With the present choice of variables \eqref{dyn_var}, this model can explain the graceful exit scenario and the late time DE era separately. Such a result is a common feature of many classes of scalar-tensor theories. Therefore, it is of interest to extend the analysis to the case where the scalar field is coupled non-minimal with gravity and matter, as discussed in \cite{Pettorino:2008ez}.  Such discussion is beyond the scope of the present work.

\acknowledgements
JD acknowledges the support of the Core research grant of SERB, Department of Science and Technology India (File No. CRG/2018/001035), and the Associate program of IUCAA.	We would like to thank the referee for the comments which help us to improve the
manuscript. Finally, the authors thank Laur J\"arv for useful discussions.

	\appendix
\section{Center manifold theory analysis for critical point $B_1$} \label{cmt_b1}
In this case, we analyze the stability behavior of this point when $\xi=\frac {3(1-{w}^{2})}{2(3\,nw-n-6\,w-6)}$ i.e., along the curve D1 of Fig. \ref{fig:region_A6_w0_pow}. Under this condition, the eigenvalues corresponds to this point are $0$ and $ \frac {3(nw-n+2\,w+2)}{2n}$. First, we make a coordinate transformation in such a way that this point is shifted to the  origin. The transformation is given by $x \rightarrow x+ \left(\frac{\sqrt{6\left( 1-{w}^{2} \right)  \left( 3\,nw-n-6\,w-6 \right)}}{(w-1)n} \right), y\rightarrow y$ and we apply to the dynamical system \eqref{eq:x_ST_pow}-\eqref{eq:y_ST_pow} (which we have not presented here).  Then, the resulting system is then transformed into a standard form upon the introduction of new variables $X, Y$ given by
\[\left(\begin{array}{c}
X\\
Y \end{array} \right)=\left(\begin{array}{cc}
1   & {\frac {1}{3\,{w}^{2}-4\,w+1}\sqrt {{\frac { 6\left( 1-{w}^{2} \right)  \left( 3\,nw-n-6\,w-6 \right) }{nw-n-6\,w-6}}}} \\
0    &1   \\ \end{array} \right) \left(\begin{array}{c}
x\\
y \end{array} \right) \,. \]
On employing these new variables, we can rewrite  the  corresponding dynamical system as
\[\left(\begin{array}{c}
X'\\
Y' \end{array} \right)=\left(\begin{array}{cc}
\frac {3(nw-n+2\,w+2)}{2n}   & 0\\
0    & 0   \\ \end{array} \right) \left(\begin{array}{c}
X\\
Y \end{array} \right) + \left(\begin{array}{c}
g_1\\
f_1 \end{array} \right) \,, \] 
where $g_1$, $f_1$ have not been presented due to their length. Then by the center manifold theory, there exist a continuously differentiable function $h: \mathbb{R}\rightarrow \mathbb{R}$ defined by $X=h(Y)=a_2 Y^2+a_3 Y^3 +\mathcal{O}(4)$, where $a_2, a_3 \in \mathbb{R}$ are determined from the quasi-linear equation
\begin{equation}\label{quasi}
Dh(Y) \left[A+f_1(Y, h(Y))\right]- B h(Y)-g_1(Y,h(Y))=0\,,
\end{equation}
where $A=0$, $B= \frac{3}{2}\frac{n (w-1)+2(w+1)}{n}  $ and $D$ denotes the derivative with respect to $Y$. On substituting the expression of $A, B, f_1, g_1$ and $h$ in \eqref{quasi}, we obtain the values of $a_2$ and $a_3$ given by
\begin{eqnarray}
a_2&=&  \frac{2\sqrt{{\frac {6 \left( 1-w^2 \right)  }{3\,nw-n-6\,w-6}}} \, \Big[ \left( w+1 \right)  \left( n+1 \right)  \left( 3\,nw-n-6\,w-6 \right)  \Big]}{\left( 1-w \right)  \left( 3\,w-1 \right) ^{2} \left( nw-n+2\,w+2
	\right) },\nonumber\\[1ex]
a_3&=& \frac{\sqrt{{\frac {6 \left( 1-w^2 \right)  }{3\,nw-n-6\,w-6}}}}{(w-1)(3 w-1)^3(n w-n+2 w+2)(-3 w+n-3)} \Big[ n^2 \nonumber\\
&&  \left( w+1 \right)  \left( 3\,nw-n-6\,w-6 \right)  \left( 12\,{n}^{3} w -33\,{n}^{2}{w}^{2}+12\,{n}^{3} \right.\nonumber\\&& \left.  -60\,{n}^{2}w -48\,n{w}^{2}-19\,{n}^{2} -88\,nw-24\,{w}^{2}-40\,n \right.\nonumber\\&& \left.  -48\,w-24 \right) \Big]\,. \nonumber
\end{eqnarray}
The following equation then gives the flow on the corresponding local center manifold
\begin{equation}
Y'=A+f_1(Y,h(Y))\,,
\end{equation}
i.e.,
\begin{eqnarray}
Y' &=& \frac{3n (w+1)}{1-3w} Y^2+\mathcal{O}(3).
\end{eqnarray}
This equation implies point $B_1$ is  always saddle when $\xi=\frac {3(1-{w}^{2})}{2(3\,nw-n-6\,w-6)}$.
\section{Center manifold theory analysis for critical point $B_4$} \label{cmt_b4}
In this case, we analyze the stability behavior of the point $B_4$ when  $\xi=\frac{6}{n^2-8n-20}$ (i.e., curves D2, D3 of Fig. \ref{fig:region_w_0_pow}) or $\xi=\frac {3(w+1)}{n^{2}-3\,nw-7\,n-6\,w-6}$  (i.e., curves D4, D5 of Fig. \ref{fig:region_w_0_pow}). Following the similar analysis as  for the point $B_1$, we obtained that the flow on the corresponding local center manifold in both the cases is given by
\begin{eqnarray}
Y' &=&- \frac{3(n+2)}{n-4} Y^2+\mathcal{O}(3)\,,
\end{eqnarray}
which  implies the saddle nature of point $B_4$  when  $\xi=\frac{6}{n^2-8n-20}$ or $\xi=\frac {3(w+1)}{{n}^{2}-3\,nw-7\,n-6\,w-6}$.\\
\section{A brief introduction to bifurcation tools} \label{soto}	
Here, we present preliminaries and two landmark theorems   required for understanding bifurcation analysis. For more details, reader can refer to Refs. \cite{kuznetsov2013elements,perko2013differential,pei}.
\begin{defn}
	Two dynamical systems are said to be locally topologically equivalent if
	there exists a homeomorphism (i.e., a continuous invertible function whose inverse is also continuous)  mapping orbits of one system onto orbits of another system preserving the direction of time.
\end{defn}

If the qualitative behavior remains topologically equivalent for all nearby vector fields, then the system or the vector field is said to be {\it structurally stable}.  For a two dimensional system, Peixoto's theorem completely characterizes the structural stability of vector fields on a compact, two-dimensional manifold. To understand the landmark theorem,  we present a definition of  non-wandering points.

\begin{defn}
	A point $x $ on a manifold is a {\bf non-wandering point}  of the flow $\phi_t$  defined by the vector field if for any neighborhood $U$ of $x$ and for any $T > 0$ there is  $t > T$ such that $\phi_t (U)\cap U $ is a non-empty set.
\end{defn}

\noindent{\bf Peixoto's Theorem}~ {\it Let ${\bf f}$ be a $C^1$-vector field on a compact, two dimensional, differentiable manifold $M$. Then $f$ is structurally stable on $M$
	if and only if
	\begin{itemize}
		\item[(i)] the number of critical points and cycles is finite and each is hyperbolic;
		\item[(ii)]  there are no trajectories connecting saddle points; and
		\item[(iii)] the set of all non-wandering points consists of critical points and limit cycles only.
\end{itemize}}

Recall that a limit  cycle  is an isolated closed path. By {\it isolated}, it  means that neighboring trajectories are either spiral toward or away from  a limit  cycle.

When the dynamical system depends on some parameters,  the system's phase portrait also varies as parameters vary. Thus, either the phase portrait remains topologically equivalent, or its topology changes as parameter changes. The occurrence of topologically inequivalent phase portraits under a change of parameters is called a {\it bifurcation.} A parameters' value at which the topology changes is called a bifurcation value.

Different types of bifurcation can occur for a  given dynamical system. One can classify different types of bifurcation using Sotomayor's theorem (see \cite{kuznetsov2013elements} for more details). Since, in our work, we obtained only transcritical bifurcation, in what follows, we state this theorem to determine the occurrence of transcritical bifurcation.

\noindent{\bf Sotomayor's theorem for transcritical bifurcation}\\
{\it  Consider the system ${\bf \dot{x}=f(x, \mu)}$ where the set of vector fields ${\bf f}$  equipped with the standard $C^1$-norm\footnote{The $C^1$-norm of a vector field ${\bf f}$ on an open subset $E$ of $\mathbb{R}^n$ can be defined as $\vert \vert {\bf f} \vert \vert_1 = \sup_{{\bf x } \in E} |{\bf f(x)}| +\sup_{{\bf x} \in E} ||D{\bf f}({\bf x})||$, where $|~\cdot~|$ denotes the Euclidean norm in $\mathbb{R}^n$ and and  $||~\cdot~||$ denotes the usual norm of the matrix $D{\bf f}({\bf x})$.} forms a Banach space such that ${\bf f(x_0},\mu_0)={\bf 0}$. Suppose the Jacobian matrix $(A \equiv D{\bf f(x_0},\mu_0))$ has a simple eigenvalue $\lambda=0$ with eigenvector ${\bf v}$ and ${\bf w}$ is an eigenvector of the transpose of the Jacobian matrix $A^T$ corresponds to the eigenvalue $\lambda=0$. Then the above system experiences a transcritical bifurcation at the equilibrium point $ {\bf x_0}$ as the  parameter $\mu$ varies through the bifurcation value $\mu=\mu_0$, if the following  three conditions hold:
	\begin{itemize}
		\item ${\bf w^T f_\mu (x_0},\mu_0)={\bf 0}$
		\item ${\bf w^T [D f_\mu (x_0},\mu_0) {\bf v}] \neq {\bf 0}$~~and 
		\item	${\bf w^T [D^2 f (x_0},\mu_0) {\bf (v,v)}] \neq {\bf 0}$
	\end{itemize}
	where
	\begin{gather*}
	{\bf D f_\mu (x_0,\mu_0)}{\bf v}=\sum_{i=1}^n \frac{\partial {\bf f_\mu (x_0},\mu_0)}{\partial x_i } v_i \,,\\
	{\bf D^2 f (x_0},\mu_0) {\bf (v,v)}=\sum_{i, j=1}^n \frac{\partial^2 {\bf f (x_0},\mu_0)}{\partial x_i x_j} v_i v_j\,,
	\end{gather*}		
	and ${\bf f}_\mu$ denotes partial derivative of ${\bf f}$ with respect to $\mu$.}

	\bibliography{bifurcation}

\begin{thebibliography}{70}%
\makeatletter
\providecommand \@ifxundefined [1]{%
 \@ifx{#1\undefined}
}%
\providecommand \@ifnum [1]{%
 \ifnum #1\expandafter \@firstoftwo
 \else \expandafter \@secondoftwo
 \fi
}%
\providecommand \@ifx [1]{%
 \ifx #1\expandafter \@firstoftwo
 \else \expandafter \@secondoftwo
 \fi
}%
\providecommand \natexlab [1]{#1}%
\providecommand \enquote  [1]{``#1''}%
\providecommand \bibnamefont  [1]{#1}%
\providecommand \bibfnamefont [1]{#1}%
\providecommand \citenamefont [1]{#1}%
\providecommand \href@noop [0]{\@secondoftwo}%
\providecommand \href [0]{\begingroup \@sanitize@url \@href}%
\providecommand \@href[1]{\@@startlink{#1}\@@href}%
\providecommand \@@href[1]{\endgroup#1\@@endlink}%
\providecommand \@sanitize@url [0]{\catcode `\\12\catcode `\$12\catcode
  `\&12\catcode `\#12\catcode `\^12\catcode `\_12\catcode `\%12\relax}%
\providecommand \@@startlink[1]{}%
\providecommand \@@endlink[0]{}%
\providecommand \url  [0]{\begingroup\@sanitize@url \@url }%
\providecommand \@url [1]{\endgroup\@href {#1}{\urlprefix }}%
\providecommand \urlprefix  [0]{URL }%
\providecommand \Eprint [0]{\href }%
\providecommand \doibase [0]{http://dx.doi.org/}%
\providecommand \selectlanguage [0]{\@gobble}%
\providecommand \bibinfo  [0]{\@secondoftwo}%
\providecommand \bibfield  [0]{\@secondoftwo}%
\providecommand \translation [1]{[#1]}%
\providecommand \BibitemOpen [0]{}%
\providecommand \bibitemStop [0]{}%
\providecommand \bibitemNoStop [0]{.\EOS\space}%
\providecommand \EOS [0]{\spacefactor3000\relax}%
\providecommand \BibitemShut  [1]{\csname bibitem#1\endcsname}%
\let\auto@bib@innerbib\@empty
\bibitem [{\citenamefont {Capozziello}\ and\ \citenamefont
  {de~Ritis}(1993)}]{Capozziello:1993tr}%
  \BibitemOpen
  \bibfield  {author} {\bibinfo {author} {\bibfnamefont {S.}~\bibnamefont
  {Capozziello}}\ and\ \bibinfo {author} {\bibfnamefont {R.}~\bibnamefont
  {de~Ritis}},\ }\bibfield  {title} {\enquote {\bibinfo {title} {{Scale factor
  duality and general transformations for string cosmology}},}\ }\href
  {\doibase 10.1142/S0218271893000258} {\bibfield  {journal} {\bibinfo
  {journal} {Int. J. Mod. Phys.}\ }\textbf {\bibinfo {volume} {D2}},\ \bibinfo
  {pages} {367--371} (\bibinfo {year} {1993})}\BibitemShut {NoStop}%
\bibitem [{\citenamefont {Cardone}\ \emph {et~al.}(2005)\citenamefont
  {Cardone}, \citenamefont {Troisi},\ and\ \citenamefont
  {Capozziello}}]{Cardone:2005aa}%
  \BibitemOpen
  \bibfield  {author} {\bibinfo {author} {\bibfnamefont {Vincenzo~F.}\
  \bibnamefont {Cardone}}, \bibinfo {author} {\bibfnamefont {A.}~\bibnamefont
  {Troisi}}, \ and\ \bibinfo {author} {\bibfnamefont {S.}~\bibnamefont
  {Capozziello}},\ }\bibfield  {title} {\enquote {\bibinfo {title}
  {{Inflessence: A Phenomenological model for inflationary quintessence}},}\
  }\href {\doibase 10.1103/PhysRevD.72.043501} {\bibfield  {journal} {\bibinfo
  {journal} {Phys. Rev.}\ }\textbf {\bibinfo {volume} {D72}},\ \bibinfo {pages}
  {043501} (\bibinfo {year} {2005})},\ \Eprint
  {http://arxiv.org/abs/astro-ph/0506371} {arXiv:astro-ph/0506371 [astro-ph]}
  \BibitemShut {NoStop}%
\bibitem [{\citenamefont {Bahamonde}\ \emph {et~al.}(2019)\citenamefont
  {Bahamonde}, \citenamefont {Marciu},\ and\ \citenamefont
  {Rudra}}]{Bahamonde:2019urw}%
  \BibitemOpen
  \bibfield  {author} {\bibinfo {author} {\bibfnamefont {Sebastian}\
  \bibnamefont {Bahamonde}}, \bibinfo {author} {\bibfnamefont {Mihai}\
  \bibnamefont {Marciu}}, \ and\ \bibinfo {author} {\bibfnamefont {Prabir}\
  \bibnamefont {Rudra}},\ }\bibfield  {title} {\enquote {\bibinfo {title}
  {{Dynamical system analysis of generalized energy-momentum-squared
  gravity}},}\ }\href {\doibase 10.1103/PhysRevD.100.083511} {\bibfield
  {journal} {\bibinfo  {journal} {Phys. Rev.}\ }\textbf {\bibinfo {volume}
  {D100}},\ \bibinfo {pages} {083511} (\bibinfo {year} {2019})},\ \Eprint
  {http://arxiv.org/abs/1906.00027} {arXiv:1906.00027 [gr-qc]} \BibitemShut
  {NoStop}%
\bibitem [{\citenamefont {Basilakos}\ \emph {et~al.}(2019)\citenamefont
  {Basilakos}, \citenamefont {Leon}, \citenamefont {Papagiannopoulos},\ and\
  \citenamefont {Saridakis}}]{Basilakos:2019dof}%
  \BibitemOpen
  \bibfield  {author} {\bibinfo {author} {\bibfnamefont {Spyros}\ \bibnamefont
  {Basilakos}}, \bibinfo {author} {\bibfnamefont {Genly}\ \bibnamefont {Leon}},
  \bibinfo {author} {\bibfnamefont {G.}~\bibnamefont {Papagiannopoulos}}, \
  and\ \bibinfo {author} {\bibfnamefont {Emmanuel~N.}\ \bibnamefont
  {Saridakis}},\ }\bibfield  {title} {\enquote {\bibinfo {title} {{Dynamical
  system analysis at background and perturbation levels: Quintessence in severe
  disadvantage comparing to $\Lambda$CDM}},}\ }\href {\doibase
  10.1103/PhysRevD.100.043524} {\bibfield  {journal} {\bibinfo  {journal}
  {Phys. Rev.}\ }\textbf {\bibinfo {volume} {D100}},\ \bibinfo {pages} {043524}
  (\bibinfo {year} {2019})},\ \Eprint {http://arxiv.org/abs/1904.01563}
  {arXiv:1904.01563 [gr-qc]} \BibitemShut {NoStop}%
\bibitem [{\citenamefont {Alho}\ \emph {et~al.}(2019)\citenamefont {Alho},
  \citenamefont {Uggla},\ and\ \citenamefont {Wainwright}}]{Alho:2019jho}%
  \BibitemOpen
  \bibfield  {author} {\bibinfo {author} {\bibfnamefont {Artur}\ \bibnamefont
  {Alho}}, \bibinfo {author} {\bibfnamefont {Claes}\ \bibnamefont {Uggla}}, \
  and\ \bibinfo {author} {\bibfnamefont {John}\ \bibnamefont {Wainwright}},\
  }\bibfield  {title} {\enquote {\bibinfo {title} {{Perturbations of the
  Lambda-CDM model in a dynamical systems perspective}},}\ }\href {\doibase
  10.1088/1475-7516/2019/09/045} {\bibfield  {journal} {\bibinfo  {journal}
  {JCAP}\ }\textbf {\bibinfo {volume} {1909}},\ \bibinfo {pages} {045}
  (\bibinfo {year} {2019})},\ \Eprint {http://arxiv.org/abs/1904.02463}
  {arXiv:1904.02463 [gr-qc]} \BibitemShut {NoStop}%
\bibitem [{\citenamefont {Dutta}\ \emph
  {et~al.}(2018{\natexlab{a}})\citenamefont {Dutta}, \citenamefont {Khyllep},\
  and\ \citenamefont {Tamanini}}]{Dutta:2017wfd}%
  \BibitemOpen
  \bibfield  {author} {\bibinfo {author} {\bibfnamefont {Jibitesh}\
  \bibnamefont {Dutta}}, \bibinfo {author} {\bibfnamefont {Wompherdeiki}\
  \bibnamefont {Khyllep}}, \ and\ \bibinfo {author} {\bibfnamefont {Nicola}\
  \bibnamefont {Tamanini}},\ }\bibfield  {title} {\enquote {\bibinfo {title}
  {{Dark energy with a gradient coupling to the dark matter fluid: cosmological
  dynamics and structure formation}},}\ }\href {\doibase
  10.1088/1475-7516/2018/01/038} {\bibfield  {journal} {\bibinfo  {journal}
  {JCAP}\ }\textbf {\bibinfo {volume} {01}},\ \bibinfo {pages} {038} (\bibinfo
  {year} {2018}{\natexlab{a}})},\ \Eprint {http://arxiv.org/abs/1707.09246}
  {arXiv:1707.09246 [gr-qc]} \BibitemShut {NoStop}%
\bibitem [{\citenamefont {Zonunmawia}\ \emph {et~al.}(2018)\citenamefont
  {Zonunmawia}, \citenamefont {Khyllep}, \citenamefont {Dutta},\ and\
  \citenamefont {Järv}}]{Zonunmawia:2018xvf}%
  \BibitemOpen
  \bibfield  {author} {\bibinfo {author} {\bibfnamefont {Hmar}\ \bibnamefont
  {Zonunmawia}}, \bibinfo {author} {\bibfnamefont {Wompherdeiki}\ \bibnamefont
  {Khyllep}}, \bibinfo {author} {\bibfnamefont {Jibitesh}\ \bibnamefont
  {Dutta}}, \ and\ \bibinfo {author} {\bibfnamefont {Laur}\ \bibnamefont
  {Järv}},\ }\bibfield  {title} {\enquote {\bibinfo {title} {{Cosmological
  dynamics of brane gravity: A global dynamical system perspective}},}\ }\href
  {\doibase 10.1103/PhysRevD.98.083532} {\bibfield  {journal} {\bibinfo
  {journal} {Phys. Rev.}\ }\textbf {\bibinfo {volume} {D98}},\ \bibinfo {pages}
  {083532} (\bibinfo {year} {2018})},\ \Eprint
  {http://arxiv.org/abs/1810.03816} {arXiv:1810.03816 [gr-qc]} \BibitemShut
  {NoStop}%
\bibitem [{\citenamefont {Carloni}\ \emph {et~al.}(2019)\citenamefont
  {Carloni}, \citenamefont {Rosa},\ and\ \citenamefont
  {Lemos}}]{Carloni:2018yoz}%
  \BibitemOpen
  \bibfield  {author} {\bibinfo {author} {\bibfnamefont {Sante}\ \bibnamefont
  {Carloni}}, \bibinfo {author} {\bibfnamefont {João~Luís}\ \bibnamefont
  {Rosa}}, \ and\ \bibinfo {author} {\bibfnamefont {José P.~S.}\ \bibnamefont
  {Lemos}},\ }\bibfield  {title} {\enquote {\bibinfo {title} {{Cosmology of
  $f(R, \Box R)$ gravity}},}\ }\href {\doibase 10.1103/PhysRevD.99.104001}
  {\bibfield  {journal} {\bibinfo  {journal} {Phys. Rev.}\ }\textbf {\bibinfo
  {volume} {D99}},\ \bibinfo {pages} {104001} (\bibinfo {year} {2019})},\
  \Eprint {http://arxiv.org/abs/1808.07316} {arXiv:1808.07316 [gr-qc]}
  \BibitemShut {NoStop}%
\bibitem [{\citenamefont {Dutta}\ \emph
  {et~al.}(2018{\natexlab{b}})\citenamefont {Dutta}, \citenamefont {Khyllep},
  \citenamefont {Saridakis}, \citenamefont {Tamanini},\ and\ \citenamefont
  {Vagnozzi}}]{Dutta:2017fjw}%
  \BibitemOpen
  \bibfield  {author} {\bibinfo {author} {\bibfnamefont {Jibitesh}\
  \bibnamefont {Dutta}}, \bibinfo {author} {\bibfnamefont {Wompherdeiki}\
  \bibnamefont {Khyllep}}, \bibinfo {author} {\bibfnamefont {Emmanuel~N.}\
  \bibnamefont {Saridakis}}, \bibinfo {author} {\bibfnamefont {Nicola}\
  \bibnamefont {Tamanini}}, \ and\ \bibinfo {author} {\bibfnamefont {Sunny}\
  \bibnamefont {Vagnozzi}},\ }\bibfield  {title} {\enquote {\bibinfo {title}
  {{Cosmological dynamics of mimetic gravity}},}\ }\href {\doibase
  10.1088/1475-7516/2018/02/041} {\bibfield  {journal} {\bibinfo  {journal}
  {JCAP}\ }\textbf {\bibinfo {volume} {1802}},\ \bibinfo {pages} {041}
  (\bibinfo {year} {2018}{\natexlab{b}})},\ \Eprint
  {http://arxiv.org/abs/1711.07290} {arXiv:1711.07290 [gr-qc]} \BibitemShut
  {NoStop}%
\bibitem [{\citenamefont {Dutta}\ \emph {et~al.}(2019)\citenamefont {Dutta},
  \citenamefont {Khyllep},\ and\ \citenamefont {Zonunmawia}}]{Dutta:2018xcz}%
  \BibitemOpen
  \bibfield  {author} {\bibinfo {author} {\bibfnamefont {Jibitesh}\
  \bibnamefont {Dutta}}, \bibinfo {author} {\bibfnamefont {Wompherdeiki}\
  \bibnamefont {Khyllep}}, \ and\ \bibinfo {author} {\bibfnamefont {Hmar}\
  \bibnamefont {Zonunmawia}},\ }\bibfield  {title} {\enquote {\bibinfo {title}
  {{Cosmological dynamics of the general non-canonical scalar field models}},}\
  }\href {\doibase 10.1140/epjc/s10052-019-6885-2} {\bibfield  {journal}
  {\bibinfo  {journal} {Eur. Phys. J.}\ }\textbf {\bibinfo {volume} {C79}},\
  \bibinfo {pages} {359} (\bibinfo {year} {2019})},\ \Eprint
  {http://arxiv.org/abs/1812.07836} {arXiv:1812.07836 [gr-qc]} \BibitemShut
  {NoStop}%
\bibitem [{\citenamefont {Khyllep}\ and\ \citenamefont
  {Dutta}(2019)}]{Khyllep:2019odd}%
  \BibitemOpen
  \bibfield  {author} {\bibinfo {author} {\bibfnamefont {Wompherdeiki}\
  \bibnamefont {Khyllep}}\ and\ \bibinfo {author} {\bibfnamefont {Jibitesh}\
  \bibnamefont {Dutta}},\ }\bibfield  {title} {\enquote {\bibinfo {title}
  {{Linear growth index of matter perturbations in Rastall gravity}},}\ }\href
  {\doibase 10.1016/j.physletb.2019.134796} {\bibfield  {journal} {\bibinfo
  {journal} {Phys. Lett. B}\ }\textbf {\bibinfo {volume} {797}},\ \bibinfo
  {pages} {134796} (\bibinfo {year} {2019})},\ \Eprint
  {http://arxiv.org/abs/1907.09221} {arXiv:1907.09221 [gr-qc]} \BibitemShut
  {NoStop}%
\bibitem [{\citenamefont {Kerachian}\ \emph {et~al.}(2019)\citenamefont
  {Kerachian}, \citenamefont {Acquaviva},\ and\ \citenamefont
  {Lukes-Gerakopoulos}}]{Kerachian:2019tar}%
  \BibitemOpen
  \bibfield  {author} {\bibinfo {author} {\bibfnamefont {Morteza}\ \bibnamefont
  {Kerachian}}, \bibinfo {author} {\bibfnamefont {Giovanni}\ \bibnamefont
  {Acquaviva}}, \ and\ \bibinfo {author} {\bibfnamefont {Georgios}\
  \bibnamefont {Lukes-Gerakopoulos}},\ }\bibfield  {title} {\enquote {\bibinfo
  {title} {{Classes of nonminimally coupled scalar fields in spatially curved
  FRW spacetimes}},}\ }\href {\doibase 10.1103/PhysRevD.99.123516} {\bibfield
  {journal} {\bibinfo  {journal} {Phys. Rev. D}\ }\textbf {\bibinfo {volume}
  {99}},\ \bibinfo {pages} {123516} (\bibinfo {year} {2019})},\ \Eprint
  {http://arxiv.org/abs/1905.08512} {arXiv:1905.08512 [gr-qc]} \BibitemShut
  {NoStop}%
\bibitem [{\citenamefont {Leon}\ \emph {et~al.}(2020)\citenamefont {Leon},
  \citenamefont {Paliathanasis},\ and\ \citenamefont
  {Velazquez~Abab}}]{Leon:2018skk}%
  \BibitemOpen
  \bibfield  {author} {\bibinfo {author} {\bibfnamefont {Genly}\ \bibnamefont
  {Leon}}, \bibinfo {author} {\bibfnamefont {Andronikos}\ \bibnamefont
  {Paliathanasis}}, \ and\ \bibinfo {author} {\bibfnamefont {Luisberis}\
  \bibnamefont {Velazquez~Abab}},\ }\bibfield  {title} {\enquote {\bibinfo
  {title} {{Stability of a modified Jordan-Brans-Dicke theory in the dilatonic
  frame}},}\ }\href {\doibase 10.1007/s10714-020-02718-7} {\bibfield  {journal}
  {\bibinfo  {journal} {Gen. Rel. Grav.}\ }\textbf {\bibinfo {volume} {52}},\
  \bibinfo {pages} {71} (\bibinfo {year} {2020})},\ \Eprint
  {http://arxiv.org/abs/1812.03830} {arXiv:1812.03830 [physics.gen-ph]}
  \BibitemShut {NoStop}%
\bibitem [{\citenamefont {Khyllep}\ \emph {et~al.}(2021)\citenamefont
  {Khyllep}, \citenamefont {Paliathanasis},\ and\ \citenamefont
  {Dutta}}]{Khyllep:2021pcu}%
  \BibitemOpen
  \bibfield  {author} {\bibinfo {author} {\bibfnamefont {Wompherdeiki}\
  \bibnamefont {Khyllep}}, \bibinfo {author} {\bibfnamefont {Andronikos}\
  \bibnamefont {Paliathanasis}}, \ and\ \bibinfo {author} {\bibfnamefont
  {Jibitesh}\ \bibnamefont {Dutta}},\ }\bibfield  {title} {\enquote {\bibinfo
  {title} {{Cosmological solutions and growth index of matter perturbations in
  $f(Q)$ gravity}},}\ }\href {\doibase 10.1103/PhysRevD.103.103521} {\bibfield
  {journal} {\bibinfo  {journal} {Phys. Rev. D}\ }\textbf {\bibinfo {volume}
  {103}},\ \bibinfo {pages} {103521} (\bibinfo {year} {2021})},\ \Eprint
  {http://arxiv.org/abs/2103.08372} {arXiv:2103.08372 [gr-qc]} \BibitemShut
  {NoStop}%
\bibitem [{\citenamefont {Paliathanasis}\ \emph {et~al.}(2021)\citenamefont
  {Paliathanasis}, \citenamefont {Leon}, \citenamefont {Khyllep}, \citenamefont
  {Dutta},\ and\ \citenamefont {Pan}}]{Paliathanasis:2021egx}%
  \BibitemOpen
  \bibfield  {author} {\bibinfo {author} {\bibfnamefont {Andronikos}\
  \bibnamefont {Paliathanasis}}, \bibinfo {author} {\bibfnamefont {Genly}\
  \bibnamefont {Leon}}, \bibinfo {author} {\bibfnamefont {Wompherdeiki}\
  \bibnamefont {Khyllep}}, \bibinfo {author} {\bibfnamefont {Jibitesh}\
  \bibnamefont {Dutta}}, \ and\ \bibinfo {author} {\bibfnamefont {Supriya}\
  \bibnamefont {Pan}},\ }\bibfield  {title} {\enquote {\bibinfo {title}
  {{Interacting quintessence in light of generalized uncertainty principle:
  cosmological perturbations and dynamics}},}\ }\href {\doibase
  10.1140/epjc/s10052-021-09362-8} {\bibfield  {journal} {\bibinfo  {journal}
  {Eur. Phys. J. C}\ }\textbf {\bibinfo {volume} {81}},\ \bibinfo {pages} {607}
  (\bibinfo {year} {2021})},\ \Eprint {http://arxiv.org/abs/2104.06097}
  {arXiv:2104.06097 [gr-qc]} \BibitemShut {NoStop}%
\bibitem [{\citenamefont {Bahamonde}\ \emph {et~al.}(2018)\citenamefont
  {Bahamonde}, \citenamefont {Böhmer}, \citenamefont {Carloni}, \citenamefont
  {Copeland}, \citenamefont {Fang},\ and\ \citenamefont
  {Tamanini}}]{Bahamonde:2017ize}%
  \BibitemOpen
  \bibfield  {author} {\bibinfo {author} {\bibfnamefont {Sebastian}\
  \bibnamefont {Bahamonde}}, \bibinfo {author} {\bibfnamefont {Christian~G.}\
  \bibnamefont {Böhmer}}, \bibinfo {author} {\bibfnamefont {Sante}\
  \bibnamefont {Carloni}}, \bibinfo {author} {\bibfnamefont {Edmund~J.}\
  \bibnamefont {Copeland}}, \bibinfo {author} {\bibfnamefont {Wei}\
  \bibnamefont {Fang}}, \ and\ \bibinfo {author} {\bibfnamefont {Nicola}\
  \bibnamefont {Tamanini}},\ }\bibfield  {title} {\enquote {\bibinfo {title}
  {{Dynamical systems applied to cosmology: dark energy and modified
  gravity}},}\ }\href {\doibase 10.1016/j.physrep.2018.09.001} {\bibfield
  {journal} {\bibinfo  {journal} {Phys. Rept.}\ }\textbf {\bibinfo {volume}
  {775-777}},\ \bibinfo {pages} {1--122} (\bibinfo {year} {2018})},\ \Eprint
  {http://arxiv.org/abs/1712.03107} {arXiv:1712.03107 [gr-qc]} \BibitemShut
  {NoStop}%
\bibitem [{\citenamefont {Seydel}(2009)}]{seydel2009practical}%
  \BibitemOpen
  \bibfield  {author} {\bibinfo {author} {\bibfnamefont {R{\"u}diger}\
  \bibnamefont {Seydel}},\ }\href@noop {} {\emph {\bibinfo {title} {Practical
  bifurcation and stability analysis}}},\ Vol.~\bibinfo {volume} {5}\ (\bibinfo
   {publisher} {Springer Science \& Business Media},\ \bibinfo {year}
  {2009})\BibitemShut {NoStop}%
\bibitem [{\citenamefont {Kuznetsov}(2013)}]{kuznetsov2013elements}%
  \BibitemOpen
  \bibfield  {author} {\bibinfo {author} {\bibfnamefont {Yuri~A}\ \bibnamefont
  {Kuznetsov}},\ }\href@noop {} {\emph {\bibinfo {title} {Elements of applied
  bifurcation theory}}},\ Vol.\ \bibinfo {volume} {112}\ (\bibinfo  {publisher}
  {Springer Science \& Business Media},\ \bibinfo {year} {2013})\BibitemShut
  {NoStop}%
\bibitem [{\citenamefont {Perko}(2013)}]{perko2013differential}%
  \BibitemOpen
  \bibfield  {author} {\bibinfo {author} {\bibfnamefont {Lawrence}\
  \bibnamefont {Perko}},\ }\href@noop {} {\emph {\bibinfo {title} {Differential
  equations and dynamical systems}}},\ Vol.~\bibinfo {volume} {7}\ (\bibinfo
  {publisher} {Springer Science \& Business Media},\ \bibinfo {year}
  {2013})\BibitemShut {NoStop}%
\bibitem [{\citenamefont {Kl\"en}\ and\ \citenamefont
  {Molina}(2020)}]{Klen:2020kdb}%
  \BibitemOpen
  \bibfield  {author} {\bibinfo {author} {\bibfnamefont {W.S.}\ \bibnamefont
  {Kl\"en}}\ and\ \bibinfo {author} {\bibfnamefont {C.}~\bibnamefont
  {Molina}},\ }\bibfield  {title} {\enquote {\bibinfo {title} {{Dynamical
  analysis of null geodesics in brane-world spacetimes}},}\ }\href {\doibase
  10.1103/PhysRevD.102.104051} {\bibfield  {journal} {\bibinfo  {journal}
  {Phys. Rev. D}\ }\textbf {\bibinfo {volume} {102}},\ \bibinfo {pages}
  {104051} (\bibinfo {year} {2020})},\ \Eprint
  {http://arxiv.org/abs/2011.03054} {arXiv:2011.03054 [gr-qc]} \BibitemShut
  {NoStop}%
\bibitem [{\citenamefont {Humieja}\ and\ \citenamefont
  {Szydłowski}(2019)}]{Humieja:2019ywy}%
  \BibitemOpen
  \bibfield  {author} {\bibinfo {author} {\bibfnamefont {Franciszek}\
  \bibnamefont {Humieja}}\ and\ \bibinfo {author} {\bibfnamefont {Marek}\
  \bibnamefont {Szydłowski}},\ }\bibfield  {title} {\enquote {\bibinfo {title}
  {{Bifurcations in Ratra–Peebles quintessence models and their
  extensions}},}\ }\href {\doibase 10.1140/epjc/s10052-019-7299-x} {\bibfield
  {journal} {\bibinfo  {journal} {Eur. Phys. J.}\ }\textbf {\bibinfo {volume}
  {C79}},\ \bibinfo {pages} {794} (\bibinfo {year} {2019})},\ \Eprint
  {http://arxiv.org/abs/1901.06578} {arXiv:1901.06578 [gr-qc]} \BibitemShut
  {NoStop}%
\bibitem [{\citenamefont {Szydlowski}\ and\ \citenamefont
  {Tambor}(2008)}]{Szydlowski:2008rz}%
  \BibitemOpen
  \bibfield  {author} {\bibinfo {author} {\bibfnamefont {Marek}\ \bibnamefont
  {Szydlowski}}\ and\ \bibinfo {author} {\bibfnamefont {Pawel}\ \bibnamefont
  {Tambor}},\ }\bibfield  {title} {\enquote {\bibinfo {title} {{Emergence and
  Effective Theory of the Universe: The Case Study of Lambda Cold Dark Matter
  Cosmological Model}},}\ }\href@noop {} {\  (\bibinfo {year} {2008})},\
  \Eprint {http://arxiv.org/abs/0805.2665} {arXiv:0805.2665 [gr-qc]}
  \BibitemShut {NoStop}%
\bibitem [{\citenamefont {Kokarev}(2009)}]{Kokarev:2008ba}%
  \BibitemOpen
  \bibfield  {author} {\bibinfo {author} {\bibfnamefont {Sergey~S.}\
  \bibnamefont {Kokarev}},\ }\bibfield  {title} {\enquote {\bibinfo {title}
  {{Structural instability of Friedmann-Robertson-Walker cosmological
  models}},}\ }\href {\doibase 10.1007/s10714-008-0748-8} {\bibfield  {journal}
  {\bibinfo  {journal} {Gen. Rel. Grav.}\ }\textbf {\bibinfo {volume} {41}},\
  \bibinfo {pages} {1777--1794} (\bibinfo {year} {2009})},\ \Eprint
  {http://arxiv.org/abs/0810.5080} {arXiv:0810.5080 [gr-qc]} \BibitemShut
  {NoStop}%
\bibitem [{\citenamefont {Uzan}(1999)}]{Uzan:1999ch}%
  \BibitemOpen
  \bibfield  {author} {\bibinfo {author} {\bibfnamefont {Jean-Philippe}\
  \bibnamefont {Uzan}},\ }\bibfield  {title} {\enquote {\bibinfo {title}
  {{Cosmological scaling solutions of nonminimally coupled scalar fields}},}\
  }\href {\doibase 10.1103/PhysRevD.59.123510} {\bibfield  {journal} {\bibinfo
  {journal} {Phys. Rev.}\ }\textbf {\bibinfo {volume} {D59}},\ \bibinfo {pages}
  {123510} (\bibinfo {year} {1999})},\ \Eprint
  {http://arxiv.org/abs/gr-qc/9903004} {arXiv:gr-qc/9903004 [gr-qc]}
  \BibitemShut {NoStop}%
\bibitem [{\citenamefont {Gunzig}\ \emph {et~al.}(2000)\citenamefont {Gunzig},
  \citenamefont {Faraoni}, \citenamefont {Figueiredo}, \citenamefont {Rocha},\
  and\ \citenamefont {Brenig}}]{Gunzig:2000ce}%
  \BibitemOpen
  \bibfield  {author} {\bibinfo {author} {\bibfnamefont {E.}~\bibnamefont
  {Gunzig}}, \bibinfo {author} {\bibfnamefont {V.}~\bibnamefont {Faraoni}},
  \bibinfo {author} {\bibfnamefont {A.}~\bibnamefont {Figueiredo}}, \bibinfo
  {author} {\bibfnamefont {T.~M.}\ \bibnamefont {Rocha}}, \ and\ \bibinfo
  {author} {\bibfnamefont {L.}~\bibnamefont {Brenig}},\ }\bibfield  {title}
  {\enquote {\bibinfo {title} {{The dynamical system approach to scalar field
  cosmology}},}\ }\href {\doibase 10.1088/0264-9381/17/8/304} {\bibfield
  {journal} {\bibinfo  {journal} {Class. Quant. Grav.}\ }\textbf {\bibinfo
  {volume} {17}},\ \bibinfo {pages} {1783--1814} (\bibinfo {year}
  {2000})}\BibitemShut {NoStop}%
\bibitem [{\citenamefont {Faraoni}\ \emph {et~al.}(2006)\citenamefont
  {Faraoni}, \citenamefont {Jensen},\ and\ \citenamefont
  {Theuerkauf}}]{Faraoni:2006sr}%
  \BibitemOpen
  \bibfield  {author} {\bibinfo {author} {\bibfnamefont {V.}~\bibnamefont
  {Faraoni}}, \bibinfo {author} {\bibfnamefont {M.~N.}\ \bibnamefont {Jensen}},
  \ and\ \bibinfo {author} {\bibfnamefont {S.~A.}\ \bibnamefont {Theuerkauf}},\
  }\bibfield  {title} {\enquote {\bibinfo {title} {{Non-chaotic dynamics in
  general-relativistic and scalar-tensor cosmology}},}\ }\href {\doibase
  10.1088/0264-9381/23/12/016} {\bibfield  {journal} {\bibinfo  {journal}
  {Class. Quant. Grav.}\ }\textbf {\bibinfo {volume} {23}},\ \bibinfo {pages}
  {4215--4230} (\bibinfo {year} {2006})},\ \Eprint
  {http://arxiv.org/abs/gr-qc/0605050} {arXiv:gr-qc/0605050 [gr-qc]}
  \BibitemShut {NoStop}%
\bibitem [{\citenamefont {Carloni}\ \emph {et~al.}(2008)\citenamefont
  {Carloni}, \citenamefont {Capozziello}, \citenamefont {Leach},\ and\
  \citenamefont {Dunsby}}]{Carloni:2007eu}%
  \BibitemOpen
  \bibfield  {author} {\bibinfo {author} {\bibfnamefont {S.}~\bibnamefont
  {Carloni}}, \bibinfo {author} {\bibfnamefont {S.}~\bibnamefont
  {Capozziello}}, \bibinfo {author} {\bibfnamefont {J.~A.}\ \bibnamefont
  {Leach}}, \ and\ \bibinfo {author} {\bibfnamefont {P.~K.~S.}\ \bibnamefont
  {Dunsby}},\ }\bibfield  {title} {\enquote {\bibinfo {title} {{Cosmological
  dynamics of scalar-tensor gravity}},}\ }\href {\doibase
  10.1088/0264-9381/25/3/035008} {\bibfield  {journal} {\bibinfo  {journal}
  {Class. Quant. Grav.}\ }\textbf {\bibinfo {volume} {25}},\ \bibinfo {pages}
  {035008} (\bibinfo {year} {2008})},\ \Eprint
  {http://arxiv.org/abs/gr-qc/0701009} {arXiv:gr-qc/0701009 [gr-qc]}
  \BibitemShut {NoStop}%
\bibitem [{\citenamefont {Szydlowski}\ and\ \citenamefont
  {Hrycyna}(2009)}]{Szydlowski:2008in}%
  \BibitemOpen
  \bibfield  {author} {\bibinfo {author} {\bibfnamefont {Marek}\ \bibnamefont
  {Szydlowski}}\ and\ \bibinfo {author} {\bibfnamefont {Orest}\ \bibnamefont
  {Hrycyna}},\ }\bibfield  {title} {\enquote {\bibinfo {title} {{Scalar field
  cosmology in the energy phase-space -- unified description of dynamics}},}\
  }\href {\doibase 10.1088/1475-7516/2009/01/039} {\bibfield  {journal}
  {\bibinfo  {journal} {JCAP}\ }\textbf {\bibinfo {volume} {0901}},\ \bibinfo
  {pages} {039} (\bibinfo {year} {2009})},\ \Eprint
  {http://arxiv.org/abs/0811.1493} {arXiv:0811.1493 [astro-ph]} \BibitemShut
  {NoStop}%
\bibitem [{\citenamefont {Maeda}\ and\ \citenamefont
  {Fujii}(2009)}]{Maeda:2009js}%
  \BibitemOpen
  \bibfield  {author} {\bibinfo {author} {\bibfnamefont {Kei-ichi}\
  \bibnamefont {Maeda}}\ and\ \bibinfo {author} {\bibfnamefont {Yasunori}\
  \bibnamefont {Fujii}},\ }\bibfield  {title} {\enquote {\bibinfo {title}
  {{Attractor Universe in the Scalar-Tensor Theory of Gravitation}},}\ }\href
  {\doibase 10.1103/PhysRevD.79.084026} {\bibfield  {journal} {\bibinfo
  {journal} {Phys. Rev.}\ }\textbf {\bibinfo {volume} {D79}},\ \bibinfo {pages}
  {084026} (\bibinfo {year} {2009})},\ \Eprint {http://arxiv.org/abs/0902.1221}
  {arXiv:0902.1221 [hep-th]} \BibitemShut {NoStop}%
\bibitem [{\citenamefont {Jarv}\ \emph {et~al.}(2010)\citenamefont {Jarv},
  \citenamefont {Kuusk},\ and\ \citenamefont {Saal}}]{Jarv:2010zc}%
  \BibitemOpen
  \bibfield  {author} {\bibinfo {author} {\bibfnamefont {Laur}\ \bibnamefont
  {Jarv}}, \bibinfo {author} {\bibfnamefont {Piret}\ \bibnamefont {Kuusk}}, \
  and\ \bibinfo {author} {\bibfnamefont {Margus}\ \bibnamefont {Saal}},\
  }\bibfield  {title} {\enquote {\bibinfo {title} {{Potential dominated
  scalar-tensor cosmologies in the general relativity limit: phase space
  view}},}\ }\href {\doibase 10.1103/PhysRevD.81.104007} {\bibfield  {journal}
  {\bibinfo  {journal} {Phys. Rev.}\ }\textbf {\bibinfo {volume} {D81}},\
  \bibinfo {pages} {104007} (\bibinfo {year} {2010})},\ \Eprint
  {http://arxiv.org/abs/1003.1686} {arXiv:1003.1686 [gr-qc]} \BibitemShut
  {NoStop}%
\bibitem [{\citenamefont {Hrycyna}\ and\ \citenamefont
  {Szyd\l{}owski}(2015)}]{Hrycyna:2015eta}%
  \BibitemOpen
  \bibfield  {author} {\bibinfo {author} {\bibfnamefont {Orest}\ \bibnamefont
  {Hrycyna}}\ and\ \bibinfo {author} {\bibfnamefont {Marek}\ \bibnamefont
  {Szyd\l{}owski}},\ }\bibfield  {title} {\enquote {\bibinfo {title}
  {{Cosmological dynamics with non-minimally coupled scalar field and a
  constant potential function}},}\ }\href {\doibase
  10.1088/1475-7516/2015/11/013} {\bibfield  {journal} {\bibinfo  {journal}
  {JCAP}\ }\textbf {\bibinfo {volume} {11}},\ \bibinfo {pages} {013} (\bibinfo
  {year} {2015})},\ \Eprint {http://arxiv.org/abs/1506.03429} {arXiv:1506.03429
  [gr-qc]} \BibitemShut {NoStop}%
\bibitem [{\citenamefont {Kohli}\ and\ \citenamefont
  {Haslam}(2018)}]{Kohli:2018lsg}%
  \BibitemOpen
  \bibfield  {author} {\bibinfo {author} {\bibfnamefont {Ikjyot~Singh}\
  \bibnamefont {Kohli}}\ and\ \bibinfo {author} {\bibfnamefont {Michael~C.}\
  \bibnamefont {Haslam}},\ }\bibfield  {title} {\enquote {\bibinfo {title}
  {{Einstein’s field equations as a fold bifurcation}},}\ }\href {\doibase
  10.1016/j.geomphys.2017.10.001} {\bibfield  {journal} {\bibinfo  {journal}
  {J. Geom. Phys.}\ }\textbf {\bibinfo {volume} {123}},\ \bibinfo {pages}
  {434--437} (\bibinfo {year} {2018})},\ \Eprint
  {http://arxiv.org/abs/1607.05300} {arXiv:1607.05300 [physics.gen-ph]}
  \BibitemShut {NoStop}%
\bibitem [{\citenamefont {Goheer}\ and\ \citenamefont
  {Dunsby}(2002)}]{Goheer:2002bq}%
  \BibitemOpen
  \bibfield  {author} {\bibinfo {author} {\bibfnamefont {Naureen}\ \bibnamefont
  {Goheer}}\ and\ \bibinfo {author} {\bibfnamefont {Peter}\ \bibnamefont
  {Dunsby}},\ }\bibfield  {title} {\enquote {\bibinfo {title} {{Brane world
  dynamics of inflationary cosmologies with exponential potentials}},}\ }\href
  {\doibase 10.1103/PhysRevD.66.043527} {\bibfield  {journal} {\bibinfo
  {journal} {Phys. Rev. D}\ }\textbf {\bibinfo {volume} {66}},\ \bibinfo
  {pages} {043527} (\bibinfo {year} {2002})},\ \Eprint
  {http://arxiv.org/abs/gr-qc/0204059} {arXiv:gr-qc/0204059} \BibitemShut
  {NoStop}%
\bibitem [{\citenamefont {Feng}\ \emph {et~al.}(2012)\citenamefont {Feng},
  \citenamefont {Li},\ and\ \citenamefont {Xi}}]{Feng:2012wx}%
  \BibitemOpen
  \bibfield  {author} {\bibinfo {author} {\bibfnamefont {Chao-Jun}\
  \bibnamefont {Feng}}, \bibinfo {author} {\bibfnamefont {Xin-Zhou}\
  \bibnamefont {Li}}, \ and\ \bibinfo {author} {\bibfnamefont {Ping}\
  \bibnamefont {Xi}},\ }\bibfield  {title} {\enquote {\bibinfo {title} {{Global
  behavior of cosmological dynamics with interacting Veneziano ghost}},}\
  }\href {\doibase 10.1007/JHEP05(2012)046} {\bibfield  {journal} {\bibinfo
  {journal} {JHEP}\ }\textbf {\bibinfo {volume} {05}},\ \bibinfo {pages} {046}
  (\bibinfo {year} {2012})},\ \Eprint {http://arxiv.org/abs/1204.4055}
  {arXiv:1204.4055 [astro-ph.CO]} \BibitemShut {NoStop}%
\bibitem [{\citenamefont {Hrycyna}\ and\ \citenamefont
  {Szyd\l{}owski}(2013)}]{Hrycyna:2013yia}%
  \BibitemOpen
  \bibfield  {author} {\bibinfo {author} {\bibfnamefont {Orest}\ \bibnamefont
  {Hrycyna}}\ and\ \bibinfo {author} {\bibfnamefont {Marek}\ \bibnamefont
  {Szyd\l{}owski}},\ }\bibfield  {title} {\enquote {\bibinfo {title}
  {{Dynamical complexity of the Brans-Dicke cosmology}},}\ }\href {\doibase
  10.1088/1475-7516/2013/12/016} {\bibfield  {journal} {\bibinfo  {journal}
  {JCAP}\ }\textbf {\bibinfo {volume} {12}},\ \bibinfo {pages} {016} (\bibinfo
  {year} {2013})},\ \Eprint {http://arxiv.org/abs/1310.1961} {arXiv:1310.1961
  [gr-qc]} \BibitemShut {NoStop}%
\bibitem [{\citenamefont {Szydlowski}\ \emph {et~al.}(2014)\citenamefont
  {Szydlowski}, \citenamefont {Hrycyna},\ and\ \citenamefont
  {Stachowski}}]{Szydlowski:2013sma}%
  \BibitemOpen
  \bibfield  {author} {\bibinfo {author} {\bibfnamefont {Marek}\ \bibnamefont
  {Szydlowski}}, \bibinfo {author} {\bibfnamefont {Orest}\ \bibnamefont
  {Hrycyna}}, \ and\ \bibinfo {author} {\bibfnamefont {Aleksander}\
  \bibnamefont {Stachowski}},\ }\bibfield  {title} {\enquote {\bibinfo {title}
  {{Scalar field cosmology - geometry of dynamics}},}\ }\href {\doibase
  10.1142/S0219887814600123} {\bibfield  {journal} {\bibinfo  {journal} {Int.
  J. Geom. Meth. Mod. Phys.}\ }\textbf {\bibinfo {volume} {11}},\ \bibinfo
  {pages} {1460012} (\bibinfo {year} {2014})},\ \Eprint
  {http://arxiv.org/abs/1308.4069} {arXiv:1308.4069 [gr-qc]} \BibitemShut
  {NoStop}%
\bibitem [{\citenamefont {Hell}\ \emph {et~al.}(2020)\citenamefont {Hell},
  \citenamefont {Lappicy},\ and\ \citenamefont {Uggla}}]{Hell:2020yyo}%
  \BibitemOpen
  \bibfield  {author} {\bibinfo {author} {\bibfnamefont {Juliette}\
  \bibnamefont {Hell}}, \bibinfo {author} {\bibfnamefont {Phillipo}\
  \bibnamefont {Lappicy}}, \ and\ \bibinfo {author} {\bibfnamefont {Claes}\
  \bibnamefont {Uggla}},\ }\bibfield  {title} {\enquote {\bibinfo {title}
  {{Bifurcations and Chaos in Ho\v rava-Lifshitz Cosmology}},}\ }\href@noop {}
  {\  (\bibinfo {year} {2020})},\ \Eprint {http://arxiv.org/abs/2012.07614}
  {arXiv:2012.07614 [gr-qc]} \BibitemShut {NoStop}%
\bibitem [{\citenamefont {Mishra}\ and\ \citenamefont
  {Chakraborty}(2019)}]{Mishra:2019vnv}%
  \BibitemOpen
  \bibfield  {author} {\bibinfo {author} {\bibfnamefont {Sudip}\ \bibnamefont
  {Mishra}}\ and\ \bibinfo {author} {\bibfnamefont {Subenoy}\ \bibnamefont
  {Chakraborty}},\ }\bibfield  {title} {\enquote {\bibinfo {title} {{Stability
  and bifurcation analysis of interacting f(T) cosmology}},}\ }\href {\doibase
  10.1140/epjc/s10052-019-6839-8} {\bibfield  {journal} {\bibinfo  {journal}
  {Eur. Phys. J.}\ }\textbf {\bibinfo {volume} {C79}},\ \bibinfo {pages} {328}
  (\bibinfo {year} {2019})}\BibitemShut {NoStop}%
\bibitem [{\citenamefont {Azim}\ \emph {et~al.}(2020)\citenamefont {Azim},
  \citenamefont {Awad},\ and\ \citenamefont {Lashin}}]{Azim:2020yce}%
  \BibitemOpen
  \bibfield  {author} {\bibinfo {author} {\bibfnamefont {Asmaa~Abdel}\
  \bibnamefont {Azim}}, \bibinfo {author} {\bibfnamefont {Adel}\ \bibnamefont
  {Awad}}, \ and\ \bibinfo {author} {\bibfnamefont {E.I.}\ \bibnamefont
  {Lashin}},\ }\bibfield  {title} {\enquote {\bibinfo {title} {{Degenerate
  Bogdanov\textendash{}Takens bifurcations in a bulk viscous cosmology}},}\
  }\href {\doibase 10.1140/epjc/s10052-020-8384-x} {\bibfield  {journal}
  {\bibinfo  {journal} {Eur. Phys. J. C}\ }\textbf {\bibinfo {volume} {80}},\
  \bibinfo {pages} {868} (\bibinfo {year} {2020})},\ \Eprint
  {http://arxiv.org/abs/2001.04981} {arXiv:2001.04981 [gr-qc]} \BibitemShut
  {NoStop}%
\bibitem [{\citenamefont {{Bergmann}}(1968)}]{1968IJTP....1...25B}%
  \BibitemOpen
  \bibfield  {author} {\bibinfo {author} {\bibfnamefont {Peter~G.}\
  \bibnamefont {{Bergmann}}},\ }\bibfield  {title} {\enquote {\bibinfo {title}
  {{Comments on the scalar-tensor theory}},}\ }\href {\doibase
  10.1007/BF00668828} {\bibfield  {journal} {\bibinfo  {journal} {International
  Journal of Theoretical Physics}\ }\textbf {\bibinfo {volume} {1}},\ \bibinfo
  {pages} {25--36} (\bibinfo {year} {1968})}\BibitemShut {NoStop}%
\bibitem [{\citenamefont {Nordtvedt}(1970)}]{Nordtvedt:1970uv}%
  \BibitemOpen
  \bibfield  {author} {\bibinfo {author} {\bibfnamefont {Kenneth}\ \bibnamefont
  {Nordtvedt}, \bibfnamefont {Jr.}},\ }\bibfield  {title} {\enquote {\bibinfo
  {title} {{PostNewtonian metric for a general class of scalar tensor
  gravitational theories and observational consequences}},}\ }\href {\doibase
  10.1086/150607} {\bibfield  {journal} {\bibinfo  {journal} {Astrophys. J.}\
  }\textbf {\bibinfo {volume} {161}},\ \bibinfo {pages} {1059--1067} (\bibinfo
  {year} {1970})}\BibitemShut {NoStop}%
\bibitem [{\citenamefont {{Wagoner}}(1970)}]{1970PhRvD...1.3209W}%
  \BibitemOpen
  \bibfield  {author} {\bibinfo {author} {\bibfnamefont {Robert~V.}\
  \bibnamefont {{Wagoner}}},\ }\bibfield  {title} {\enquote {\bibinfo {title}
  {{Scalar-Tensor Theory and Gravitational Waves}},}\ }\href {\doibase
  10.1103/PhysRevD.1.3209} {\bibfield  {journal} {\bibinfo  {journal} {\prd}\
  }\textbf {\bibinfo {volume} {1}},\ \bibinfo {pages} {3209--3216} (\bibinfo
  {year} {1970})}\BibitemShut {NoStop}%
\bibitem [{\citenamefont {Capozziello}\ \emph {et~al.}(2006)\citenamefont
  {Capozziello}, \citenamefont {Nojiri}, \citenamefont {Odintsov},\ and\
  \citenamefont {Troisi}}]{Capozziello:2006dj}%
  \BibitemOpen
  \bibfield  {author} {\bibinfo {author} {\bibfnamefont {Salvatore}\
  \bibnamefont {Capozziello}}, \bibinfo {author} {\bibfnamefont
  {S.}~\bibnamefont {Nojiri}}, \bibinfo {author} {\bibfnamefont {S.~D.}\
  \bibnamefont {Odintsov}}, \ and\ \bibinfo {author} {\bibfnamefont
  {A.}~\bibnamefont {Troisi}},\ }\bibfield  {title} {\enquote {\bibinfo {title}
  {{Cosmological viability of f(R)-gravity as an ideal fluid and its
  compatibility with a matter dominated phase}},}\ }\href {\doibase
  10.1016/j.physletb.2006.06.034} {\bibfield  {journal} {\bibinfo  {journal}
  {Phys. Lett.}\ }\textbf {\bibinfo {volume} {B639}},\ \bibinfo {pages}
  {135--143} (\bibinfo {year} {2006})},\ \Eprint
  {http://arxiv.org/abs/astro-ph/0604431} {arXiv:astro-ph/0604431 [astro-ph]}
  \BibitemShut {NoStop}%
\bibitem [{\citenamefont {Capozziello}\ \emph {et~al.}(1993)\citenamefont
  {Capozziello}, \citenamefont {de~Ritis},\ and\ \citenamefont
  {Rubano}}]{Capozziello:1993vs}%
  \BibitemOpen
  \bibfield  {author} {\bibinfo {author} {\bibfnamefont {S.}~\bibnamefont
  {Capozziello}}, \bibinfo {author} {\bibfnamefont {R.}~\bibnamefont
  {de~Ritis}}, \ and\ \bibinfo {author} {\bibfnamefont {C.}~\bibnamefont
  {Rubano}},\ }\bibfield  {title} {\enquote {\bibinfo {title} {{String dilaton
  cosmology with exponential potential}},}\ }\href {\doibase
  10.1016/0375-9601(93)90365-7} {\bibfield  {journal} {\bibinfo  {journal}
  {Phys. Lett.}\ }\textbf {\bibinfo {volume} {A177}},\ \bibinfo {pages} {8--12}
  (\bibinfo {year} {1993})}\BibitemShut {NoStop}%
\bibitem [{\citenamefont {Brans}\ and\ \citenamefont
  {Dicke}(1961)}]{Brans:1961sx}%
  \BibitemOpen
  \bibfield  {author} {\bibinfo {author} {\bibfnamefont {C.}~\bibnamefont
  {Brans}}\ and\ \bibinfo {author} {\bibfnamefont {R.~H.}\ \bibnamefont
  {Dicke}},\ }\bibfield  {title} {\enquote {\bibinfo {title} {{Mach's principle
  and a relativistic theory of gravitation}},}\ }\href {\doibase
  10.1103/PhysRev.124.925} {\bibfield  {journal} {\bibinfo  {journal} {Phys.
  Rev.}\ }\textbf {\bibinfo {volume} {124}},\ \bibinfo {pages} {925--935}
  (\bibinfo {year} {1961})}\BibitemShut {NoStop}%
\bibitem [{\citenamefont {Esposito-Farese}\ and\ \citenamefont
  {Polarski}(2001)}]{EspositoFarese:2000ij}%
  \BibitemOpen
  \bibfield  {author} {\bibinfo {author} {\bibfnamefont {Gilles}\ \bibnamefont
  {Esposito-Farese}}\ and\ \bibinfo {author} {\bibfnamefont {D.}~\bibnamefont
  {Polarski}},\ }\bibfield  {title} {\enquote {\bibinfo {title} {{Scalar tensor
  gravity in an accelerating universe}},}\ }\href {\doibase
  10.1103/PhysRevD.63.063504} {\bibfield  {journal} {\bibinfo  {journal} {Phys.
  Rev.}\ }\textbf {\bibinfo {volume} {D63}},\ \bibinfo {pages} {063504}
  (\bibinfo {year} {2001})},\ \Eprint {http://arxiv.org/abs/gr-qc/0009034}
  {arXiv:gr-qc/0009034 [gr-qc]} \BibitemShut {NoStop}%
\bibitem [{\citenamefont {Capozziello}\ \emph {et~al.}(1997)\citenamefont
  {Capozziello}, \citenamefont {de~Ritis},\ and\ \citenamefont
  {Marino}}]{Capozziello:1996xg}%
  \BibitemOpen
  \bibfield  {author} {\bibinfo {author} {\bibfnamefont {S.}~\bibnamefont
  {Capozziello}}, \bibinfo {author} {\bibfnamefont {R.}~\bibnamefont
  {de~Ritis}}, \ and\ \bibinfo {author} {\bibfnamefont {Alma~Angela}\
  \bibnamefont {Marino}},\ }\bibfield  {title} {\enquote {\bibinfo {title}
  {{Some aspects of the cosmological conformal equivalence between `Jordan
  frame' and `Einstein frame'}},}\ }\href {\doibase
  10.1088/0264-9381/14/12/010} {\bibfield  {journal} {\bibinfo  {journal}
  {Class. Quant. Grav.}\ }\textbf {\bibinfo {volume} {14}},\ \bibinfo {pages}
  {3243--3258} (\bibinfo {year} {1997})},\ \Eprint
  {http://arxiv.org/abs/gr-qc/9612053} {arXiv:gr-qc/9612053 [gr-qc]}
  \BibitemShut {NoStop}%
\bibitem [{\citenamefont {Fang}\ \emph {et~al.}(2009)\citenamefont {Fang},
  \citenamefont {Li}, \citenamefont {Zhang},\ and\ \citenamefont
  {Lu}}]{Fang:2008fw}%
  \BibitemOpen
  \bibfield  {author} {\bibinfo {author} {\bibfnamefont {Wei}\ \bibnamefont
  {Fang}}, \bibinfo {author} {\bibfnamefont {Ying}\ \bibnamefont {Li}},
  \bibinfo {author} {\bibfnamefont {Kai}\ \bibnamefont {Zhang}}, \ and\
  \bibinfo {author} {\bibfnamefont {Hui-Qing}\ \bibnamefont {Lu}},\ }\bibfield
  {title} {\enquote {\bibinfo {title} {{Exact Analysis of Scaling and Dominant
  Attractors Beyond the Exponential Potential}},}\ }\href {\doibase
  10.1088/0264-9381/26/15/155005} {\bibfield  {journal} {\bibinfo  {journal}
  {Class. Quant. Grav.}\ }\textbf {\bibinfo {volume} {26}},\ \bibinfo {pages}
  {155005} (\bibinfo {year} {2009})},\ \Eprint {http://arxiv.org/abs/0810.4193}
  {arXiv:0810.4193 [hep-th]} \BibitemShut {NoStop}%
\bibitem [{\citenamefont {Zhou}(2008)}]{Zhou:2007xp}%
  \BibitemOpen
  \bibfield  {author} {\bibinfo {author} {\bibfnamefont {Shuang-Yong}\
  \bibnamefont {Zhou}},\ }\bibfield  {title} {\enquote {\bibinfo {title} {{A
  New Approach to Quintessence and Solution of Multiple Attractors}},}\ }\href
  {\doibase 10.1016/j.physletb.2007.12.020} {\bibfield  {journal} {\bibinfo
  {journal} {Phys. Lett.}\ }\textbf {\bibinfo {volume} {B660}},\ \bibinfo
  {pages} {7--12} (\bibinfo {year} {2008})},\ \Eprint
  {http://arxiv.org/abs/0705.1577} {arXiv:0705.1577 [astro-ph]} \BibitemShut
  {NoStop}%
\bibitem [{\citenamefont {Xiao}\ and\ \citenamefont {Zhu}(2011)}]{Xiao:2011nh}%
  \BibitemOpen
  \bibfield  {author} {\bibinfo {author} {\bibfnamefont {Kui}\ \bibnamefont
  {Xiao}}\ and\ \bibinfo {author} {\bibfnamefont {Jian-Yang}\ \bibnamefont
  {Zhu}},\ }\bibfield  {title} {\enquote {\bibinfo {title} {{Stability analysis
  of an autonomous system in loop quantum cosmology}},}\ }\href {\doibase
  10.1103/PhysRevD.83.083501} {\bibfield  {journal} {\bibinfo  {journal} {Phys.
  Rev. D}\ }\textbf {\bibinfo {volume} {83}},\ \bibinfo {pages} {083501}
  (\bibinfo {year} {2011})},\ \Eprint {http://arxiv.org/abs/1102.2695}
  {arXiv:1102.2695 [gr-qc]} \BibitemShut {NoStop}%
\bibitem [{\citenamefont {Nunes}\ and\ \citenamefont
  {Mimoso}(2000)}]{Nunes:2000yc}%
  \BibitemOpen
  \bibfield  {author} {\bibinfo {author} {\bibfnamefont {Ana}\ \bibnamefont
  {Nunes}}\ and\ \bibinfo {author} {\bibfnamefont {Jose~P.}\ \bibnamefont
  {Mimoso}},\ }\bibfield  {title} {\enquote {\bibinfo {title} {{On the
  potentials yielding cosmological scaling solutions}},}\ }\href {\doibase
  10.1016/S0370-2693(00)00919-9} {\bibfield  {journal} {\bibinfo  {journal}
  {Phys. Lett. B}\ }\textbf {\bibinfo {volume} {488}},\ \bibinfo {pages}
  {423--427} (\bibinfo {year} {2000})},\ \Eprint
  {http://arxiv.org/abs/gr-qc/0008003} {arXiv:gr-qc/0008003} \BibitemShut
  {NoStop}%
\bibitem [{\citenamefont {Perivolaropoulos}(2005)}]{Perivolaropoulos:2005yv}%
  \BibitemOpen
  \bibfield  {author} {\bibinfo {author} {\bibfnamefont {Leandros}\
  \bibnamefont {Perivolaropoulos}},\ }\bibfield  {title} {\enquote {\bibinfo
  {title} {{Crossing the phantom divide barrier with scalar tensor
  theories}},}\ }\href {\doibase 10.1088/1475-7516/2005/10/001} {\bibfield
  {journal} {\bibinfo  {journal} {JCAP}\ }\textbf {\bibinfo {volume} {0510}},\
  \bibinfo {pages} {001} (\bibinfo {year} {2005})},\ \Eprint
  {http://arxiv.org/abs/astro-ph/0504582} {arXiv:astro-ph/0504582 [astro-ph]}
  \BibitemShut {NoStop}%
\bibitem [{\citenamefont {Fujii}\ and\ \citenamefont
  {Maeda}(2007)}]{Fujii:2003pa}%
  \BibitemOpen
  \bibfield  {author} {\bibinfo {author} {\bibfnamefont {Y.}~\bibnamefont
  {Fujii}}\ and\ \bibinfo {author} {\bibfnamefont {K.}~\bibnamefont {Maeda}},\
  }\href {\doibase 10.1017/CBO9780511535093} {\emph {\bibinfo {title} {{The
  scalar-tensor theory of gravitation}}}},\ Cambridge Monographs on
  Mathematical Physics\ (\bibinfo  {publisher} {Cambridge University Press},\
  \bibinfo {year} {2007})\BibitemShut {NoStop}%
\bibitem [{\citenamefont {Capozziello}\ \emph {et~al.}(1996)\citenamefont
  {Capozziello}, \citenamefont {De~Ritis}, \citenamefont {Rubano},\ and\
  \citenamefont {Scudellaro}}]{Capozziello:1996bi}%
  \BibitemOpen
  \bibfield  {author} {\bibinfo {author} {\bibfnamefont {S.}~\bibnamefont
  {Capozziello}}, \bibinfo {author} {\bibfnamefont {R.}~\bibnamefont
  {De~Ritis}}, \bibinfo {author} {\bibfnamefont {C.}~\bibnamefont {Rubano}}, \
  and\ \bibinfo {author} {\bibfnamefont {P.}~\bibnamefont {Scudellaro}},\
  }\bibfield  {title} {\enquote {\bibinfo {title} {{Noether symmetries in
  cosmology}},}\ }\href {\doibase 10.1007/BF02742992} {\bibfield  {journal}
  {\bibinfo  {journal} {Riv. Nuovo Cim.}\ }\textbf {\bibinfo {volume} {19N4}},\
  \bibinfo {pages} {1--114} (\bibinfo {year} {1996})}\BibitemShut {NoStop}%
\bibitem [{\citenamefont {Paliathanasis}\ \emph {et~al.}(2014)\citenamefont
  {Paliathanasis}, \citenamefont {Tsamparlis}, \citenamefont {Basilakos},\ and\
  \citenamefont {Capozziello}}]{Paliathanasis:2014rja}%
  \BibitemOpen
  \bibfield  {author} {\bibinfo {author} {\bibfnamefont {Andronikos}\
  \bibnamefont {Paliathanasis}}, \bibinfo {author} {\bibfnamefont {Michael}\
  \bibnamefont {Tsamparlis}}, \bibinfo {author} {\bibfnamefont {Spyros}\
  \bibnamefont {Basilakos}}, \ and\ \bibinfo {author} {\bibfnamefont
  {Salvatore}\ \bibnamefont {Capozziello}},\ }\bibfield  {title} {\enquote
  {\bibinfo {title} {{Scalar-Tensor Gravity Cosmology: Noether symmetries and
  analytical solutions}},}\ }\href {\doibase 10.1103/PhysRevD.89.063532}
  {\bibfield  {journal} {\bibinfo  {journal} {Phys. Rev. D}\ }\textbf {\bibinfo
  {volume} {89}},\ \bibinfo {pages} {063532} (\bibinfo {year} {2014})},\
  \Eprint {http://arxiv.org/abs/1403.0332} {arXiv:1403.0332 [astro-ph.CO]}
  \BibitemShut {NoStop}%
\bibitem [{\citenamefont {Finelli}\ \emph {et~al.}(2008)\citenamefont
  {Finelli}, \citenamefont {Tronconi},\ and\ \citenamefont
  {Venturi}}]{Finelli:2007wb}%
  \BibitemOpen
  \bibfield  {author} {\bibinfo {author} {\bibfnamefont {F.}~\bibnamefont
  {Finelli}}, \bibinfo {author} {\bibfnamefont {A.}~\bibnamefont {Tronconi}}, \
  and\ \bibinfo {author} {\bibfnamefont {Giovanni}\ \bibnamefont {Venturi}},\
  }\bibfield  {title} {\enquote {\bibinfo {title} {{Dark Energy, Induced
  Gravity and Broken Scale Invariance}},}\ }\href {\doibase
  10.1016/j.physletb.2007.11.053} {\bibfield  {journal} {\bibinfo  {journal}
  {Phys. Lett. B}\ }\textbf {\bibinfo {volume} {659}},\ \bibinfo {pages}
  {466--470} (\bibinfo {year} {2008})},\ \Eprint
  {http://arxiv.org/abs/0710.2741} {arXiv:0710.2741 [astro-ph]} \BibitemShut
  {NoStop}%
\bibitem [{\citenamefont {Umiltà}\ \emph {et~al.}(2015)\citenamefont
  {Umiltà}, \citenamefont {Ballardini}, \citenamefont {Finelli},\ and\
  \citenamefont {Paoletti}}]{Umilta:2015cta}%
  \BibitemOpen
  \bibfield  {author} {\bibinfo {author} {\bibfnamefont {C.}~\bibnamefont
  {Umiltà}}, \bibinfo {author} {\bibfnamefont {M.}~\bibnamefont {Ballardini}},
  \bibinfo {author} {\bibfnamefont {F.}~\bibnamefont {Finelli}}, \ and\
  \bibinfo {author} {\bibfnamefont {D.}~\bibnamefont {Paoletti}},\ }\bibfield
  {title} {\enquote {\bibinfo {title} {{CMB and BAO constraints for an induced
  gravity dark energy model with a quartic potential}},}\ }\href {\doibase
  10.1088/1475-7516/2015/08/017} {\bibfield  {journal} {\bibinfo  {journal}
  {JCAP}\ }\textbf {\bibinfo {volume} {08}},\ \bibinfo {pages} {017} (\bibinfo
  {year} {2015})},\ \Eprint {http://arxiv.org/abs/1507.00718} {arXiv:1507.00718
  [astro-ph.CO]} \BibitemShut {NoStop}%
\bibitem [{\citenamefont {Ballardini}\ \emph {et~al.}(2016)\citenamefont
  {Ballardini}, \citenamefont {Finelli}, \citenamefont {Umiltà},\ and\
  \citenamefont {Paoletti}}]{Ballardini:2016cvy}%
  \BibitemOpen
  \bibfield  {author} {\bibinfo {author} {\bibfnamefont {Mario}\ \bibnamefont
  {Ballardini}}, \bibinfo {author} {\bibfnamefont {Fabio}\ \bibnamefont
  {Finelli}}, \bibinfo {author} {\bibfnamefont {Caterina}\ \bibnamefont
  {Umiltà}}, \ and\ \bibinfo {author} {\bibfnamefont {Daniela}\ \bibnamefont
  {Paoletti}},\ }\bibfield  {title} {\enquote {\bibinfo {title} {{Cosmological
  constraints on induced gravity dark energy models}},}\ }\href {\doibase
  10.1088/1475-7516/2016/05/067} {\bibfield  {journal} {\bibinfo  {journal}
  {JCAP}\ }\textbf {\bibinfo {volume} {05}},\ \bibinfo {pages} {067} (\bibinfo
  {year} {2016})},\ \Eprint {http://arxiv.org/abs/1601.03387} {arXiv:1601.03387
  [astro-ph.CO]} \BibitemShut {NoStop}%
\bibitem [{\citenamefont {Alho}\ \emph {et~al.}(2016)\citenamefont {Alho},
  \citenamefont {Carloni},\ and\ \citenamefont {Uggla}}]{Alho:2016gzi}%
  \BibitemOpen
  \bibfield  {author} {\bibinfo {author} {\bibfnamefont {Artur}\ \bibnamefont
  {Alho}}, \bibinfo {author} {\bibfnamefont {Sante}\ \bibnamefont {Carloni}}, \
  and\ \bibinfo {author} {\bibfnamefont {Claes}\ \bibnamefont {Uggla}},\
  }\bibfield  {title} {\enquote {\bibinfo {title} {{On dynamical systems
  approaches and methods in $f(R)$ cosmology}},}\ }\href {\doibase
  10.1088/1475-7516/2016/08/064} {\bibfield  {journal} {\bibinfo  {journal}
  {JCAP}\ }\textbf {\bibinfo {volume} {08}},\ \bibinfo {pages} {064} (\bibinfo
  {year} {2016})},\ \Eprint {http://arxiv.org/abs/1607.05715} {arXiv:1607.05715
  [gr-qc]} \BibitemShut {NoStop}%
\bibitem [{\citenamefont {Coley}(2003)}]{Coley:2003mj}%
  \BibitemOpen
  \bibfield  {author} {\bibinfo {author} {\bibfnamefont {A.A.}\ \bibnamefont
  {Coley}},\ }\href {\doibase 10.1007/978-94-017-0327-7} {\emph {\bibinfo
  {title} {{Dynamical systems and cosmology}}}},\ Vol.\ \bibinfo {volume}
  {291}\ (\bibinfo  {publisher} {Kluwer},\ \bibinfo {address} {Dordrecht,
  Netherlands},\ \bibinfo {year} {2003})\BibitemShut {NoStop}%
\bibitem [{\citenamefont {Garcia-Bellido}\ and\ \citenamefont
  {Quiros}(1990)}]{GarciaBellido:1990jz}%
  \BibitemOpen
  \bibfield  {author} {\bibinfo {author} {\bibfnamefont {J.}~\bibnamefont
  {Garcia-Bellido}}\ and\ \bibinfo {author} {\bibfnamefont {M.}~\bibnamefont
  {Quiros}},\ }\bibfield  {title} {\enquote {\bibinfo {title} {{Extended
  Inflation in Scalar - Tensor Theories of Gravity}},}\ }\href {\doibase
  10.1016/0370-2693(90)90954-5} {\bibfield  {journal} {\bibinfo  {journal}
  {Phys. Lett.}\ }\textbf {\bibinfo {volume} {B243}},\ \bibinfo {pages}
  {45--51} (\bibinfo {year} {1990})}\BibitemShut {NoStop}%
\bibitem [{\citenamefont {Damour}\ and\ \citenamefont
  {Nordtvedt}(1993)}]{Damour:1993id}%
  \BibitemOpen
  \bibfield  {author} {\bibinfo {author} {\bibfnamefont {T.}~\bibnamefont
  {Damour}}\ and\ \bibinfo {author} {\bibfnamefont {K.}~\bibnamefont
  {Nordtvedt}},\ }\bibfield  {title} {\enquote {\bibinfo {title} {{Tensor -
  scalar cosmological models and their relaxation toward general
  relativity}},}\ }\href {\doibase 10.1103/PhysRevD.48.3436} {\bibfield
  {journal} {\bibinfo  {journal} {Phys. Rev.}\ }\textbf {\bibinfo {volume}
  {D48}},\ \bibinfo {pages} {3436--3450} (\bibinfo {year} {1993})}\BibitemShut
  {NoStop}%
\bibitem [{\citenamefont {Mimoso}\ and\ \citenamefont
  {Nunes}(1998)}]{Mimoso:1998dn}%
  \BibitemOpen
  \bibfield  {author} {\bibinfo {author} {\bibfnamefont {J.~P.}\ \bibnamefont
  {Mimoso}}\ and\ \bibinfo {author} {\bibfnamefont {A.~M.}\ \bibnamefont
  {Nunes}},\ }\bibfield  {title} {\enquote {\bibinfo {title} {{General
  relativity as a cosmological attractor of scalar tensor gravity theories}},}\
  }\href {\doibase 10.1016/S0375-9601(98)00724-5} {\bibfield  {journal}
  {\bibinfo  {journal} {Phys. Lett.}\ }\textbf {\bibinfo {volume} {A248}},\
  \bibinfo {pages} {325--331} (\bibinfo {year} {1998})}\BibitemShut {NoStop}%
\bibitem [{\citenamefont {Serna}\ \emph {et~al.}(2002)\citenamefont {Serna},
  \citenamefont {Alimi},\ and\ \citenamefont {Navarro}}]{Serna:2002fj}%
  \BibitemOpen
  \bibfield  {author} {\bibinfo {author} {\bibfnamefont {A.}~\bibnamefont
  {Serna}}, \bibinfo {author} {\bibfnamefont {J.~M.}\ \bibnamefont {Alimi}}, \
  and\ \bibinfo {author} {\bibfnamefont {A.}~\bibnamefont {Navarro}},\
  }\bibfield  {title} {\enquote {\bibinfo {title} {{Convergence of scalar
  tensor theories toward general relativity and primordial nucleosynthesis}},}\
  }\href {\doibase 10.1088/0264-9381/19/5/302} {\bibfield  {journal} {\bibinfo
  {journal} {Class. Quant. Grav.}\ }\textbf {\bibinfo {volume} {19}},\ \bibinfo
  {pages} {857--874} (\bibinfo {year} {2002})},\ \Eprint
  {http://arxiv.org/abs/gr-qc/0201049} {arXiv:gr-qc/0201049 [gr-qc]}
  \BibitemShut {NoStop}%
\bibitem [{\citenamefont {Jarv}\ \emph {et~al.}(2011)\citenamefont {Jarv},
  \citenamefont {Kuusk},\ and\ \citenamefont {Saal}}]{Jarv:2010xm}%
  \BibitemOpen
  \bibfield  {author} {\bibinfo {author} {\bibfnamefont {Laur}\ \bibnamefont
  {Jarv}}, \bibinfo {author} {\bibfnamefont {Piret}\ \bibnamefont {Kuusk}}, \
  and\ \bibinfo {author} {\bibfnamefont {Margus}\ \bibnamefont {Saal}},\
  }\bibfield  {title} {\enquote {\bibinfo {title} {{Scalar-tensor cosmologies
  with a potential in the general relativity limit: time evolution}},}\ }\href
  {\doibase 10.1016/j.physletb.2010.09.029} {\bibfield  {journal} {\bibinfo
  {journal} {Phys. Lett.}\ }\textbf {\bibinfo {volume} {B694}},\ \bibinfo
  {pages} {1--5} (\bibinfo {year} {2011})},\ \Eprint
  {http://arxiv.org/abs/1006.1246} {arXiv:1006.1246 [gr-qc]} \BibitemShut
  {NoStop}%
\bibitem [{\citenamefont {Jarv}\ \emph {et~al.}(2012)\citenamefont {Jarv},
  \citenamefont {Kuusk},\ and\ \citenamefont {Saal}}]{Jarv:2011sm}%
  \BibitemOpen
  \bibfield  {author} {\bibinfo {author} {\bibfnamefont {Laur}\ \bibnamefont
  {Jarv}}, \bibinfo {author} {\bibfnamefont {Piret}\ \bibnamefont {Kuusk}}, \
  and\ \bibinfo {author} {\bibfnamefont {Margus}\ \bibnamefont {Saal}},\
  }\bibfield  {title} {\enquote {\bibinfo {title} {{Scalar-tensor cosmologies
  with dust matter in the general relativity limit}},}\ }\href {\doibase
  10.1103/PhysRevD.85.064013} {\bibfield  {journal} {\bibinfo  {journal} {Phys.
  Rev.}\ }\textbf {\bibinfo {volume} {D85}},\ \bibinfo {pages} {064013}
  (\bibinfo {year} {2012})},\ \Eprint {http://arxiv.org/abs/1112.5308}
  {arXiv:1112.5308 [gr-qc]} \BibitemShut {NoStop}%
\bibitem [{\citenamefont {Quartin}\ \emph {et~al.}(2008)\citenamefont
  {Quartin}, \citenamefont {Calvao}, \citenamefont {Joras}, \citenamefont
  {Reis},\ and\ \citenamefont {Waga}}]{Quartin:2008px}%
  \BibitemOpen
  \bibfield  {author} {\bibinfo {author} {\bibfnamefont {Miguel}\ \bibnamefont
  {Quartin}}, \bibinfo {author} {\bibfnamefont {Mauricio~O.}\ \bibnamefont
  {Calvao}}, \bibinfo {author} {\bibfnamefont {Sergio~E.}\ \bibnamefont
  {Joras}}, \bibinfo {author} {\bibfnamefont {Ribamar R.~R.}\ \bibnamefont
  {Reis}}, \ and\ \bibinfo {author} {\bibfnamefont {Ioav}\ \bibnamefont
  {Waga}},\ }\bibfield  {title} {\enquote {\bibinfo {title} {{Dark Interactions
  and Cosmological Fine-Tuning}},}\ }\href {\doibase
  10.1088/1475-7516/2008/05/007} {\bibfield  {journal} {\bibinfo  {journal}
  {JCAP}\ }\textbf {\bibinfo {volume} {0805}},\ \bibinfo {pages} {007}
  (\bibinfo {year} {2008})},\ \Eprint {http://arxiv.org/abs/0802.0546}
  {arXiv:0802.0546 [astro-ph]} \BibitemShut {NoStop}%
\bibitem [{\citenamefont {Aghanim}\ \emph {et~al.}(2020)\citenamefont {Aghanim}
  \emph {et~al.}}]{Aghanim:2018eyx}%
  \BibitemOpen
  \bibfield  {author} {\bibinfo {author} {\bibfnamefont {N.}~\bibnamefont
  {Aghanim}} \emph {et~al.} (\bibinfo {collaboration} {Planck}),\ }\bibfield
  {title} {\enquote {\bibinfo {title} {{Planck 2018 results. VI. Cosmological
  parameters}},}\ }\href {\doibase 10.1051/0004-6361/201833910} {\bibfield
  {journal} {\bibinfo  {journal} {Astron. Astrophys.}\ }\textbf {\bibinfo
  {volume} {641}},\ \bibinfo {pages} {A6} (\bibinfo {year} {2020})},\ \Eprint
  {http://arxiv.org/abs/1807.06209} {arXiv:1807.06209 [astro-ph.CO]}
  \BibitemShut {NoStop}%
\bibitem [{\citenamefont {Pettorino}\ and\ \citenamefont
  {Baccigalupi}(2008)}]{Pettorino:2008ez}%
  \BibitemOpen
  \bibfield  {author} {\bibinfo {author} {\bibfnamefont {Valeria}\ \bibnamefont
  {Pettorino}}\ and\ \bibinfo {author} {\bibfnamefont {Carlo}\ \bibnamefont
  {Baccigalupi}},\ }\bibfield  {title} {\enquote {\bibinfo {title} {{Coupled
  and Extended Quintessence: theoretical differences and structure
  formation}},}\ }\href {\doibase 10.1103/PhysRevD.77.103003} {\bibfield
  {journal} {\bibinfo  {journal} {Phys. Rev. D}\ }\textbf {\bibinfo {volume}
  {77}},\ \bibinfo {pages} {103003} (\bibinfo {year} {2008})},\ \Eprint
  {http://arxiv.org/abs/0802.1086} {arXiv:0802.1086 [astro-ph]} \BibitemShut
  {NoStop}%
\bibitem [{\citenamefont {Peixoto}(1962)}]{pei}%
  \BibitemOpen
  \bibfield  {author} {\bibinfo {author} {\bibfnamefont {M.M.}\ \bibnamefont
  {Peixoto}},\ }\bibfield  {title} {\enquote {\bibinfo {title} {Structural
  stability on two-dimensional manifolds},}\ }\href {\doibase
  https://doi.org/10.1016/0040-9383(65)90018-2} {\bibfield  {journal} {\bibinfo
   {journal} {Topology}\ }\textbf {\bibinfo {volume} {1}},\ \bibinfo {pages}
  {101--120} (\bibinfo {year} {1962})}\BibitemShut {NoStop}%
\end{thebibliography}%
\end{document}